\documentclass[sn-basic]{sn-jnl}

\usepackage{graphicx}%
\usepackage{multirow}%
\usepackage{amsmath,amssymb,amsfonts}%
\usepackage{amsthm}%
\usepackage{mathrsfs}%
\usepackage[title]{appendix}%
\usepackage[table]{xcolor}%
\usepackage{textcomp}%
\usepackage{manyfoot}%
\usepackage{booktabs}%
\usepackage{algorithm}%
\usepackage{algorithmicx}%
\usepackage{algpseudocode}%
\usepackage{listings}%

\usepackage{colortbl}%

\usepackage{microtype}%

\usepackage{orcidlink}

\newcommand*\rot{\rotatebox{90}}

\newcommand{\footnoteremark}{\addtocounter{footnote}{-1}\footnotemark}

\newcommand{\tfnh}[1]{\footnotesize{\textit{\textbf{\textsuperscript{#1}}}}}
\newcommand{\tfnt}[2]{\footnotesize{\textit{\textbf{\textsuperscript{#1}}}#2}}

\newcommand{\etal}{~et al.}%
\newcommand{\citetp}[1]{\citeauthor{#1}'s \citeyearpar{#1}}
\newcommand{\citeeg}[1]{\citep[e.g.][]{#1}}
\newcommand{\citeta}[1]{\citetalias{#1}~\citeyearpar{#1}}
\newcommand{\citetap}[1]{\citetalias{#1}'s~\citeyearpar{#1}}

\begin{document}

\title[A Critical Review of AI Apology]{AI Apology: A Critical Review of Apology in AI Systems}


\author*[1,2]{\fnm{Hadassah}~\sur{Harland}\orcidlink{0000-0002-3425-7846}}\email{h.harland@research.deakin.edu.au} 
\author[1]{ \fnm{Richard}~\sur{Dazeley}\orcidlink{0000-0002-6199-9685}}\email{richard.dazeley@deakin.edu.au} 
\author[2]{ \fnm{Hashini}~\sur{Senaratne}\orcidlink{0000-0001-5203-3793}}\email{Hashini.Senaratne@data61.csiro.au} 
\author[3]{ \fnm{Peter}~\sur{Vamplew}\orcidlink{0000-0002-8687-4424}}\email{p.vamplew@federation.edu.au} 
\author[4,5]{ \fnm{Francisco}~\sur{Cruz}\orcidlink{0000-0002-1131-3382}}\email{f.cruz@unsw.edu.au}
\author[1]{ \fnm{Bahareh}~\sur{Nakisa}\orcidlink{0000-0003-2211-2997}}\email{bahar.nakisa@deakin.edu.au} 
\affil*[1]{\orgname{Deakin University}, \orgaddress{\city{Geelong}, \state{Victoria}, \country{Australia}}}
\affil[2]{\orgdiv{Data61}, \orgname{Commonwealth Scientific and Industrial Research Organization}, \orgaddress{\city{Pullenvale}, \state{Queensland}, \country{Australia}}} 
\affil[3]{\orgname{Federation University}, \orgaddress{\city{Ballarat}, \state{Victoria}, \country{Australia}}}
\affil[4]{\orgname{University of New South Wales}, \orgaddress{\city{Sydney}, \state{New South Wales}, \country{Australia}}}
\affil[5]{\orgdiv{Escuela de Ingenieria}, \orgname{Universidad Central de Chile}, \orgaddress{\city{Santiago}, \country{Chile}}}


\abstract{
Apologies are a powerful tool used in human-human interactions to provide affective support, regulate social processes, and exchange information following a trust violation. The emerging field of AI apology investigates the use of apologies by artificially intelligent systems, with recent research suggesting how this tool may provide similar value in human-machine interactions. Until recently, contributions to this area were sparse, and these works have yet to be synthesised into a cohesive body of knowledge. This article provides the first synthesis and critical analysis of the state of AI apology research, focusing on studies published between 2020 and 2023. We derive a framework of attributes to describe five core elements of apology: outcome, interaction, offence, recipient, and offender. With this framework as the basis for our critique, we show how apologies can be used to recover from misalignment in human-AI interactions, and examine trends and inconsistencies within the field. Among the observations, we outline the importance of curating a human-aligned and cross-disciplinary perspective in this research, with consideration for improved system capabilities and long-term outcomes. 
} %





\maketitle


\section{Introduction}\label{Int} 
In recent years, the use of apology in Artificially Intelligent (AI)\footnote{Here, we use the term ``AI'' to describe the various forms of artificial agents in its broadest definition: intelligent machines, robots, computer programs and similar systems that behave with some perceived sense of autonomy. Social robots, mental health chatbots, and disaster response drones are a few examples where interactions with human users form an inherent part of the system's role.} systems has seen growing interest, driven by the increased presence of these systems in interactive roles with human users~\citep{Dautenhahn2015}. For example, as a barista robot~\citep{Wright2022}, a home support companion~\citep{Tewari2022}, or a customer complaint management chatbot~\citep{Liu2020}. Such roles often encounter challenges arising from a misalignment between the system and its user~\citep{Borboni2023, Semeraro2022, Shneiderman2020, Taylor2020}. Misalignment between the agent and the user's individual goals, or in the approach applied in pursuit of a mutual goal, may result in conflict~\citep{Babel2021}. Misalignment between the user's intentions and the agent's interpretation of the directions provided may cause the agent to exhibit unexpected and undesirable behaviour~\citep{Amodei2016}. Misalignment between the system's capabilities and the user's expectations may cause frustration and disappointment~\citep{Tolmeijer2020}, or in the case of anthropomorphic cues, psychological discomfort~\citep{Ho2017}. The success of AI systems within such roles relies on the ability of the system to address these challenges and restore appropriate alignment with the needs and values of its human users~\citep{Glikson2020}. 

\textit{Alignment} in AI refers to the body of research that seeks to address how increasingly advanced intelligent systems can align with their users, including how to address challenges with misalignment that impede effective collaboration and cohabitation~\citep{Soares2014}. This requires the system to demonstrate tactful navigation of the user's needs, within its limitations, to fulfil its core purpose~\citep{Vamplew2018}. In the dynamic context of human interaction, the system needs the capability to respond to changing circumstances. This includes an awareness of the user, and the ability to address and recover from issues when they occur~\citep{Mutlu2016}.

Apology is a social exchange that is pivotal in human-human interactions~\citep{Smith2008}, playing an affective, regulatory and informative role to enable recovery following a trust violation. In human-AI interactions, when used appropriately, it represents a proactive and user-centric approach to resolve issues and realign the system to the user. A suitably applied apology demonstrates a persuasive affective influence to soothe disharmony and support social relationships in both human~\citep{Schlenker1981} and sociotechnical systems~\citep{Park2012}. An AI system that apologises following undesirable behaviour may be seen as more trustworthy, credible and easy to work with~\citep{Lohan2014}. Furthermore, apology is a dynamic communication tool through which a system may demonstrate transparency and accountability, as well as provide practical support for future interactions~\citep{Harland2023}. It provides an effective framework for presenting situationally relevant information to the user, such as the extent of the system's understanding of the impacts of its behaviour, its functional limitations, and any other information of which the user may be unaware~\citep{Honig2018}. In summary, apology represents both a catalyst for and manifestation of the pursuit of human-alignment in AI systems.

Foundational research in the exploration of AI apology emerged in the 2000s within the field of Human-Robot Interaction (HRI)~\citep{Lee2010}, as a recovery strategy to mitigate the adverse effects of breakdowns by service robots. Prior work had laid the foundations for considering apology for managing imperfect systems by identifying the need for these systems to enable an appropriate level of user reliance~\citep{Lee2004} and proactive transparency~\citep{Kim2006}. Subsequent early research found that AI apology can be an effective approach for repairing trust following a trust violation~\citep{Hamacher2015, Hamacher2016, Shiomi2013, Lohan2014, Stowers2017} and began investigating the effects of its varying properties, such as the timing of the apology~\citep{Robinette2015, Robinette2017}. However, this research encountered mixed effects, such as a limited restoration of the agent's perceived intelligence despite effective recovery of its likeability~\citep{Engelhardt2017}. Preliminary studies indicate the presence of complex nuances and interaction effects between the needs and preferences of the user, and the specific features of the apology~\citeeg{Lee2011, Mutlu2016, Sebo2019}. Since then, interest in this work has escalated with investigations seeking to understand these effects. Further encouraged by a number of calls for research to advance the capabilities of AI systems like robots in managing increasingly complex and dynamic interactions~\citep{Baker2018, deVisser2018}, there has been substantial growth in the field in recent years. 

Studies in the fields of HRI and Human-Computer Interaction (HCI) are central to the study of apology in technology systems: investigating the use of robotics and other human–machine interfaces, and the design features and behaviour of those interactive technologies. However, a comprehensive implementation of apology in an autonomous system involves many broad-reaching considerations, including areas such as identification (to decide when to initiate an apology), reasoning and explanation (to construct the apology itself), and system behaviour (to correct the issue)~\citep{Mutlu2016}. Thus, a multidisciplinary approach to incorporate and integrate social, theoretical, and technical contributions is essential. Other less represented features of this field of research span broad topics, including algorithm design, causal reasoning, system personalisation, as well as ethics via accountability and transparency, and interpretability. As research regarding the use of apology in AI systems is still emerging, it is yet to be considered from a holistic standpoint. As such, between-study comparisons of the effectiveness of various apologetic approaches are limited~\citep{Esterwood2022b, Song2023}, and contributions towards technical realisation of associated system capabilities are sparse~\citep{Harland2023}. 

This paper seeks to address this discordance and bring these various disjoint features into focus. The application of human-oriented social behaviours in HCI contexts is underwritten by the Computers-are-Social-Actors paradigm~\citep{deVisser2018, Lee2010}, which maintains that humans subconsciously apply similar social scripts to their interactions with technology as they do with other humans~\citep{Nass1994}. Applying this theory in social technologies research enables existing knowledge of human-human behaviours to provide insight into human-AI interactions, acting as a baseline for this research and simplifying many broader questions into the easier task of identifying similarities or differences from the human case~\citep{Baker2018, Madhavan2007}. On this premise, we draw upon the established literature on human apology and integrate this with HCI theory to establish a baseline to enable our critique. 

In this work, we present a framework for AI apology, survey and critique the present literature, and provide suggestions for future work in the field. First, we establish our framework to encompass the core concepts and associated vocabulary for AI apology (Section \ref{def:apol}). Next, we conduct a comprehensive review of studies that contribute to the AI apology literature, including experimental studies, theory development, and technical implementation (Section \ref{meth}). We provide a critique by examining each of the proposed attributes of AI apology in the literature, reporting upon the contributing theories and identifying discrepancies in how features have been described and measured, for a thorough exposition of the field (Section \ref{res}). The work concludes with a discussion on gaps and opportunities for further technical, experimental and theoretical development, presenting four key areas where further research is needed alongside some recommendations to facilitate improved internal alignment in the field (Section \ref{disc}). 


\section{What is an apology?}\label{def:apol}
\begin{figure}[!h]
\centering
\includegraphics[width=57.12mm]{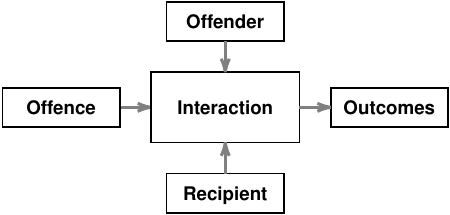} 
\caption[Elements of an apology]{An apology is described by five elements: the \textit{interaction} between the \textit{offender} and the \textit{recipient} regarding an \textit{offence} towards an \textit{outcome}. Arrows describe how these elements influence each other}
\label{facets}
\end{figure}

An apology is a social exchange that features prominently in human-human interactions as a means to transform and regulate relationships between engaging parties following harmful behaviour~\citep{Aydin2013}. In its natural form, it is an \textit{interaction} (or \textit{remedy}) undertaken by an \textit{offender}, following and with regard to an \textit{offence}, in the pursuit of reconciliation with the \textit{recipient} (or \textit{victim} or \textit{offended})~\citep{Almuttalibi2016, Searle1969, Tavuchis1993}. These represent the four key elements that collectively describe an apology. The apology itself is an expression\footnote{Specifically, an apology is a speech act known as an illocution~\citep{Leech2002, Searle1969}; meaning that its utterance is the performance of its function.}~\citep{Leech2002, Searle1969} that conveys the intentions, emotions, actions and reactions~\citep{Homans1961} of the \textit{offender} to the \textit{recipient}, as an acknowledgement and renunciation of the \textit{offence} that was committed against them. The \textit{outcomes} of the apology can be regarded as a fifth element to enable a complete description. These elements and their influence on each other can be visualised in an interaction diagram, as shown in Figure \ref{facets}. 

While there is no singularly agreed-upon definition for what constitutes an apology, the exchange can be functionally described through the attributes of each of the core elements. Human apology research is an extensive and multidisciplinary field, encompassing various areas of ethics, linguistics and social science. The associated research methodologies include both theoretical and empirical perspectives, utilising philosophical~\citep{Gill2000, Kort1975, Lazare2004, Smith2008, Tavuchis1993}, quantitative~\citep{Cohen1981, Lewicki2016, Scher1997}, qualitative~\citep{Blum-kulka1984, Slocum2011}, and mixed-method~\citep{Aydin2013} analyses. Two paradigms emerge from this research: a prescriptive view that converges towards a compositional definition, and a descriptive view that emphasises the influence of contextual variations~\citep{Smith2008}. 

\begin{figure}[!h]
\centering
\includegraphics[width=55.44mm]{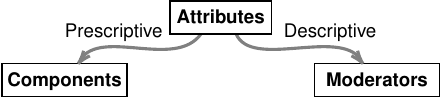} 
\caption[Attributes of apology]{Attributes of an apology consist of prescriptive components and descriptive moderators}
\label{fig:att}
\end{figure}

We draw upon these two paradigms collectively to describe the attributes of an apology. The prescriptive view suggests that an apology can be described as a combination of distinct parts that catalogue what is communicated during the \textit{interaction}~\citep{Smith2008}. We refer to them as \textit{components} hereinafter. The descriptive view suggests that an apology is described by a set of divergent properties, related to the \textit{interaction}, \textit{offence}, \textit{recipient}, and \textit{offender}, that influence its suitability and reception~\citep{Tavuchis1993}. We refer to them as \textit{moderators} hereinafter. Thus, we can describe an apology according to its attributes as being the combination of the prescriptive \textit{components} and descriptive \textit{moderators} (Figure \ref{fig:att}).

We consult this research to describe the \textit{outcomes} of an apology, and catalogue the key attributes of each of the four remaining elements: \textit{interaction}, \textit{offence}, \textit{recipient}, and \textit{offender}. We provide consideration for the \textit{offender} in the AI case by incorporating findings from the HCI literature, including a brief derivation of the necessary capabilities for an AI system to be able to apologise. We present these results as a framework of proposed considerations relevant to the study of AI apology. 


\subsection{Outcomes of apology} \label{apol:out}
The prospective \textit{outcomes} of an apology are what motivates its use. They arise from within a broad set of functions that can be summarised according to three core areas of effect: \textit{affective}, \textit{regulatory}, and \textit{informative}. 

A suitably applied apology can provide \textit{affective} support by meeting the psychological needs of the victim, such as giving assurance, restoring dignity, and providing an avenue for remediation~\citep{Lazare2004}. It can contribute towards feelings of satisfaction, foster a sense of validation, comfort, and of being understood for the victim, and promote the restoration of trust in the relationship~\citep{Allan2006, Bachman2006}. 

An apology may also \textit{regulate} the social and moral standing of the offender, as a demonstration of their suitability to engage in society. Through the presentation of their account, the offender may articulate their understanding of the inappropriateness of their behaviour and their reformed attitude~\citep{Gill2000, Scher1997}. This process can be used to restore and maintain a collective social standard of conduct~\citep{Eslami2010, Scher1997}, as a demonstration of polite behaviour~\citep{Aydin2013}, to relieve social friction~\citep{Wagatsuma1986}, and promote forgiveness~\citep{Allan2006} in the victim. Furthermore, it can act as a test and demonstration of the offender's moral capacity~\citep{Bennet2008}, through the recognition, admission and renunciation of harmful behaviour arising from a moral deficit~\citep{Gill2000}. 

Apologising also provides a mechanism through which \textit{information} can be exchanged~\citep{Leech2002, Searle1969}. In the apology, the offender may reveal their underlying knowledge, beliefs, and intents by expressing their understanding of the preceding events and the motivations for their behaviour. The apology also indicates the offender's perception of the recipient's wants and needs, both in the impression of what harm was caused through which behaviours and the remedial actions selected~\citep{Homans1961}. The recipient can use this information to help identify where the misalignment occurred, calibrate their trust in the offender, and refine their expectations for future behaviour~\citep{Lewicki2012, Scher1997}. 

These functions may be understood as acts of realignment. The affective function acts to restore emotional alignment between the offender and the victim: lowering the affective standing of the former through the demonstration of humility, raising the affective standing of the latter by providing support, and using empathy to share in the burden of the offence. The regulatory function acts as a defence or reinstatement of the alignment of the offender's behaviour with the recipient's existing expectations. Conversely, the informative function aligns the recipient's expectations with the offender's behaviour. These processes act in parallel; an apology that addresses how the cause of the offence will be realigned is both regulatory and informative. 

A related frame to describe the function of an apology is based on the premise of trust, defined as the willingness to be vulnerable to a trustee based on positive expectations of their behaviour and intentions~\citep{Colquitt2007}. Apology is one way in which an offender may seek to restore trust that has been violated by the offence for which the offender is apologising~\citep{Kim2004}. Trust is described by two dimensions: affect-based and cognition-based~\citep{Cacioppo1986}, and three trustworthiness antecedents: benevolence, integrity, and ability~\citep{Mayer1995}. While these concepts and the functions of apology presented do not directly align, there are correlations between these three frames~\citep{Tomlinson2020}. Cognitive trust arises from knowledge~\citep{Cacioppo1986} and is most strongly predicted by ability and integrity~\citep{Tomlinson2020}, corresponding to informative and regulatory functions, respectively. Affective trust is most strongly predicted by benevolence, influenced by affective functions. In this way, apology may also be described as a trust repair strategy~\citep{Lewicki2012}. 

The concept of trust is also emphasised in HCI literature regarding alignment in AI systems. \citet{Lee2004} suggest that an appropriate level of trust is essential for effective interactions between humans and agents~\citep[see also][]{Krausman2022}. Too little trust (described as \textit{undertrust}) impedes user acceptance, while too much trust (described as \textit{overtrust}) may lead to inappropriate reliance that puts the user at risk~\citep[see also][]{Ullrich2021}. Trust in technological systems is recognised as a multifaceted sociotechnical construct~\citep{Kok2020, Milanovic2021}, broadly used to describe concepts such as users' reliance on the system~\citep{Esterwood2022b, Kok2020}. While dynamic models of human-AI trust are beginning to emerge~\citep{deVisser2020}, they maintain a likeness to the human-human case. 


\subsection{Attributes of the interaction} 

\begin{table}
\setlength{\tabcolsep}{2.5pt}%
\caption[Components of Apology]{A Description of each of 12 Components of AI Apology, with an example \\ \\
\textbf{Example Brief } Ashley has just ordered a coffee at a local café and is settling in to read a book. The server, Taylor, approaches the table carrying the coffee as ordered. On arrival, however, Taylor becomes distracted by a cell phone notification and carelessly tips the serving tray. The coffee cup tips over, and hot coffee spills out over Ashley's book. Taylor's subsequent actions and speech are as described in the rightmost column: \textit{Contextual Example}. The text within the shaded highlight is intended to be read as a continuous quote}\label{tab:comp}
\begin{tabular*}{\textwidth}{p{7pt} p{65pt} p{135pt} @{\hskip10pt} p{135pt}} 

\toprule
\multicolumn{2}{l}{Component} & Description & Contextual Example \\%
\midrule%
 & & & \hspace{-4pt} \colorbox{blue!8}{\makebox(135pt,195pt)[l]{\strut}} \\[-200pt]%
\multirow{9}{7pt}{\rot{Explicit}}%
 & Cue
 & Indicative words or phrases such as ``sorry'' or ``apologise''%
 & ``I am so \textbf{sorry}! \ldots%
 \\\cmidrule(l{2.5pt}r{2.5pt}){2-3}
 & Responsibility 
 & Appropriate self-assignment of blame%
 & ``I was being incredibly careless \ldots 
 \\\cmidrule(l{0pt}r{7.5pt}){2-3}
 & Affirmation
 & Demonstrate awareness of the issue and its impact%
 & ``I've damaged your book, and disrupted your reading time. \ldots%
 \\\cmidrule(l{2.5pt}r{7.5pt}){2-3}
 & Explanation 
 & Provide further context to clarify the behaviour 
 & ``I wasn't watching where I was going because I was checking my phone. \ldots
 \\\cmidrule(l{0pt}r{7.5pt}){2-3}
 & Moral Admission 
 & Recognition of any moral aspects of the harmful behaviour%
 & ``It was irresponsible of me to be distracted while working\ldots %
 \\\cmidrule(l{0pt}r{7.5pt}){2-3}
 & Regret 
 & Express remorse and desire that the offence had not occurred%
 & ``\ldots and I feel truly awful about what happened. \ldots%
 \\\cmidrule(l{0pt}r{7.5pt}){2-3}
 & Reform%
 & Express the resolve to improve and prevent the offence repeating%
 & ``I will keep my phone away so won't happen again. \ldots 
 \\\cmidrule(l{0pt}r{7.5pt}){2-3}
 & Repair%
 & Propose actions to counteract the offence%
 & ``Please let me replace your book''%
 \\\cmidrule(l{0pt}r{7.5pt}){2-3}
 & Petition 
 & Request for forgiveness and to put the transgression to rest
 & ``\ldots and forgive my carelessness \ldots''%
 \\\midrule%
\multirow{3}{7pt}{\rot{Implicit}}%
 & Dialogue 
 & Interacting with the victim: listening, reciprocating and accommodating 
 & Taylor uses prompts to facilitate Ashley's engagement, and pays attention to their reactions%
 \\\cmidrule(l{0pt}r{7.5pt}){2-3}
 & Engagement 
 & Expressing interest in the victim's well-being: empathy, respect, emotive language
 & The apology is personable and indicates concern for the impact on Ashley 
 \\\cmidrule(l{0pt}r{7.5pt}){2-3}
 & Demonstration
 & Following through with the commitments of reform and repair%
 & Taylor replaces the book and puts the phone away. 
 \\%
\botrule
\end{tabular*}
\end{table}

\subsubsection{Components of an apology}\label{apol:intcomp}
We have identified twelve components within the human apology literature associated with effective apologies (see Table \ref{tab:comp}). These components can be further described according to their presentation as either explicitly present, or implied through ancillary communication modes such as emotional expression and body language. \textit{Explicit components} are clearly articulated within the spoken or written body of the apology. Conversely, \textit{implicit components} might be communicated within the presentation of explicit components or be inherent to the shared mutual understanding between the engaging parties~\citep{Cohen2020}. However, the fluidity of human communication dictates that this is not a strict designation~\citep{Kort1975}. 

Nine of these components have been described with an explicit form, given their common representations. Five prominent instances include a statement of apology or \textit{cue}\footnote{In linguistics, this is formally defined as an \textit{illocutionary force indicating device} (IFID)~\citep{Austin1975, Blum-kulka1984, Scher1997}.}~\citep[e.g.][]{Aydin2013, Blum-kulka1984, Cohen1981, Gill2000, Scher1997}, acceptance of \textit{responsibility}~\citep[e.g.][]{Blum-kulka1984, Cohen1981, Gill2000, Kort1975, Lazare2004, Lewicki2012, Lewicki2016, Scher1997, Slocum2011, Smith2005, Smith2008, Tavuchis1993}, an \textit{explanation} or account~\citep[e.g.][]{Blum-kulka1984, Lazare2004, Lewicki2012, Lewicki2016, Scher1997, Slocum2011}, a promise of \textit{reform} or forbearance~\citep[e.g.][]{Blum-kulka1984, Cohen1981, Gill2000, Lazare2004, Lewicki2012, Lewicki2016, Scher1997, Slocum2011, Smith2008, Tavuchis1993}, and an offer of \textit{repair}~\citep[e.g.][]{Blum-kulka1984, Cohen1981, Lewicki2012, Lewicki2016, Scher1997, Slocum2011}. An expression of \textit{regret} for the harm is also often included~\citep[e.g.][]{Blum-kulka1984, Cohen1981, Gill2000, Kort1975, Lazare2004, Lewicki2012, Lewicki2016, Scher1997, Slocum2011, Smith2005, Smith2008, Tavuchis1993}, whereas a \textit{petition} for forgiveness was comparatively less common~\citep[e.g.][]{Allan2006, Lewicki2012, Lewicki2016, Tavuchis1993}. Where relevant, an apology may also involve a \textit{moral admission}~\citep[e.g.][]{Choi2009, Gill2000, Kort1975, Lazare2004, Smith2005, Smith2008, Tavuchis1993}.

The last of these nine is a collation of ideas based on another popular theme arising in the literature, focused on prescribing features of a comprehensive apology is of explicitly addressing the details of the offence~\citep[e.g.][]{Choi2009, Gill2000, Kort1975, Slocum2011, Smith2005, Smith2008, Tavuchis1993}. Extended analyses such as that by \citet{Smith2008} distinguish components such as a corroboration of the facts, and identification and validation of the transgression. In reasonable use, however, the division between these three components can be indistinct, and can be more succinctly described as an \textit{affirmation}~\citep{Slocum2011}. 

The three remaining components have been described with an implicit form, as this is the form in which they are likely to present. These are an interactive \textit{dialogue} with the victim~\citep[e.g.][]{Bachman2006, Frantz2005, Slocum2011, Smith2005, Smith2008, Wilson2020}, affective \textit{engagement}~\citep[e.g.][]{Bachman2006, Blum-kulka1984, Kort1975, Lazare2004, Roschk2013, Slocum2011, Wilson2020}, and the \textit{demonstration} of what was promised through changed behaviour~\citep[e.g.][]{Bennet2008, Gill2000, Hui2011, Kador2009, Lazare2004, Schweitzer2006, Slocum2011}. 

The example in Table \ref{tab:comp} presents each component of apology through explicit and segregated text to clearly articulate the associated concept. As a result, it is more extensive and formal than would be expected for the circumstance. In practice, an apology will likely only include a subset of these components. The relevance of each component for a given apology depends on both how it is applied, and the context of the offence and the circumstances leading to it~\citep{Lazare2004}. Furthermore, the receiver may emphasise different components~\citep{Fehr2010}. In some circumstances, shared context can be leveraged to alter or simplify the present apology. For example, an affirmation may be unnecessary if the harm is obvious, and social roles may dictate that it is not useful for the victim to understand why the mistake occurred (explanation) or how it will be avoided (reform), so long as the present issue is resolved. The offender may also substitute vocalisations with actions, such as expressing regret and shame through chastened body language, and proactively demonstrating repair and reform without express articulation~\citep{Lazare2004}. 
In this context, a more succinct apology might be:

\begin{quote}
``I am so sorry, that was careless of me. I will replace this [the book].''
\end{quote}
A further consideration regarding these components is that they can often be imprecisely defined in the literature, without a clear consensus as to the requirements for its designation. This has recognised implications for human apology research~\citep{Slocum2011}, reflecting inconsistent representations of these concepts within the literature. A notable example is \textit{regret}, often described as a psychological state reflecting the desire to have made a prior decision differently, as to have brought about a different result~\citep{Gill2000}. As a component of apology, it can be conflated with a cue~\citep{Blum-kulka1984}, an expression of responsibility~\citep{Kort1975}, or a petition~\citep{Kort1975}. While not all expressions of regret constitute an apology, Kort insists that it is an inherent and necessary component of apology, and that the apology is specific as to what the regret is for~\citep{Kort1975, Lazare2004}. However, despite extensive consultation with relevant literature on apology, we cannot provide a clear and precise prescription for what constitutes regret in an apology, and acknowledge the impact of this imprecision on our present analysis.

The implicit component of \textit{engagement} is another such case. Here, it describes an aggregation of empathetic, emotional, and expressive characteristics of an apology, as distinct from the explicit articulate components otherwise described. The role of engagement is to connect with the recipient on an affective level, to demonstrate care and concern for their well-being, to understand and value their needs~\citep{Slocum2011}, and to affirm that they are entitled to respect~\citep{Kort1975}. Humans communicate engagement in an apology through verbal content~\citep{Roschk2013} as well as non-verbal cues such as body language, gestures, and facial expressions~\citep{Kort1975, Lazare2004, Slocum2011}. 

\subsubsection{Moderators of the interaction}
Moderators of the interaction describe attributes related to the delivery. We have distinguished six such moderators, encompassing concepts ranging from highly abstract to distinct and measurable. 

The first of these are the perceived \textit{sincerity} of the apology, describing the perceived authenticity, honesty, and genuineness of the expression, indicative of the alignment between the actions of the offender and their intentions, values and beliefs~\citep{Allan2006, Lazare2004, Liddle2017, Smith2005, Smith2008}. The second is the \textit{intensity} of an apology~\citep{Blum-kulka1984, Cohen1981, Roschk2013, Smith2008}, which describes the strength of illocutionary force that it imparts~\citep{Fedoryuk2019}. It can be influenced by aspects of phrasing, delivery, and length of an expression~\citep{Blum-kulka1984, Bowers1964}. For example, it can be increased through the use of amplifiers (``I am \textit{really} sorry'') or decreased via downtoners (``I \textit{may have} made a mistake'')~\citep{Bowers1964, Fedoryuk2019}. It is also positively correlated with both the use of expressive language, such as metaphor, and more formal language~\citep{Bowers1964}. These abstract aspects of delivery are susceptible to being influenced by other contextual elements. 

Two further moderators are associated with the cost invested in recovery. The \textit{specificity} of an apology describes its applicability to the present offence and the quality of the information that it contains~\citep{Joyce1999, Lazare2004, Slocum2011}. It is a reflection of the depth to which the offender has engaged. Similarly, \textit{compensation} describes what was offered to restore equity and bring balance to the relationship~\citep{Fehr2010, Liddle2017, Slocum2011}. Compensation is not limited to repair actions, but rather describes a much broader set of ideas and remedial measures that the offender might employ, both material (i.e. economic recovery, e.g. reimbursement of costs) and immaterial (i.e. socio-emotional recovery, e.g. a public display of respect)~\citep{Cohen2016}. Descriptions of these moderators encompassing considerations of form, applicability, and magnitude. 

The final two moderators related to the interaction are the \textit{timing}~\citep{Frantz2005, Im2021, Lazare2004, Roschk2013, Tavuchis1993} and \textit{mode}~\citep{Joyce1999, Lazare2004, Slocum2011} of the apology. These aspects of delivery are changeable by the offender, subject to limitations of the available resources and extent of control the offender has over the interaction, respectively. These moderators have been further described in Table \ref{mod}.

\begin{table}
\setlength{\tabcolsep}{2.5pt}%
\caption[Descriptive Attributes of Apology: Moderators]{Descriptive Attributes of Apology: Moderators, with examples}\label{mod}
\begin{tabular*}{\textwidth}{p{7pt} p{65pt} p{135pt} @{\hskip10pt} p{135pt}} 
\toprule%
\multicolumn{2}{l}{Moderator} & Description & Examples \\%
\midrule%
\multirow{6}{7pt}{\rot{Interaction}}%
 & Sincerity%
 & Authenticity and genuineness, as interpreted through social cues%
 & Open body language, eye contact, intonations for delivery%
 \\\cmidrule(l{2.5pt}r{7.5pt}){2-3}%
 & Intensity%
 & Applying diminutive or intensifying language to adjust emphasis
 & as in: ``I feel \textit{truly} awful''%
 \\\cmidrule(l{2.5pt}r{7.5pt}){2-3}%
 & Specificity%
 & Relevance and applicability of the described features or applied repair 
 & Meaningful explanation, economic repair for a financial loss%
 \\\cmidrule(l{2.5pt}r{7.5pt}){2-3}
 & Compensation%
 & The presence and extent of a financial, social or temporal investment 
 & Financial compensation, public acknowledgement%
 \\\cmidrule(l{2.5pt}r{7.5pt}){2-3}%
 & Timing%
 & Response time appropriately calibrated to the circumstances%
 & Urgency, allowing adequate time for contemplation
 \\\cmidrule(l{2.5pt}r{7.5pt}){2-3}%
 & Mode%
 & Manner of delivery 
 & Spoken, written, personable or through a messenger%
 \\
\midrule
\multirow{2}{7pt}{\rot{Offence}}%
 & Offence Type%
 & What underlying deficiency allowed the offence to occur 
 & Negligence, lack of integrity, incompetence %
 \\\cmidrule(l{2.5pt}r{7.5pt}){2-3}%
 & Severity%
 & The extent of the danger or inconvenience of the offence%
 & Damages, frustration, risk of serious injury or threat to life
 \\
\midrule
\multirow{2}{7pt}{\rot{Recipient}}%
 & Disposition%
 & The influence of an individual person's attitudes, values and beliefs%
 & self-esteem, present mood, patterns of behaviour%
 \\\cmidrule(l{2.5pt}r{7.5pt}){2-3}%
& Identity%
 & Group-based features of self: cultural nuances in language and behaviour, and demographics
 & Semantic variations, preferences for politeness, formality 
 \\
\midrule
\multirow{2}{7pt}{\rot{Offender}}%
 & Embodiment\tfnh{a}
 & Agent's physical form 
 & Physical robot or physically intangible program interface%
 \\\cmidrule(l{2.5pt}r{7.5pt}){2-3}
 & Anthropomor- phism\tfnh{a} 
 & Application of human-like features 
 & Humanoid form, naturalistic voice%
 \\%
\botrule%
\end{tabular*}%
\tfnt{a}{Adapted to more descriptive of the AI case}
\end{table}


\subsection{Attributes of the recipient, offence, and offender} 

\subsubsection{Moderators of the offence}
The context of the offence influences what characteristics of an apology are suitable to address the harm that occurred. We have distinguished two moderators that regard properties of the offence: \textit{offence type}~\citep{Joyce1999, Kador2009, Roschk2013} and \textit{severity}~\citep{Smith2008, Slocum2011}. The \textit{offence type} describes the form and context of the harm, such as why the offence occurred: whether due to a task being completed incorrectly due to incompetence or negligence, or a service interaction where the server was rude, or something else. The \textit{severity} is a subjective measure of its impact. 

\subsubsection{Moderators of the recipient}
Beyond the characteristics of the offence and the apology offered in reconciliation, the effectiveness of an apology is also influenced by the beliefs, biases, and attitudes of the recipient. We have distinguished two moderators to encapsulate considerations associated with the recipient of the apology, which we have described as \textit{disposition}~\citep{Fehr2010} and \textit{identity}~\citep{Borkin1978, Eslami2010, Smith2008, Lazare2004, Liddle2017, Wagatsuma1986}. The \textit{disposition} of the recipient regards the personal and non-permanent internal factors underpinned by an individual's attitudes, values and beliefs. These sub-features are generally difficult to explicitly label, although methods for categorising them have been developed~\citep{Gosling2003, Jayawickreme2021}. Conversely, the recipient's \textit{identity} encapsulates the largely static and group-based factors, such as demographic information and cultural or linguistic influences. While the classification of personality as either an aspect of disposition or identity is dependent on subscription to a subset of the contemporary research or conformity with a traditional view of personality as a measure of static traits, we recognise that these measurements are a simplification of a more complex system~\citep{Jayawickreme2021} and thus consider personality as a sub-feature of disposition.

\subsubsection{Moderators of the offender}
Similarly to the \textit{recipient}, the characteristics of the offender influence how an apology is perceived~\citep{Fehr2010, Lewicki2012}. \citet{Lewicki2012} describe this phenomenon through \textit{mental models} held by the recipient that enables the individual to quickly pass judgement on the trustworthiness of another based on a both preexisting and evolving perception of them. Some of these are already encapsulated by previously described moderators; the offender's \textit{role} is inherently needed to meaningfully describe the \textit{type} of \textit{offence}, and the \textit{interaction} is thoroughly represented above. The offender's \textit{reputation} comprises both an evolving aspect based on their \textit{performance} (i.e. \textit{demonstration}), and the underlying bias of the recipient (i.e. \textit{disposition}). A characteristic not captured elsewhere in this catalogue thus far is the \textit{appearance} of the offender, including age and gender. 

The characteristics required to describe the appearance of an AI system are distinctive from the human case due to the vastly variable ways in which AI systems may present, both in terms of physical presence and depiction. For example, an AI system may be an embodied as a humanoid robot, or might be represented as a program interface on a computer screen. These presentation features are recognised as being influential in the manner that the behaviour of an AI system is interpreted~\citep{Jensen2021b}. Thus, we designate two moderators to describe these features: the \textit{embodiment} of the AI system is its tangible form~\citep{Bainbridge2008, Bainbridge2011}, and the extent to which its characteristics mimic those of a human is described as its \textit{anthropomorphism}~\citep{Benitez2017, Duffy2002, Jensen2021b}. 

We have segregated the shared attribute of the interactive \textit{mode} from the AI-specific attribute of \textit{embodiment} to distinguish between \textit{having a physical presence} and \textit{being physically present}. A human offender possesses some physical body and so may apologise in person, or else use some alternative communication mode, such as a phone call or a written letter. However, an AI system may take a broad variety of forms, and may be restricted in the modes through which it can communicate. While a text or audio-only interaction would be considered an alternative communication mode in the human case, it may be the most direct approach for a system that lacks any physical presence. Conversely, any communication by an AI system could be interpreted as an alternative mode, regardless of whether the system presents itself in a physical embodiment or as an application on a computer screen. Furthermore, this interpretation is dependent on not just the \textit{mode} of the apology and the \textit{embodiment} of the offender, but also the recipient's perception of the offender as seen through the lens of the interaction. While the specifics of how these features interact are not well understood, this distinction allows for these cases to be meaningfully distinguished as this knowledge is further developed.

The influence of anthropomorphism also provides additional consideration to the distinction between the human and AI case with regard to the concept of \textit{common ground}: implicit, mutual knowledge about the shared social context between the interacting parties. Common ground enables humans to communicate with each other more efficiently by compensating for missing information. This efficiency is highly valued in human interactions with both human and artificial agents~\citep{Miller2017}. Yet, it is unknown to what extent this shared social context could be similarly leveraged in interactions with AI systems~\citep{Gambino2020}. Moderators such as the perceived anthropomorphism of the system, influenced by aspects such as visual and audio presentation and human-like behavioural cues, can influence the extent to which human users will be willing to accept common ground~\citep{Gambino2020}. 


\subsection{Capabilities for apology} \label{Res:syscapa}
For an AI system to be reasonably capable of apologising, the system must possess a sufficient understanding of itself, its user, its objective, and the effects of its actions. An apology is an act of deliberate engagement that requires and indicates that the offender is accountable for and understanding of the nature of the offence~\citep{Kador2009, Lazare2004}. From this, we can derive two distinct criteria for an offender to present an apology: both their being a suitable representative for the blame and possessing the reasoning capability for its construction. 

The first criterion refers to what \citet{Smith2005} describes as \textit{blameworthiness}: being positioned to reasonably assume responsibility for the harm. Responsibility is a concept accompanied by some challenging nuance, comprised of both an objective requirement of causal effects and a subjective requirement of accountability~\citep{Strawson1974}. A detailed exposition is not suitable for the present text, given that the focus is a broad compilation of these concepts, but we recommend the following papers as an introduction to the topic~\citep{Coeckelbergh2020, Gogoshin2021, Strawson1974, Tigard2021}. In brief, attributing causal responsibility for an offence to an agent requires that the offence was, in some way, brought about by the actions (or inaction) of that agent. For an agent to be accountable requires that they understand and control their actions. A valid admission of responsibility requires both causal responsibility and accountability. 

The second criterion regards how the offender can access and make use of knowledge to express an apology for the harm. Having established the causal link between the action taken and the resulting offence, the offender may construct an apology to address it. The apology consists of components that each contribute towards providing affective support, fulfilling a regulatory function, or providing information to facilitate recovery from the offence. To achieve this, its content should be a meaningful and accurate representation of what the offender believes~\citep{Lazare2004}, suited to the intended outcome. This process hinges on the ability to explain the reasoning behind various decisions, interpretations, and perspectives. 

The third criterion can be derived from \citetp{Kador2009} specification of apology as a \textit{disposition to act}; a commitment to do something to resolve the discrepancy between the desired and existing state. It is an expression of future intent, leveraging this understanding of the consequences of past behaviour to inform better future behaviour. Thus, having established the cause of the offence and expressed self-reproach, the offender should demonstrate this intent. 

\begin{figure}[!h]
\centering
\includegraphics[width=80.64mm]{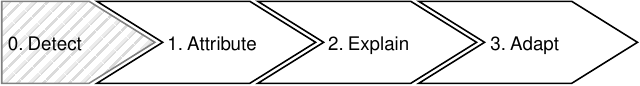} 
\caption[Capabilities required for apology]{Capabilities used in an apology include (1) attribute responsibility, (2) explain what has happened, and (3) adapt future behaviour. Initiation of an apology additionally may require the capability to (0) detect the offence, although this may be circumvented if the offender is informed}
\label{fig:capa}
\end{figure}

Following this reasoning, we can derive the key capabilities for an autonomous apology as consisting of (Figure \ref{fig:capa}):
\begin{enumerate}
\item attribution of fault through an understanding of the impact of the AI system's behaviour and how it is unaligned with the user's needs and priorities,
\item an articulate comprehension of the cause for the behaviour through an explainable decision process that reveals the discrepancy, 
thus facilitating 
\item adaption of future behaviour for repair or reform, demonstrating the intent not to repeat the offence~\citet{Harland2023, Pereira2022}.
\end{enumerate}

These three capabilities; attribution, explanation, and adaptation; encompass the process of creating and delivering an apology for a known offence. These capabilities are sufficient under the assumption that the requirement for an apology has already been established, such as when the user has reported a failure. Alternatively, a highly autonomous system may be capable of recognising and initiating an apology without the need for a user's report. This presents an additional item to be prepended to this list of capabilities: 
\begin{enumerate}\addtocounter{enumi}{-1}
\item detection and identification of a failure for which one should apologise, through ongoing active contextual awareness of the self, the user, and the shared environment. 
\end{enumerate}


\subsection{Framework for AI apology} 
\begin{figure}[!h]
\centering
\includegraphics[width=115.92mm]{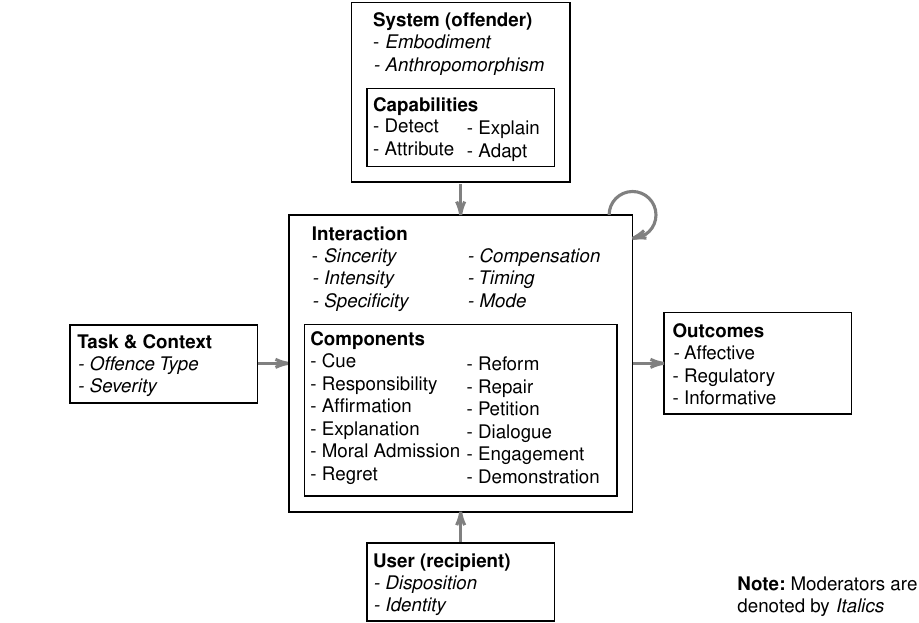} 
\caption[Framework of AI apology]{The {\textbf{Framework of AI apology}} transposes the five elements of apology into the human-AI case: an \textit{interaction} between a \textit{system} (\textit{offender}) and its \textit{user} (\textit{recipient}) regarding some \textit{task and context} (\textit{offence}), resulting in an \textit{outcome(s)}. Each of the components, moderators and capabilities described are represented in this framework according to their associated element, with arrows describing how these influence each other}\label{framework}
\end{figure}

A framework describing the combination of these concepts can be visualised by an interaction diagram (Figure \ref{framework}), based on the five elements of apology previously described (Figure \ref{facets}). This framework has been derived from the \textit{Framework of Human-Computer Interaction} first described by \citet{Zhang2005} and adapted by \citet{Rzepka2018}. According to the original work, HCI research also has five core elements: an \textit{interaction} that occurs between a \textit{system} and a \textit{user}, with respect to a \textit{task and context}, resulting in an \textit{outcome}~\citep{Rzepka2018}. The derived \textit{framework of AI apology} consolidates these elements of apology and HCI as five elements of AI Apology: an \textit{interaction} between a \textit{system} (\textit{offender}) and its \textit{user} (\textit{recipient}) regarding some \textit{task and context} (\textit{offence}), resulting in some \textit{outcome(s)}. 

The classical HCI view of an interaction positions the user at the top to indicate the user's role in initiating the interaction with the system. In this view, however, we have positioned the system at the top. This has been done to emphasise that an apology is an offender-driven interaction and that we are presenting the case of AI apology wherein the offender is the system. The horizontal view describes the progression of the complete process: from the initial offence, through the interaction and concluding with the outcomes. 

Each of these elements exhibits characteristics that influence the interaction and the subsequent outcomes, described by the components, moderators, and capabilities of the apology. The \textit{system} is characterised by its presentation, described by moderators \textit{embodiment} and \textit{anthropomorphism}, and its capabilities. The \textit{task and context} refer to the specifics of the \textit{offence}, described by the moderators of \textit{offence type} and \textit{severity}. The \textit{user}'s characteristics are described by the moderators of \textit{disposition} and \textit{identity}. The \textit{interaction} regards the expression of the apology, characterised by the full set of components in addition to the subset of moderators that can be controlled through actions. These are \textit{sincerity}, \textit{intensity}, \textit{compensation}, \textit{specificity}, \textit{timing}, and \textit{mode}. Finally, the \textit{outcomes} of the apology are categorised into \textit{affective}, \textit{regulatory}, or \textit{informative} effects. We postulate that the interaction is influenced both externally by the characteristics of the system, user, task, and context, as well as internally through interdependencies between the characteristics of the interaction itself. The final outcome is a product of these combined effects. These interaction effects are described by the arrows in Figure \ref{framework}. 

We examine the presentation and interactions between these features as described by the existent literature on AI apology by way of a critical review. The \textit{framework of AI apology} has been developed to serve as a basis for this review, which will be reported in the following sections. 


\section{Methods}\label{meth}
We conducted an in-depth review of the literature using a systematic approach. The methodology for this work was informed by published protocols for rigorous systematic and comprehensive reviews~\citep{Okoli2015, Schnable2021, Stratton2016}. Multiple approaches were combined to contend with the challenges arising from the imprecise and widely varied terminology used in the related literature, much of which is heavily derived from common language descriptions. The retrieval and screening methods were modelled after \citetp{Schnable2021} targeted approach for interdisciplinary scholarship. Additionally, we used a recursively-applied forward-backward snowballing approach, recommended when the literature keywords contain general terms~\citep{Badampudi2015}. The complete review process consisted of five steps: 1) preliminary exploration, 2) literature retrieval, 3) abstract screening and coding, 4) full-text screening and eligibility check, and 5) review and inclusion. 


\subsection{Preliminary exploration} 
Prior to the primary literature retrieval process, a series of scoping searches were undertaken. Relevant articles were retrieved through exploration of academic databases, primarily using Google Scholar and EBSCO Host databases, with a variety of targeted search terms. The results of this exploration were used to clarify the review scope, refine the search terms, and develop the final search query that was applied in the following steps. Relevant articles located during this step were included in the review as \textit{prior known} studies, alongside those identified during earlier research on the topic, recommended by colleagues, or suggested by automated systems.

This preliminary work highlighted the gap in existing meta-analysis and integration between related bodies of work, which the present review was designed to address. This review aims to provide comprehensive coverage of all contemporary contributions to scientific knowledge regarding the use of apologies by AI systems. A four-year search window of 2020-2023 was selected to provide a bounded region for extensive interrogation and a contemporary perspective on the active evolution of the field\footnote{We acknowledge that this bounding excludes early references to AI apology from the 2000's~\citep{Lee2004, Tzeng2004, Madhavan2007}, early-mid 2010's~\citep{Akgun2010, Hamacher2016, Lee2010, Marinaccio2015, Park2012, Salem2015, Robinette2015}, and late 2010's~\citep{Baker2018, Correia2018, deVisser2018, Honig2018, Lucas2018, Nayyar2018, Robinette2017, Sebo2019, Wang2018}. While these works form part of the background of this field, they are not reviewed in this article and we encourage the reader to consult these works independently if needed.}. The scope was limited to those studies that referenced apology or its directly related ideas to provide a targeted synthesis on the topic. The search query was designed accordingly, as follows: (``apology'' OR ``apologise'' OR ``apologize'' OR ``sorry'' OR ``apologies'' OR ``apologetic'') AND (``artificial intelligence'' OR ``AI'' OR ``autonomous'' OR ``agent'' OR ``human.robot'' OR ``social robot'' OR ``human.computer'' OR ``human.agent'')\footnote{The effectiveness of the literature retrieval process was estimated by assessing the overlap between each literature source. The majority of eligible records were retrieved from multiple sources, with very few records having only been found in the prior known literature, and none of which indicated methodological gaps to be addressed. The information used for this analysis is available in Appendix \ref{records}.}. 


\subsection{Literature retrieval}
A \textit{database search} was conducted to identify relevant studies, using the search query designed during the preparatory phase. Queries were run between January 2023 and January 2024 on the ACM Digital Library, ArXiv, IEEE Xplore, Scopus, and Web of Science databases. Google Scholar was not used in this step due to the comparatively extensive volume of false positive results. Search tools were used to enforce the inclusion criteria, which required that the article be available in English and published between 2020 and 2023. Articles were required to be academic in nature: excluding informal articles but allowing for thesis documents, pre-prints, and articles awaiting peer review, provided that they were coherent and presented in a research format (see Section~\ref{meth:inc} for inclusions). Paper retractions and amendments were excluded. 

Parallel literature retrieval using a recursive forward-backward \textit{snowballing} approach was also performed to further broaden this coverage. The approach involved retrieving any relevant literature that was cited by or is citing those studies identified as relevant to this review. The initial set included both database results and prior known literature, and was iterated until the process revealed no new results. 

\begin{figure}[ht!]
\centering
\includegraphics[width=130.48mm]{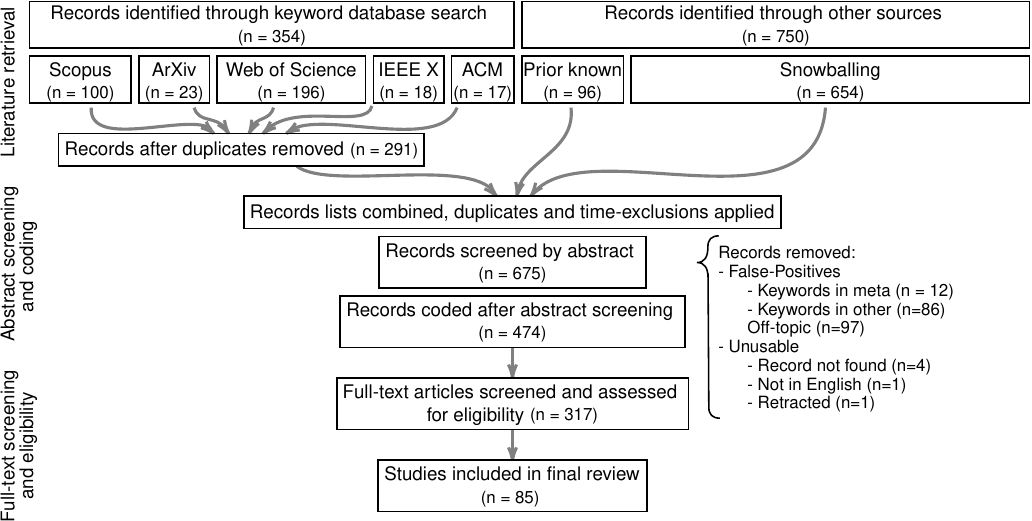} 
\caption[Literature Retrieval Flow Diagram]{
The steps applied in the retrieval, screening and eligibility assessment of the literature for this review}\label{flow}
\end{figure}

The full list of retrieved records comprised the literature items obtained through any of these three sources: \textit{prior known}, \textit{database search} and \textit{snowballing}. The processing steps to compile these results into a single database involved enforcing the time-based exclusion, merging of any duplicate records, and labelling each by source\footnoteremark. Process flow and retrieval counts are described in Figure \ref{flow}. 

\subsection{Abstract screening and coding}
The remaining records were then screened by title and abstract. 
False-positive results (i.e. irrelevant use of keywords, off-topic) were identified manually and excluded, as well as any remaining results that were irretrievable, retracted, or non-English. A coding scheme was developed to provide an objective measure of relevance and maintain consistency in record classification, and was applied to all remaining records. The scheme consisted of two axes: to specify the research format of each paper (i.e. record classification) and its topic in relation to apology (i.e. objective measure of relevance). 

For the former, four categories of research format were specified. 
\begin{description}
 \item[\textbf{Theory:}] Theoretical models, frameworks, taxonomies, or philosophical positions developed through written argument, 
 \item[\textbf{Reviews:}] Compilations of associated literature items and meta-analytical works, 
 \item[\textbf{Human Research:}] Empirical research with human participants, referring to or engaging in real and simulated human-AI interactions, 
 \item[\textbf{Technical Works:}] Implementation or analysis of an algorithm, system or method in-context.
\end{description}

For the latter, a ten-level relevancy specification was developed as to fully encapsulate and position the topics encountered\footnote{The full coding scheme has been provided in Appendix \ref{code}.}. Ultimately, only the three most relevant tiers were consulted in depth, as follows. 
\begin{description}
 \item[\textbf{Direct:}] explicitly and directly addressing apology in AI systems, 
 \item[\textbf{Indirect:}] indirectly addressing the use of apology in AI systems, 
 \item[\textbf{Analogous:}] addressing the use of apology components (Table \ref{tab:comp}) or capabilities (Figure \ref{fig:capa}) in AI systems similarly to apology, without direct reference. 
\end{description}


\subsection{Full-text screening and eligibility} 
The remaining literature items were individually assessed via a full-text screening process based on their coding classification and assessed for eligibility for the present review. Records classified as direct, indirect or analogous progressed to full-text review\footnote{Lower classifications describe broader research in areas such as non-apologetic service recovery and human-robot trust, and are beyond the scope of the present review. While this research does contribute towards broadening knowledge relevant to AI apology, it has been omitted here as to not dilute the focus of the current research. However, the topic-based discussion does draw upon a selection of highly interesting supporting works to illustrate some ideas. Furthermore, a record of all the retrieved studies has been provided in Appendix \ref{allrecords}.}. The final eligibility criteria required that the studies addressed apologies be offered by the AI system, in reference to an offence. Some notable exclusions were studies that explored user-generated apologies towards AI systems \citeeg{Kuttal2021}, or towards other humans on behalf of a robot avatar the user had embodied \citeeg{Aymerich-Franch2020}. Empathetic expressions offered to provide emotional support presented as an apology but not in reference to any discord in the interaction \citeeg{Winkle2022} were also excluded. 


\subsection{Review and inclusion} \label{meth:inc}
The formal bounds of this review included only those records classified as having \textit{direct} relevance to AI apology, as defined by the final eligibility criteria. In total, the literature search returned 85 articles by this description. Of these articles, 46 were published in academic journals, 29 were conference proceedings, 6 were pre-prints, 3 were theses, and 1 was a book chapter. While a majority of these articles arise from fields specific to the study of human-AI interactions and computer science, $21\%$ were published in venues specific to marketing and hospitality research. Additionally, a selection of applicable studies from the indirect and analogous classifications have been referenced as supporting literature. A complete list of the included studies and supporting works, categorised according to their study type, is given in Table \ref{inc}. 


\section{Results and critique} \label{res} 
We present the results and critique the findings of this review in this section. First, we provide an overview of the reviewed literature, introducing explicitly reporting on other existent reviews. Following, we synthesise the reviewed works with an evaluation of the relevant themes according to the \textit{Framework of AI Apology} (Section \ref{def:apol}, Figure \ref{framework}), identify gaps, and summarise the main findings. The proposed motivations for these studies and the measurements used in empirical research are contrasted alongside the \textit{outcomes} of AI apology in Section \ref{res:out}. Section \ref{res:int} analyses each component of apology that arises in the literature as part of the \textit{interaction}, in addition to a critique of the related moderators: sincerity, intensity, specificity, compensation, timing, and mode. The results from research investigating the \textit{task and context} in Section \ref{res:tac} includes the synthesis of a significant body of work regarding the categorisation and influence of the type of offence. Section \ref{res:user} reports the findings related to the \textit{User} (recipient), including considerations regarding the recruitment of study participants. Finally, Section \ref{res:sys} critiques research regarding the \textit{system} (offender), which includes a review of embodiment and anthropomorphism, and, moreover, encompasses the analysis of system capabilities described by relevant technical works. 


\subsection{Overview}\label{over} 
In collating the literature, we identified four main research formats encompassing the contributions to AI apology research: theory, reviews, human research, and technical works. A substantial volume of current research in AI apology is focused on experimental human research studies. Of the 85 articles identified as directly addressing AI apology, this format accounted for a total count of 66, representing 78\% of the included works. The remaining 19 articles were distributed between theory (16), reviews (9) and technical works (2), inclusive of some overlap. However, each of these research formats has a distinct role to fulfil towards the establishment and evolution of the field. 

\begin{table}[h]
\setlength{\tabcolsep}{2.5pt}%
\caption[Included Studies]{This table presents a categorised listing of all studies formally considered in this review. Secondary listings for cross-disciplinary studies and supporting works are denoted by \textit{italics}}\label{inc}
\begin{tabular*}{\textwidth}{@{\extracolsep\fill} p{65pt} p{300pt}}

\toprule%
Study Type & Studies \\
\midrule
%
Theory
 &{\citet{deVisser2020, Galdon2020, Galdon2020b, Kim2022, Kureha2023, Lajante2023, Pak2023, Pereira2020, Porra2020, Rakova2023, Tolmeijer2020, Wright2022} \newline
 \textit{See also:~\citet{LiM2023}\tfnh{b}; \citet{Morimoto2020}\tfnh{b}; \citet{Tewari2022}\tfnh{b}; \citet{Zhang2023a}\tfnh{b}; \citet{Ahmad2022}\tfnh{c}; \citet{Azaria2023}\tfnh{c}; \citet{Chandra2022}\tfnh{c}; \citet{Dazeley2021a}\tfnh{c}; \citet{Kraus2022}\tfnh{c}; \citet{Krausman2022}\tfnh{c}; \citet{Marinaccio2015}\tfnh{c}; \citet{Stock-Homburg2021}\tfnh{c}; \citet{Tsakalakis2022}\tfnh{c}; \citet{Wester2023}\tfnh{c})}}
 \\\cmidrule{2-2}
%
Reviews
 &{\citet{Benner2021, Esterwood2022b, Khavas2021, Liew2021, Rebensky2021, Wischnewski2023} \newline
 \textit{See also~\citet{Pak2023}\tfnh{b}; \citet{Lajante2023}\tfnh{b}; \citet{Tolmeijer2020}\tfnh{b}; \citet{Ahmad2017}\tfnh{c}; \citet{Almeida2022}\tfnh{c}; \citet{Cheng2021a}\tfnh{c}; \citet{Marinaccio2015}\tfnh{c}}}
 \\\cmidrule{2-2}
%
Human Research
 &{\citet{Ahn2023, Albayram2020, Aliasghari2021, Babel2021, Babel2022, Cahya2021, Cameron2021, Choi2021, Esterwood2021, Esterwood2022, Esterwood2023, Esterwood2023c, Fan2020, Feng2022, Fota2022, Fratczak2021, Harris2023, Hu2021, Jelinek2023, Jensen2022, Karli2023, Kim2021, Kim2023, Kox2021, Kox2022, Kraus2023, Kreiter2023, Liu2020, Liu2023, Lu2022, Lv2021, Lv2022, LiM2023, LvL2022, Mahmood2022, Mahmood2023, Na2023, Natarajan2020, Nesset2023, Okada2023, Pei2023, Perkins2021, Perkins2022, Pompe2022, Rogers2023, Schelble2022, SharifHeravi2020, Shen2022, Song2023, Tewari2022, Textor2022, vanOver2020, Wang2021, Weiler2022, XuJ2022, XuX2022, Xu2023, Yang2022, Yang2023, YuS2022, Zhang2023, Zhang2023a, ZhangX2023b, Zhu2023} \newline 
 \textit{See also:~\citet{Galdon2020}\tfnh{b};~\citet{Galdon2020b}\tfnh{b}; \citet{Alarcon2020}\tfnh{c}; \citet{Brown-Devlin2022}\tfnh{c}; \citet{Glikson2023}\tfnh{c}; \citet{Pitardi2020}\tfnh{c}}}
\\\cmidrule{2-2}
%
Technical Works
 &{\citet{Esterwood2023b, Harland2023} 
 \newline\textit{See also:~\citet{Koc2023}\tfnh{c}; \citet{Pereira2022}\tfnh{c}; \citet{Yu2023}\tfnh{c}}}
 \\
\botrule
\end{tabular*}
\tfnt{b}{Cross-disciplinary works}
\tfnt{c}{Supporting works}
\end{table}

Theoretical works describe studies that make a direct contribution to theory through the development of frameworks, taxonomies, or philosophical positions via written argument. In AI apology, these studies have begun to identify and unravel many of the foundational concepts and motivating factors for this research, towards understanding the role and position of apology in AI systems. A substantial focus in this work regards the user's trust towards a system, and its violation, repair, and calibration. Less prominent considerations regard the rights of human users in the present and evolving AI-landscape and the moral faculties of these systems. While a united theory of AI apology remains elusive, these arguments encompass a broad scope of relevant considerations. 

Reviews play a similar and critical role in collating and documenting these contributions, often alongside their own contributions to theory. The present review represents the first focused and extensive analysis of the AI apology literature. However, this literature body does include related reviews that clarify connections between AI apology and broader fields of research. In addition to their contributions presented alongside relevant topics in the latter discussion, we specifically acknowledge and distinguish these works in Section \ref{rev}. 

Human research experiments represent the largest body of work in AI apology. This classification describes studies reporting empirical research with human participants, referring to or engaging in real and simulated human-AI interactions. This form of investigation is crucial for translating theory into practice, especially in regard to understanding the influence of a system's behaviour on its human users. Factors that impact reception and acceptance of an apology by an AI agent towards a user for a predetermined offence are of particular interest in this research. The current literature focuses on the value of apology in repairing trust in these interactions, with considerations for aspects of presentation and delivery. This body of work represents a broad set of contexts, and the varied results imply that outcomes are highly context-dependent. 

Technical works, describing the implementation or analysis of an algorithm or system in the context of AI apology, are the least represented format. Current works include explorations of mechanisms for technical implementation, assessment of current system capabilities, and the development of technical simulation environments to support user research. Many gaps in the literature with respect to explicit technical implementation remain. However, based on the four capabilities that we have identified (detection, attribution, explanation, and adaptation: see also Figure \ref{fig:capa}), we propose consideration of some supporting works from broader research bodies such as behavioural adaptation and explainability. 


\subsection{Existing reviews recognising AI apology}\label{rev} 
To date, there is no complete review that directly focuses on and encompasses AI apology, as is the purpose of the present article. However, AI apology has arisen as a topic of discussion in a few reviews in related areas. Here, we present these studies and acknowledge their contributions to the field. We also explicitly contrast the scope and intent of these reviews with this present article, thereby emphasising its necessity. 

A review by \citet{Liew2021} sought studies using social cues to indicate expertise or competence in AI systems, and briefly reported on two studies also relevant to the present review~\citep{Cameron2021, Choi2021}. They describe these cues using a classification model, `DASSCI', with axes of demographics, appearance, social prestige, specialisation, communication style, and information quality. The communication style class regarded how information was conveyed to users, and was split between verbal cues; including apology, dialogue, blame attribution, and explanations; and non-verbal cues; including gestures, voice characteristics, and emotional displays.

\citet{Benner2021} reviewed strategies for recovering from communication breakdowns in conversational agents. The identified recovery strategies allowed for the designation of six descriptive categories: confirmation, information, disclosure, social, solve and ask. The social category, under which apology was specified, was described as the introduction of typically human behaviour to the dialogue. The review reports on early work in AI apology~\citep{Engelhardt2017}, but the topic was not represented in the contemporary inclusions. In their analysis, the authors described apology as a form of `active' recovery, and noted that the social strategy was rarely used alone, but rather in combination with an additional communicative element (e.g. information).

\citet{Rebensky2021} provided a summary of factors impacting trust development and repair strategies in human-agent teams (HAT). Key attributes impacting the development of trust in HATs include tangibility, transparency, reliability and task suitability, while tactics for effective trust repair and calibration include apology, denial and context-specific approaches, including compensation. Notably, this work acknowledged the compositional nature of apologies, describing components including explanation, responsibility, repair, regret, reform and petition, but did not provide an analysis. The article briefly reported that apology is expected to be an effective trust repair approach in the appropriate context, although the research was limited. It also proposed potential interactions between repair strategies and context factors including severity of risk, time criticality, the team, environment, and mission. There was only a small cross-over in scope between this article and the present review, including three theoretical works~\citep{Glikson2020, Wagner2020, deVisser2020}, as the article did not report on any recent human research. 

With a similar aim, \citet{Esterwood2022b} undertook a brief review of experimental studies in HRI investigating trust repair strategies following trust violations, identifying apology as one of four prominent approaches beside explanations, promises, and denials. The review reports mixed results across the board, including ineffective or even damaging outcomes and issues with non-significant results. The authors noted the lack of theoretical discussion and its consequences for generalisation, and proposed that future research might benefit from considering theories from human-human literature to inspire further investigation, as well as exploring violation type and severity as moderators. While the central topic of interest of this work shares commonality with the present review, differences in scope leave only six studies in overlap~\citep{Albayram2020, Cameron2021, Fratczak2021, Kox2021, Natarajan2020, Zhang2023}. Studies within \citeauthor{Esterwood2022b}'s scope but not addressed in the present article consist of three that were excluded during the full-text screening as not relevant to apology, with the remainder excluded due to publication date. The referenced work considers a more targeted use case and type of study, that being trust repair in robot-based interactions as investigated through human research experiments, and thus has not sought as extensive an investigation on the topic as presented herein. 

The topic of AI apology was also briefly acknowledged as part of broader reviews on trust in HRI~\citep{Khavas2021}, trust calibration in automated systems~\citep{Wischnewski2023}, and on robot service recovery~\citep{Lajante2023}. The first of these described `apology for failure'~\citep{Lucas2018, Natarajan2020, Sebo2018, Sebo2019} as a behaviour-related factor for the development of trust, alongside factors related to performance, appearance, the user, and the task and environment. The second described apology and reported on a selection of relevant papers~\citep{Esterwood2021, Esterwood2022, Kim2021, Kox2021, XuJ2022}, with note of the inconsistent results, as part of a brief summary on trends within trust calibration. The third proposed apology as being one of six dimensions of interactional justice: the component of service recovery that addresses the emotional needs of the customer. The review consulted two relevant papers~\citep{Choi2021, Fan2020} in its assessment of the suitability of using interactional justice in robot service recovery and uncovered a number of considerations in the form of moderators, including service failure severity, customer characteristics, service context, and interactions with distributive and procedural justice elements.

Finally, \citet{Pak2023} developed a theoretical model based on social cognition theory as an explanation for the mixed effects of trust repair strategies by autonomous systems, including apology. The model is an applied extension of the elaboration likelihood model (ELM)~\citep{Cacioppo1986}, which proposes that there are two distinct routes to persuasion determined via either an informative or an affective emphasis. The model proposes that an emotional appeal lacking in any substantial information is less cognitively engaging for the recipient, and thus less effective at repairing trust long-term. To demonstrate the reasoning power of this model, the article examined a selection of the literature considered herein~\citep{Esterwood2021, Schelble2022, XuJ2022} and related works~\citep{Lyons2023} that report unexpected low effects. We address specific observations from this model in the discussion to follow. 


\subsection{Outcomes}\label{res:out} 
The study of AI apology is motivated by the intent to develop and maintain productive working relationships between interactive AI systems and their users. The prospective \textit{outcomes} of an apology can support these relationships by aligning the system and the user through affective, regulatory, and informative effects. The representation of the concepts used to evaluate of the outcomes of AI apology in the literature can vary greatly, and can often be described using higher-level constructs of trust and service satisfaction. Distinguishing the affective, regulatory, and informative aspects of these concepts allows us to describe their anatomy with greater specificity and thus better evaluate them in the context of the AI case. However, measures of trust and service satisfaction are not always described to an extent that allows for this designation, so we have also addressed them in aggregate here. 

Trust recovery is the most popular motivation and evaluated outcome in AI apology research. When the system behaves in a manner that is inconsistent with the expectations of the user, the user's trust in the system is violated and may lapse. Trust recovery describes the process of minimising the loss as it occurs and of regaining trust after it is lost, for which apology is one possible intervention~\citep{deVisser2020}. AI apology is sometimes described as a socio-cognitive trust repair strategy~\citep{Cameron2021, Kox2022}, and has been found to have a positive effect on trust recovery~\citep{Kraus2023, Kreiter2023}. Trust is a latent variable and so its measurement is often indirect, such as through the compilation of perceived trustworthiness measures of benevolence, integrity, and ability~\citep{Kok2020}. Questionnaire instruments for measuring trust may assess each of these separately \citeeg{Bartneck2009}, combine them as an aggregate score \citeeg{Flavian2006}, or use other heuristic measures \citeeg{Lee1992}. They may also be specifically designed to address trust in automation or within a team (Table~\ref{tab:meas}). 

Satisfaction is a similar concept that arises in service contexts. It is described by a broad compilation of contributing sub-processes, including sense of contentedness \citeeg{Tax1998}, perception of failure severity \citeeg{Hess2008, Weun2004}, and usage intentions \citeeg{Kim2009}. The literature suggests that AI apology can support customer satisfaction when used appropriately~\citep{Wang2021}. Related concepts include dissatisfaction, recovery, and evaluation of service (Table~\ref{tab:meas}).

\begin{table}[ht!]
\setlength{\tabcolsep}{2.0pt}%
\caption[Outcome Measurements]{Concepts used to evaluate the outcome of an apology in AI apology research} \label{tab:meas}
\begin{tabular*}{\textwidth}{p{15pt}p{350pt}}
\toprule
\multicolumn{2}{l}{Concepts Investigated} \\ \midrule
 \multicolumn{2}{l}{Trust} \\
& - {\citet{Aliasghari2021, Babel2021, Esterwood2023, Esterwood2021, Fratczak2021, Jelinek2023, Jensen2022, Kim2021, Kim2023, Kox2021, Kox2022, Kreiter2023, Liu2020, Lu2022, Lv2022, Mahmood2022, Morimoto2020, Shen2022, Weiler2022, Wischnewski2023, Xu2023, YuS2022}} \\\cmidrule{2-2}
& Trust in automation: {\citet{Babel2021, Babel2022, Cahya2021, Cameron2021, Feng2022, Karli2023, Kim2023, Kox2022, Kraus2023, Natarajan2020, Nesset2023, Pei2023, Pompe2022, Rogers2023, SharifHeravi2020, Wischnewski2023, XuJ2022, Zhang2023a}} \\\cmidrule{2-2}
& Team/ collaborative trust: {\citet{Kox2022, Schelble2022, Wischnewski2023}} \\\midrule
 \multicolumn{2}{l}{Satisfaction} \\
& - {\citet{Choi2021, Shen2022}} \\\cmidrule{2-2}
& Dissatisfaction: {\citet{Fan2020, Wang2021}} \\\cmidrule{2-2}
& Evaluation: {\citet{Albayram2020, Fota2022, LvL2022, Song2023, Yang2022}} \\\cmidrule{2-2}
& Recovery: {\citet{Hu2021, Mahmood2022, Song2023, Zhang2023, Zhu2023}} \\ \midrule
 \multicolumn{2}{l}{Affective} \\ 
& Anthropomorphism/ appeal: {\citet{Ahn2023, Babel2021, Babel2022, Fan2020, Fota2022, Jensen2022, Kim2021, Kox2021, Mahmood2023, Natarajan2020, Yang2022}} \\\cmidrule{2-2}
& Benevolence: {\citet{Jensen2022, LiM2023, Okada2023, Wischnewski2023, Zhang2023a}} \\\cmidrule{2-2}
& Comfort/ performance: {\citet{Fratczak2021, Pompe2022, Schelble2022, SharifHeravi2020, Xu2023}} \\\cmidrule{2-2}
& Emotional response: {\citet{Babel2021, Lu2022, Lv2021, Lv2022, Na2023, Pei2023, XuX2022, Yang2022}} \\\cmidrule{2-2}
& Likeability: {\citet{Mahmood2022, Kox2021, Cameron2021}} \\\cmidrule{2-2}
& Sincerity/ authenticity: {\citet{Albayram2020, Hu2021, Wang2021, Yang2022, Yang2023, Zhang2023}} \\\cmidrule{2-2}
& Warmth/empathy (system): {\citet{Choi2021, Lv2022, Mahmood2023, Wang2021, XuX2022, Zhang2023}} \\\cmidrule{2-2}
& Warmth/empathy (user): {\citet{Harris2023, Lv2021, Shen2022}} \\ \midrule
 \multicolumn{2}{l}{Regulatory} \\ 
& Acceptance: {\citet{Ahn2023, Babel2021, Babel2022, Kraus2023, SharifHeravi2020, Yang2023}} \\\cmidrule{2-2}
& Blame/ retribution: {\citet{Fan2020, Perkins2022}} \\\cmidrule{2-2}
& Ethicality/ moral trust: {\citet{Ahn2023, LiM2023, Okada2023, Schelble2022}} \\

\botrule
\end{tabular*}
\hfill (Table \ref{tab:meas} Continued on next page)
\end{table}

\begin{table}
\setlength{\tabcolsep}{2.0pt}%
(Table \ref{tab:meas} Continued)
\begin{tabular*}{\textwidth}{p{15pt}p{350pt}}
\toprule

& Forgiveness: {\citet{Kreiter2023, LvL2022, Okada2023, Shen2022}} \\\cmidrule{2-2}
& Integrity: {\citet{Jensen2022, Perkins2022, Wischnewski2023, Zhang2023a, ZhangX2023b}} \\\cmidrule{2-2}
& Perceived threat/ safety: {\citet{Kreiter2023}} \\\cmidrule{2-2}
& Reliability/ stability: {\citet{Harris2023, LiM2023, Okada2023}} \\\cmidrule{2-2}
& Tolerance/ severity of failure: {\citet{Hu2021, Lv2021, Lv2022, Song2023, XuX2022, Yang2022, Zhang2023, Zhang2023a}} \\\cmidrule{2-2}
& Usability/ intention to use: {\citet{Albayram2020, Babel2021, Babel2022, Cameron2021, Feng2022, Fota2022, Kraus2023, Lv2022, Liu2020, Liu2023, Na2023, Okada2023, Wang2021, Weiler2022, Wischnewski2023, Xu2023}} \\\cmidrule{2-2}
& Use/ compliance: {\citet{Albayram2020, Babel2022, Karli2023, Kim2021, Kim2023, Kox2021, Perkins2021, Perkins2022, SharifHeravi2020, Weiler2022, XuJ2022}} \\ \midrule
 \multicolumn{2}{l}{Informative} \\
& Competence/ ability: {\citet{Cameron2021, Choi2021, Jensen2022, LiM2023, Mahmood2023, Na2023, Okada2023, Wischnewski2023, XuX2022, Zhang2023, Zhang2023a, ZhangX2023b}} \\\cmidrule{2-2}
& Intelligence: {\citet{Kox2021, Liu2023, Mahmood2022, Pompe2022, Yang2023}} \\\cmidrule{2-2}
& Performance expectation: {\citet{Lv2021}} \\\cmidrule{2-2}
& Transparency: {\citet{LiM2023, Okada2023}} \\\cmidrule{2-2}
& Trust appropriateness: {\citet{Jensen2022}} \\ \botrule
\end{tabular*}
\end{table}

\paragraph{Affective} \label{out:aff} 
The \textit{affective} role of AI apology has been investigated in the AI apology literature as a means of symbolic recovery~\citep{Zhang2023}. Studies that employed questionnaire-based evaluations of this effect included measures of users' emotional responses and perception of the system (Table~\ref{tab:meas}). An AI system that apologises has been found to be beneficial to perceptions of likeability and anthropomorphism~\citep{Cahya2021, Kox2021, Mahmood2022}. It can also evoke an empathetic response from the user that enhances their satisfaction~\citep{Feng2022, Lv2021, Shen2022}. A positive effect of apology on users' emotional states has also been demonstrated using behavioural and biometric data. For example, \citet{Pei2023} measured respondents' physiological stress using a biometric marker known as the galvanic skin response~\citep{Hirshfield2014}. 

The affective function of apology is not only relevant for emotions towards the offender, but also for an individual's performance. A user's emotional state can impact their ability to complete necessary tasks. One study explored this effect in an industrial context, following an offence in which a robot performed a sudden unexpected movement, resulting in a collision with the user. \citet{Fratczak2021} measured the respondent's stress response and recovery via their postural displacement and their accuracy in completing the assigned task. When the robot apologised, the user recovered more quickly and had a lower stress response when the robot exhibits sudden unexpected movements at a later point. 

However, unregulated and uninformed positive affect may artificially inflate performance expectations and willingness to use, leading to overtrust~\citep{Porra2020}. This can lead to issues such as inappropriate compliance that enables possibly harmful user behaviour~\citep{Karli2023, Natarajan2020}. 

\paragraph{Regulatory} \label{out:reg} 
The \textit{regulatory} role of AI apology explores the social and moral standing of AI systems as participants within a shared society. One measure that clearly illustrates the regulatory function is of forgiveness, in which apology was both found to be a preferred and effective approach in some studies~\citep{LvL2022, Okada2023}. Apologies have been found to support user acceptance~\citep{Kraus2023}, tolerance~\citep{Lv2021}, and willingness to use~\citep{Mahmood2022}. 

Morality, in the context of the cognition and behaviour of artificially intelligent agents, is a related topic of rising interest in the AI apology discourse. Moral principles are the subjective set of rules that govern the behaviour of individuals interacting within a society~\citep{sep-morality-definition}, and are a key mechanism used to articulate standards and boundary conditions for appropriate behaviour in human interactions (Section \ref{def:apol}). The act of apologising for violating these principles functions as a test of the offender's moral capacity~\citep{Bennet2008, Gill2000}. However, AI systems are generally recognised as lacking moral agency~\citep{CarstenStahl2004}. The AI apology literature similarly finds that users have mixed beliefs regarding whether machines can have moral reasoning~\citep{Textor2022}. Yet, other recent works indicate active work towards artificial moral autonomy~\citep{Pereira2020, Pereira2022}. If realised, this could theoretically enable AI systems to demonstrate moral self-regulation through apology in a manner akin to human moral apologies.

\paragraph{Informative} \label{out:inf} 
The \textit{informative} function of apology is an aspect that is most critical in the context of AI systems as it regards how an apology facilitates knowledge of the system's perspective and capabilities. AI apology provides a frame for proactive communication: to dynamically update a user's expectations~\citep{Kraus2022}, convey uncertainty~\citep{Kox2022, Kreiter2023}, request help~\citep{Xu2023}, point out weaknesses~\citep{Jensen2022}, and identify failures~\citep{Albayram2020, Aliasghari2021}. Moreover, it provides a window through meta-commentary into the subjective knowledge and reasoning processes that brought about the behaviour, revealing internal limitations and misconceptions. In this regard, it has a similar goal to eXplainable AI (XAI) research. 

Some measurements that reflect the informative role of AI apology include perceived competency~\citep{Cahya2021, Okada2023}, intelligence~\citep{Kox2021}, and performance expectations~\citep{Lv2021, Lv2022}. Apologies can improve the predictability of the system~\citep{Fratczak2021} and, given preemptively, have been shown to support user co-operation~\citep{Babel2022}. However, the informative function of an apology is not only needed for improving the user's perception of capability and intelligence in the system. Rather, successful recovery with respect to the information available to the recipient would be more accurately measured by the alignment between the user's understanding and the system's capability. None of the described measurements have sought to capture this, but one study devised a `trust appropriateness' measure~\citep{Jensen2022} based on the alignment between behavioural trust and system capability to good effect. 

Trust calibration is the process of aligning the user's trust to accurately reflect the system's performance. \Citet{deVisser2020}, among others~\citep{Wischnewski2023}, argue that autonomous trust calibration is crucial for developing and maintaining long-term relationships between human and robot collaborators. The authors proposed a theory to address the dynamics of trust development and calibration within a longitudinal interaction based on the concept of \textit{relationship equity}: a measure of the equilibrium of emotional resources invested (that may need to be released) or available for investment (required for maintenance) in the relationship. AI apology has been explored as a means of facilitating trust calibration in a number of studies~\citep{Albayram2020, deVisser2020, Jensen2022, Kraus2023, Kreiter2023, Natarajan2020}. Used in this manner, it has been found as an effective means for mitigating inappropriate compliance while reducing user frustration~\citep{Albayram2020}. 

Finally, the informative function of apology in AI may help to fulfil its ethical obligations for transparency~\citep{Harland2023, Smith2005}. AI ethics is a field of philosophical inquiry concerned with the development and application of principles intrinsic to the responsible design and use of AI systems~\citep{Cheng2021b}, as to minimise the harm that these systems pose to their human users~\citep{Huang2023, Siau2020}. Transparency is one of these principles~\citep{Zhou2020}, and has been addressed in the AI apology literature in theoretical examinations relating to the rights of the users~\citep{Galdon2020, Galdon2020b, Kim2022, Porra2020}. The research has shown that effective use of AI apology can enhance the perceived transparency of the system~\citep{Okada2023}. However, the relationship between AI apology and transparency is underexposed in the literature. 


\subsection{The interaction} \label{res:int} 
In AI apology, the \textit{interaction} is the core act in which the \textit{system} (\textit{offender}) apologises to the \textit{user} (\textit{recipient}). The framework of AI apology describes eighteen attributes related to this element, consisting of twelve components and six moderators. The AI apology literature has examined many of these attributes directly through user studies exploring the effects of different apology compositions~\cite[e.g.][]{Kraus2023, XuJ2022} and varied moderators~\cite[e.g.][]{Esterwood2021, Kim2021, Mahmood2022}, or indirectly in contrast to distinct, alternative, often non-apologetic strategies~\cite[e.g.][]{Esterwood2021, Esterwood2022, Esterwood2022b, Esterwood2023, Cahya2021, LvL2022, Song2023} (Table \ref{IntStudies}). Further insights have been provided through conceptual analyses~\cite[e.g.][]{Rebensky2021, Wright2022}. We provide a synthesis and critique of each of these attributes, herein. 

\newenvironment{tabularprespec}{\begin{tabular*}{\textwidth}{
p{140pt}
| 
p{5pt}
p{5pt}
p{5pt}
p{5pt}
p{5pt}
p{5pt}
p{5pt}
p{5pt}
p{5pt}
p{5pt}
p{5pt}
p{7pt}
| 
p{5pt}
p{5pt}
p{5pt}
p{5pt}
p{5pt}
p{5pt}
p{5pt}
p{5pt}
p{5pt}
p{5pt}
p{5pt}
p{7pt}
|}}{\end{tabular*}}

\renewcommand*\rot{\scshape\footnotesize\rotatebox{90}}

\newcommand*\y{{\centering\textopenbullet}} 
\newcommand*\yf{{\centering\textbullet}} 
\newcommand*\yc{\y} 
\newcommand*\yfc{\yf} 
\newcommand*\addyear[1]{(\citeyear{#1})}
\newcommand*\citeb[1]{\footnotesize\citeauthor{#1} \addyear{#1}}
\newcommand*\citeba[1]{\footnotesize\citetalias{#1} \addyear{#1}}

\defcitealias{Liu2023}{Liu, D.\etal}
\defcitealias{Liu2020}{Liu, M.\etal}
\defcitealias{Liu2021}{Liu, P.\etal}
\defcitealias{Lv2021}{Lv, X.\etal}
\defcitealias{Lv2022}{Lv, X.\etal}
\defcitealias{LvL2022}{Lv, L.\etal}
\defcitealias{XuJ2022}{Xu, J.\etal}
\defcitealias{XuX2022}{Xu, X.\etal}
\defcitealias{Xu2023}{Xu, Z.\etal}
\defcitealias{Yang2022}{Yang, H.\etal}
\defcitealias{Yang2023}{Yang, Z.\etal}
\defcitealias{Yang2022c}{Yang, Y.\etal}
\defcitealias{YuS2022}{Yu, S.\etal}
\defcitealias{Zhang2023}{Zhang, J.\etal}
\defcitealias{Zhang2023a}{Zhang, X.\etal}
\defcitealias{ZhangX2023b}{Zhang, X.\etal}
\defcitealias{Zhu2023}{Zhu, Y.\etal}

\newcommand{\HRtablebreak}[1]{
 \botrule
 \end{tabularprespec}
 \hfill (Table \ref{#1} Continued on next page)
 \end{table}

 \begin{table}
 (Table \ref{#1} Continued)
 \setlength{\tabcolsep}{2.0pt}
 \rowcolors[]{2}{gray!10}{white}
 \begin{tabularprespec}
 \toprule%
 Study
 & \multicolumn{12}{@{}c@{}}{Apology Composition}
 & \multicolumn{12}{@{}c@{}}{Moderators Considered}
 \\
Reference
 & \rot{Cu} 
 & \rot{Rs} 
 & \rot{Af} 
 & \rot{Ex} 
 & \rot{Mo} 
 & \rot{Rg} 
 & \rot{Rf} 
 & \rot{Rp} 
 & \rot{Pt} 
 & \rot{Di} 
 & \rot{Eg} 
 & \rot{Dm} 
 & \rot{\textit{Si}} 
 & \rot{\textit{In}} 
 & \rot{\textit{Sp}} 
 & \rot{\textit{Co}} 
 & \rot{\textit{Ti}} 
 & \rot{\textit{Md}} 
 & \rot{\textit{Ot}} 
 & \rot{\textit{Sv}} 
 & \rot{\textit{Ds}} 
 & \rot{\textit{Id}} 
 & \rot{\textit{Em}} 
 & \rot{\textit{An}}
 \\
 \midrule
 }

\begin{table}
\rowcolors[]{2}{gray!10}{white}
\setlength{\tabcolsep}{2.0pt}%
\caption[Human research studies]{Human research studies identified in the present review, described according to the apology composition and moderators considered} \label{IntStudies}
\begin{tabularprespec}

\toprule%
Study 
& \multicolumn{12}{@{}c@{}}{Apology Composition}
& \multicolumn{12}{@{}c@{}}{Moderators Considered}
\\
Reference 
& \rot{Cue (Cu)} 
& \rot{Responsibility (Rs)} 
& \rot{Affirmation (Af)} 
& \rot{Explanation (Ex)} 
& \rot{Moral Admission (Mo)} 
& \rot{Regret (Rg)} 
& \rot{Reform (Rf)} 
& \rot{Repair (Rp)} 
& \rot{Petition (Pt)} 
& \rot{Dialogue (Di)} 
& \rot{Engagement (Eg)} 
& \rot{Demonstration (Dm)} 
& \rot{Sincerity (Si)} 
& \rot{Intensity (In)} 
& \rot{Specificity (Sp)} 
& \rot{Compensation (Co)} 
& \rot{Timing (Ti)} 
& \rot{Mode (Md)} 
& \rot{Offence Type (Ot)} 
& \rot{Severity (Sv)} 
& \rot{Disposition (Ds)} 
& \rot{Identity (Id)} 
& \rot{Embodiment (Em)} 
& \rot{Anthropomorphism (An)} 
\\
\midrule
\citeb{Ahn2023} & \y & & & & & & & & & & & & & & & & & & & & & & & \yf \\ 
\citeb{Albayram2020} & \y & \yc & \y & & & & \yf & & & & & \yf & \y & & & & & & & & & & & \\ 
\citeb{Aliasghari2021} & \yf & \y & \yf & \y & & & & & & & \yf & \yf & & & & & & & & & \yf & \y & & \\ 
\citeb{Babel2021} & \y & & & \y & & & & & & & \y & & & & & & \yc & & & & \yfc & & \y & \yf \\ 
\citeb{Babel2022} & \y & & & \y & & & & & & & \y & \yc & & \yf & & & \yc & & & & & & & \\ 
\citeb{Cahya2021} & \y & \yf & & & & \yf & & \yf & & & \yf & & & & & & & & \yf & & \yf & & & \\ 
\citeb{Cameron2021} & \yf & & \y & \yfc & & & & & & & & & & & & & & & & & & & & \\ 
\citeb{Choi2021} & \y & & \y & \yf & & \yc & & & \yc & & \yc & & & & & & & & \yf & & & & & \yf \\ 
\citeb{Esterwood2021} & \y & \yc & \y & \yc & & & \yc & & & & & \y & & & \y & & & & & & & & & \yf \\ 
\citeb{Esterwood2022} & \y & \yc & \y & \yc & & & \yc & & & & & \y & & & \y & & & & & & \yf & & & \\ 
\citeb{Esterwood2023} & \y & \yc & \y & \yc & & & \yc & & & & & \yf & & & \y & & & & & & & & & \\ 
\citeb{Esterwood2023c} & \y & \yc & \y & & & & \yc & & & & & \y & & & & & & & & & \yf & & & \\ 
\citeb{Fan2020} & \y & & \y & & & & & \y & & & & & & & & & & & & & \yf & & & \yf \\ 
\citeb{Feng2022} & \y & \yf & & \y & & \yc & \yf & & & & \yf & & & & & & & & & & & & & \\ 
\citeb{Fota2022} & & & & & & & & & & & & & & & & \yf & & & & & & & & \yf \\ 
\citeb{Fratczak2021} & \y & \y & \y & \y & & & \yc & & & \y & \y & \y & & & & & & & & & \y & & & \\ 
\citeb{Galdon2020} & & \yf & & & & & & & & & & & & & & \yf & & & \y & \y & & & & \\ 
\citeb{Galdon2020b} & & \yf & & & & & & & & & & & & & & \yf & & & \y & \y & & & & \\ 
\citeb{Harris2023} & & \yf & & & & & \y & & & & & & & & & & & & \yf & & & & & \\
\citeb{Hu2021} & \y & & \y & & & & & \y & & & & & \yc & & & & & \yf & & & \y & & & \yc \\ 
\citeb{Jelinek2023} & & & & & & & & & & & & \y & & & & & & & & & & & \yc & \\ 
\citeb{Jensen2022} & \y & \y & \y & & & & \yfc & & & & & \yc & & & \yc & & & & & & & & & \yf \\ 
\citeb{Karli2023} & & & \y & \yf & & & \yf & \yc & & & & \yc & & & & & & & & & \yf & & \yc & \\ 
\citeb{Kim2021} & \y & \yf & \y & \y & & & \y & & & & & & & & & & & & & & & & & \yf \\ 
\citeb{Kim2023} & \y & \yf & \y & \y & & & \y & & & & & & & & & & & & & & \yf & & & \\ 
\citeb{Kox2021} & \yc & \y & \y & \yf & & \yf & & & & & & \yc & & \yc & & & & & & & & & & \\ 
\citeb{Kox2022} & \y & & \y & \y & & \yc & & & & & & \yc & & & & & & & & & \yf & \yf & & \\ 
\citeb{Kraus2023} & \yf & \yc & \y & \yf & & & & & & & & & & & \y & & & & & & \yf & & & \yc \\ 
\citeb{Kreiter2023} & \y & & \y & \y & & & & & & & & \yc & & & \yc & & & & & & \yf & & & \\ 
\citeb{LiM2023} & \yf & & \y & \yf & & & \yf & \y & & & & & & & & & & & \y & & & & & \\
\citeba{Liu2023} & \y & & \y & \y & & & & & & & \yf & & & \yc & & & & & & & \yf & & & \\ 
\citeba{Liu2020} & \y & & \yf & & & & & & & & \yf & & & & & & & & & & & & & \\ 
\citeb{Lu2022} & \y & \yc & \y & & & & & \yc & & & & & & & & & & & & \yf & \yf & \yf & & \yf \\ 
\citeba{LvL2022} & \y & & \y & & & & \y & & & & & & & & & & & & \yf & & & & & \\ 
\citeba{Lv2021} & \y & \y & \y & & & & & \y & & & & & & & & & & \y & \y & \yf & & & & \yf \\ 
\citeba{Lv2022} & \y & & & & & & & \y & & \y & \yf & & & \yc & \yc & & & \yf & \yf & & & & & \\ 
\citeb{Mahmood2022} & \y & \yf & \y & \y & & & & & & & & & \yf & \y & \y & & & & & & & & & \\ 
\citeb{Mahmood2023} & \yf & \y & \y & \y & & & & \y & & & & & & & & \yf & & & & & & \yf & & \yf \\ 
\citeb{Morimoto2020} & \y & & \y & \y & & & & & & \yf & & & & & & & \yf & & & & & & & \\ 
\citeb{Na2023} & & \yf & & \y & & & \y & \y & & & & & & & & & & & & & & & \y & \\ 
\citeb{Natarajan2020} & \y & \yf & \y & & & & & & & & & & & & & & & & & & & & \yf & \yf \\ 
\citeb{Nesset2023} & \yf & & \y & \yf & & & \yf & & & & & \y & & & & & & & & & & & & \\ 
\HRtablebreak{IntStudies}
\citeb{Okada2023} & \y & & & & & & & \y & & & \y & & & & & \yf & & & & & & & & \\ 
\citeb{Pei2023} & & & & \y & & & & & & & \yf & & & & & & & \yf & & & & & \yf & \\ 
\citeb{Perkins2021} & \y & \y & & \y & & & \yf & & & & & & & & & & & & & & & & & \\ 
\citeb{Perkins2022} & \y & \yf & \y & \y & \yc & & \y & & & & & \yc & & & & & & & \yf & & & & & \\ 
\citeb{Pompe2022} & \y & \y & & \y & & \yf & & & & & \yc & & \yf & & & & & & & & & & & \\ 
\citeb{Rogers2023} & \y & & & \yf & \yf & \y & & & \y & & \yf & & \y & & & & & & \yc & & & & & \\ 
\citeb{Schelble2022} & \yf & \yf & & \y & \yc & & & & & & & & & & & & & & \yf & & & & & \\ 
\citeb{SharifHeravi2020} & \y & \y & & \y & & \y & \y & & & & \y & & & & & & & \y & & & & & & \yf \\ 
\citeb{Shen2022} & \y & & \y & & & \y & & & & & \yf & & & \yc & & & & & & & \yf & & & \\ 
\citeb{Song2023} & \y & & \y & & & & \y & \y & & & \y & & & & & & & & & \yf & & & & \\ 
\citeb{Tewari2022} & \y & & \y & \y & & & & \y & & \yf & & & & & & & & & \yf & & & \yf & & \\ 
\citeb{Textor2022} & \yf & \yf & & \y & \yc & & & & & & & & & & & & & & \yf & & & & & \\ 
\citeb{vanOver2020} & \y & \y & \y & & & & & \y & & \y & \y & & & & & & & & \y & & & & & \\
\citeb{Wang2021} & \yf & & \y & & & & \y & \y & & & & & & & & & & & & & & \yf & & \yc \\ 
\citeb{Weiler2022} & \y & & \y & & & & & & & & & & & & & & \yc & & & & & & & \\
\citeba{XuJ2022} & \y & \y & & \yf & & \yf & & & \y & & \yf & & & & & & & & & & & & & \\ 
\citeba{XuX2022} & \y & \y & \y & & & \y & & \y & & & \y & & \yfc & & & & & \y & \y & \yf & & & & \\ 
\citeba{Xu2023} & \y & & & \y & & & \y & & & & & & & & & & & & & & & & & \yf \\ 
\citeba{Yang2022} & \y & \y & \y & \y & & & & & & & \y & & \yf & & & & & & & \yf & & & & \yf \\ 
\citeba{Yang2023} & \y & & \y & & & & \y & & & & \y & & \yf & & & & & & \yf & & \yf & & & \\ 
\citeba{YuS2022} & \y & & \yf & \y & & \yf & & & & & \yf & & & \y & & & & & & & & & & \yc \\ 
\citeba{Zhang2023} & \y & \yc & \y & & & & & & \y & & \y & & \yf & & & & & & & \y & \yf & & & \\ 
\citeba{Zhang2023a} & \y & \yf & \y & \yc & & \y & \y & & & & & & & & & & & & \yf & \yc & & & & \\ 
\citeba{ZhangX2023b} & \y & \yf & \y & \yc & & \y & \y & & & & & & & & & & & & \yf & \yc & \y & & & \\ 
\citeba{Zhu2023} & \y & & & & & & \y & \y & & & & & & & & \yf & \yf & & & & & & & \yf \\ 
\botrule
\end{tabularprespec}
\footnotesize{\yf\ attributes directly investigated, \y\ other attributes demonstrated or discussed in study}
\end{table}

\subsubsection{Components of an apology} \label{res:comp} 

\paragraph{Cue} \label{cue} 
An apologetic \textit{cue} is a key term used to designate an expression as an apology (e.g. ``I am \textit{sorry}'', ``I must \textit{apologise}'')~\citep{Blum-kulka1984}. It generally forms part of a broader expression comprising a selection of additional components as is appropriate for the \textit{offence}~\citep{Cohen1981}. Its effects have been demonstrated in the AI apology literature via direct comparison through toggled usage of the cue within a broader expression, as well as indirect comparison within an apology condition compared to an alternative. In direct comparison, an expression containing a cue (e.g.``Sorry, an error occurred'') has been shown to have an advantage over its non-apologetic counterpart (``An error occurred'') in terms of subsequently reported trust~\citep{Kraus2023, Pompe2022} and acceptance~\citep{Kraus2023}, as well as trusting behaviours~\citep{Pompe2022}. However, indirect comparisons have shown less distinct results. For example, the difference between a cue condition (``My apologies'') and a rejection of responsibility (``I am not responsible'') within a broader expression was found to be non-significant in some studies~\citep{Schelble2022, Wang2021}. Others found the cue-based approach preferable~\citep{Cahya2021, Natarajan2020} or reported mixed success, possibly due to interference from other contextual factors~\citep{Aliasghari2021, Rogers2023} (e.g. as part of an `expressive' condition that was sometimes disruptive). 

A cue can also be used in isolation~\citep{Blum-kulka1984}, affirmed for the AI context by \citet{Rebensky2021}, but this case does not appear in the user research. Conversely, the same language may also be used in contexts other than apology, such as condolences (e.g. ``I am sorry to hear of your bad news''), or more ambiguous cases, such as preceding an interjection (e.g. ``Sorry, what was your name?'')~\citep{Gill2000, Lazare2004}. \Citet{vanOver2020} describes how these uses similarly appear in technological contexts, through a phenomenon they describe as `cross-pollination', in adherence to social norms without relation to a task. \citet{Gill2000} distinguishes that for a cue to denote an expression as an apology requires the speaker to be morally responsible (i.e., identifiable as the \textit{offender} for a relevant \textit{offence}). In the case of AI systems, this may explain some of the negative responses observed, due to interference with the participant's pre-perceptions of the system's capabilities~\citep{Zhu2023}. \citet{Cameron2021} argue that apologising draws attention to the mistake, thus can reduce perceptions of capability. If this effect cannot be counteracted by positive perceptions associated with demonstrating empathy or self-awareness, it may have a net negative result. In some studies that collected unstructured, open-ended responses, some respondents expressed disapproval of the agent's use of the cue~\citep{Lu2022, Rogers2023, SharifHeravi2020, Textor2022}. While this attitude is not so widely represented in the data as to discredit an apology-based approach, it warrants direct examination. \citet{Esterwood2022} reasoned that this objection might be explained through cognitive dissonance theory, but were unable to obtain sufficient statistical evidence to fully support their hypothesis. Yet, the trends observable in the literature as a whole, of mixed results linked by possible common themes, encourage further research of a similar vein. 

Furthermore, it may be possible to apologise without using a cue~\citep{Blum-kulka1984}. In this case, the configuration of other components would then determine whether it is an apology or something else, such as an excuse or a denial~\citep{Schelble2022}. At least one study investigated AI apology without a cue (``Oops, I think I made a mistake'')~\citep{Karli2023}. While this study does not provide clear insight on the effect of the cue, it does report the apologetic approach as favourable to silence. Furthermore, it does not report any respondent complaints regarding the agent's use of apology, despite collecting unstructured, open-ended responses. However, some studies in Table~\ref{IntStudies} that are not marked for having included a cue are blank as they did not provide a complete specification of the apology, and thus it is not clear if a cue was used. 

Overall, the distinct effect of a cue within an apology is unclear. As it has not been a focus, the existing research presents insufficient evidence to draw out any clear patterns. However, we have noted some potential trends that future research could investigate. Some studies focusing on service recovery included a measure of perceived apology as a dependent variable~\citep{LvL2022, Song2023}. In these works, the measure was associated with fulfilment of the user's efficacy needs under comprehension errors or time stress. However, it could be repurposed to aid in investigating the importance of including a cue in recovery communication, towards both the intended outcomes of the apology and the perception of the apology as such, as well as the relationship between the two.

\paragraph{Responsibility} \label{resp} 
Taking \textit{responsibility} in AI apology is the acknowledgement and ownership of the causal relationship between the actions of the \textit{system} (\textit{offender}) and the subsequent \textit{offence}. The effects of taking responsibility have been substantially explored in AI apology as a strategy for trust repair and service recovery. The expression is largely presented in the form of self-attributed statements via first-person pronouns (i.e. ``I'')~\citep{Jensen2022} or possessive determination of the identified issue (e.g.``My mistake'')~\citep{Shen2022}. Findings suggest that a misaligned AI system that takes responsibility via self-directed blame as part of a sincere apology can recover more effectively than comparative non-responsibility strategies~\citep{Mahmood2022, Zhang2023a}, especially in higher severity contexts (e.g. the offence had serious financial consequences)~\citep{XuX2022}. Apology recipients comment positively on this approach~\citep{Albayram2020, Textor2022}, and it has been shown to lead to increases in perception of likeability~\citep{Mahmood2022}, intelligence~\citep{Mahmood2022}, competence~\citep{Zhang2023a, ZhangX2023b}, and integrity~\citep{Textor2022, Zhang2023a, ZhangX2023b} in the agents. 

The effects of not taking responsibility have also been explored as a trust repair strategy. Alternative strategies include redirecting responsibility externally, such as by blaming the system for the fault~\citep{Kim2021, Schelble2022, Textor2022}, putting the responsibility on the user~\citep{Cahya2021, Natarajan2020}, blaming some other party~\citep{Mahmood2022, Zhang2023}, blaming the environment~\citep{Feng2022}, deflecting responsibility with an empty expression~\citep{Albayram2020, Kraus2023, XuJ2022}, feigned confusion~\citep{Perkins2022}, or outright denial~\citep{Kraus2023, Schelble2022, Textor2022}. Studies that investigated the effects of externally-directed blame on reception of an apology in comparison to a self-attribution of responsibility mostly support the latter. Strategies that redirected responsibility towards the user were not favourably received~\citep{Cahya2021} and made for an ineffective trust repair strategy~\citep{Natarajan2020}. However, \citet{Natarajan2020} found that this approach did aid in mitigating overtrust and inappropriate compliance. Self-directed blame was the preferred strategy for smart speaker voice assistant AI systems~\citep{Mahmood2022} and robot pharmacy assistants~\citep{Zhang2023a, ZhangX2023b}. However, \citet{Kim2021, Kim2023} found that anthropomorphic agents repaired trust more effectively with internally-directed blame statements, but that externally-directed statements were preferable in machine-presenting agents. Others found the distinction insignificant~\citep{Feng2022}. A proposed explanation for this mixed effect is that the suitability of an expression of responsibility relies on the correctness of the attribution according to the user's perception~\citep{Alarcon2020}, but does not appear to have been directly investigated yet. 

A similar approach to externally-directed blame is outright denial, which was found to be less effective for repairing trust than most other compared approaches, such as apologies, explanations, and promises~\citep{Esterwood2023, Kraus2023, Zhang2023a, ZhangX2023b}. \citet{Kraus2023} found that denials returned the lowest results in likeability, acceptance and intention to use among an assortment of apology conditions. \citet{Esterwood2022b} noted that findings regarding the use of denials by robots are mixed, and can contradict the expected results from the human literature. They proposed a possible reason might be due to the limited plausibility of a denial in the study designs, having left little opportunity for doubt. However, when the authors investigated the approach in a later study, denials still appeared to be the least effective strategy~\citep{Esterwood2023}. Other studies also found denials to be ineffective: one investigation finding that the agent was perceived as thoughtless~\citep{Textor2022}, and another reporting similarly insubstantive results to a non-responsibility apology condition~\citep{Schelble2022}. 

Some complications in the evaluation of these approaches may stem from confounding variables, due to influential, unreported differences or similarities between the conditions. For example, in a study that compared an apology condition with a denial, a robot teammate makes decisions that are misaligned with the team's overall goal~\citep{Perkins2022}. In the denial condition, the agent stated that it was not doing what it had in fact done, and feigned confusion (``I don't know what happened''). In the apology condition, the agent explicitly admits fault but also expresses confusion, which may have contributed to why this study found support for some of its hypotheses but insignificant results for others. 

Beyond the social implications of an AI system that self-ascribes or deflects responsibility for its actions, it is also important to consider whether it is correct for the system to do so. AI systems inherently make decisions that continuously shape their behaviour throughout their interactions~\citep{Wright2022}. However, it is not always clear who is truly responsible for an offence when it occurs~\citep{Wright2022}. The premise of a `responsibility gap' describes the distance between the control of the system designer over the future behaviour of the system and the level of oversight that a user has over the technology that they operate~\citep{Matthias2004}. The decision-making processes of highly autonomous, complex AI systems lie in the space between these two boundaries~\citep{Galdon2020b}. This discrepancy violates the minimum requirements for traditional methods of ascribing responsibility, being an understanding of the consequences of those actions, for actions taken by an AI system~\citep{Matthias2004}. \citet{Galdon2020, Galdon2020b} propose that users have a `right to reparation', as an assurance of accountability despite this incomplete knowledge. This right requires that there is some party who is accountable for the behaviour of an AI system, and thus answerable to provide reparations for offences that may occur. However, a complete articulation of this proposed right and an extrapolation as to how it may be applied to overcome this uncertainty of responsibility are not present in the literature. Conversely, \citet{Kureha2023} asserts that, for this reason, it is morally impermissible for AI to apologise. They argue that allowing AI systems to apologise encourages misattribution of responsibility towards the agent and thus away from the creator, who is the responsible party. However, this argument does not hold true under the premise of the responsibility gap; that there is a responsible human behind an offence by an AI for which it might apologise is not the case for the advanced AI systems in which this functionality is being theoretically explored \citeeg{Galdon2020, Galdon2020b, Kim2022, Wright2022}. Furthermore, it unduly discredits the awareness of users of present technologies to distinguish system misalignment problems for which AI apology is proposed, from the systematic unethical behaviour of their parent company~\citep{Pitardi2020}. 

\paragraph{Affirmation} \label{Aff} 
An \textit{affirmation} is an expression that describes the \textit{offence} for which the \textit{system} (\textit{offender}) is apologising. This component is often presented in the AI apology literature as an accompaniment to a cue to form a simple baseline apology~\citep{Esterwood2021, Esterwood2023, Kraus2023}. Intrinsically, it is a statement describing where the system has become unaligned with the user's expectations. It may manifest as an articulation of what the apology is for~\citep{Choi2021, LvL2022, Mahmood2022, XuX2022}, the identification of a mistake~\citep{Fan2020, Karli2023, Kox2021, Jensen2022, Shen2022}, or a reiteration of the events that have occurred~\citep{Liu2020, Yang2023}. 

The AI apology literature presents substantial support for the use of an affirmation. The expression of an affirmation differed between apology conditions in \citetp{Mahmood2022}, which found the most effective strategy to be one in which the agent acknowledged the impact on the user (``I am sorry \textit{for the inconvenience}…''), rather than one that substituted this for an empty expression (``Sorry \textit{for the mishap}…''). \citet{Aliasghari2021} investigated a student-robot with a human-teacher and found that when the robot was expressive about recognising its mistakes, it was perceived to have shown greater improvement than without, despite ending with an error. Similarly, \citeta{Liu2020} described affirmative expressions as part of emotional strategies including \textit{affect reflection} and \textit{content reflection}, defined as identifying and summarising user emotions and key information, respectively. These strategies were used in conjunction with a brief apology in a conversational exchange between a customer and a virtual chatbot. The study found the inclusion of these emotional strategies to be highly preferable, especially in the case of a social media exchange, resulting in higher reported satisfaction and usability. In contrast, \citet{Schelble2022} compared an apology and denial condition wherein both expressions used a non-affirmative statement of effect (``…for any negative outcomes'') and found that the influence and distinction between the conditions were insignificant. 

\paragraph{Explanation} \label{Expl} 
In AI apology, an \textit{explanation} is an expression that goes beyond an \textit{affirmation} to convey novel information that describes or justifies the reason why an \textit{offence} occurred~\citep{Smith2005}. The development and use of explanations in AI is its own extensive field of research~\citep{Dazeley2021a} to which the informative function of AI apology is intrinsically linked (Section~\ref{out:inf}). The theory suggests that a genuine apology includes an explanation~\citep{Smith2005}, and that it is a necessary component for effective trust repair~\citep{Tolmeijer2020}, important for enabling a long-lasting effect~\citep{Pak2023}, and facilitating understanding and calibration for the user~\citep{Harland2023, Rebensky2021}. 

Further justification for the need for an explanation in an apology is given by \citet{Kim2022}. Similarly to the ``right to reparation'' (see Section~\ref{resp})~\citep{Galdon2020, Galdon2020b}, the article proposes that users have a ``right to explanation'' under the premise of informed consent as an ongoing transaction of trust. Ensuring the user is informed is a fundamental challenge for human-alignment in decision-making algorithms, as the combined effect of high complexity and low interpretability of these raw systems means that it is often not possible to facilitate a complete understanding~\citep{Matthias2004}. However, \citet{Kim2022} propose that the user's right to informed consent may be fulfilled if the user is sufficiently and appropriately informed as to enable an ongoing process of trust calibration. In this case, the obligation of the system is to remain accountable to the user and facilitate this process by providing appropriate, proactive, and informative explanations as required. When an imposition on the trust that defines this dynamic informed consent occurs, the user is owed a remedial explanation, such as that which might form part of an apology. 

Human research within the AI apology literature demonstrates some support for explanations, but lacks a complete consensus. \citet{Pei2023} report that apology including an explanation is more effective at reducing distrust than an apology without explanation and also corresponds with lower physiological stress, and \citet{Rogers2023} found that it was the only approach in which the system admitted to lying that gained advantage over non-admission. Similarly, a study by \citeta{XuJ2022} found that including an explanation as part of an apology appeared to reduce perceptions of deceit, although limitations in the study meant that statistical significance could not be established. Conversely, \citet{Karli2023} found that an explanation was less effective at restoring trust than an expression of reform. Others reported the restorative effect was only significant in particular contexts, such as when used in conjunction with an expression of regret~\citep{Kox2021}, or when used by a humanoid (vs. non-humanoid) robot~\citep{Choi2021}. \citet{Esterwood2021} found that explanations were specifically effective for repairing perceptions of integrity compared to other approaches (baseline apology, promises, denials) and other dimensions of trustworthiness (ability, benevolence). However, this effect is not consistently identified even in highly similar studies~\citep{Esterwood2022, Esterwood2023}, reflecting the mixed results in the literature~\citep{Esterwood2022b}. Further research specifically targeting these interactions is needed if we wish to understand the underlying relationships in play. 

Beyond just the presence of an explanation, the content of the expression itself influences its reception. \citet{Kraus2023} investigated three forms of explanation, considering measures of overall trust repair, likeability, acceptance and intention to use. The results found the \textit{empty} explanation (``something went wrong'') was indistinct from the no-explanation case, whereas the \textit{anthropomorphic} explanation (``I was confused for a moment'') performed significantly worse. Only the \textit{technical} explanation (``my sensor information was unclear'') presented an improvement, which was significant for each of these measures. The quality and appropriateness of the information conveyed by an explanation are described by the moderator \textit{specificity}. A possible reason for this poor reception of an empty explanation, as well as the insignificant findings of \citet{Schelble2022}, could be explained by quotes provided by the participants criticising the approach in the study by \citet{Textor2022}. Participants in the apology condition that saw the agent behave unethically found the apology unsatisfying, stating ``I think it all ties back to just not even being [sic] explaining anything. There was no true explanation and no reasoning that [the agent] gave us.'' Another variation on the form was demonstrated by \citet{Cameron2021}. This study used a strategy labelled as an explanation, described by the robot having `identifie[d] cues that it is on the incorrect level', differing from the causal explanation style generally referenced in the literature. The study reported significant effects aligned with their hypothesis that explanations of this type facilitated higher perceptions of capability. However, as the actual apology conditions used in the study were not clearly declared, it is difficult to compare and synthesise this work with the broader field.

\paragraph{Moral Admission} \label{MoA} 
A \textit{moral admission} is a specialist component of an apology enacted in response to an \textit{offence} that constitutes a moral violation. The component of \textit{moral admission} is a distinct but derivative concept from broader analyses on the morality of AI apology \citeeg{Kureha2023}, representing a demonstration of the system's (even superficial) moral capacity. 

In a philosophical exploration of the development of human moral principles, \citet{Pereira2020} traverse the framing of morality through religious precursors, group dynamics and humanist ideals, until arriving at a premise derived from evolutionary psychology. They establish the case of Evolutionary Game Theory (EGT) as a scientific approach to describing morality, and subsequently derive foundational ideas of morality in artificial agents. The analysis provides several practical contributions to the AI apology literature. First, it proposes ability criteria for moral decision-making in learning systems, in descending importance; equal treatment of similar cases, the imagination of possible solutions to novel problems, and counterfactual reasoning. Furthermore, it presents a practical scenario in a game format that demonstrates the process of learning morals though consequences. It also identifies the role of guilt and apology in moral behaviour, with specific acknowledgement of quality assurances; cost, sincerity, and intent to collaborate. Later supporting work has sought to understand human morals using models developed through AI as a step towards developing benevolent and self-regulating AI~\citep{Pereira2022}, exploring the way apologies are used and associated concepts of intent recognition, forgiveness, and counterfactual reasoning. 

Regarding user research, the component of moral admission is largely unaddressed in the AI apology literature. Mostly, this appears to be due to the low representation of morally-relevant offences in the evaluated scenarios. However, a few studies specifically investigated cases of unethical agent behaviour~\citep{Rogers2023, Schelble2022, Textor2022} with limited acknowledgement of the moral imposition in the expressed apologies. \citet{Rogers2023} explored a scenario in which an autonomous vehicle voice assistant gave dishonest advice to the user with the intent to mitigate their likelihood of engaging in risky behaviour during an emergency (``My sensors detect police up ahead. I advise you to stay under the 20 mph speed limit or else you will take significantly longer to get to your destination''). However, the study found that the apology condition that included a moral admission was a poor trust recovery strategy, representing the worst performance of all the investigated strategies, the best of which was for no admission or apology at all. The authors suggest that this outcome might arise from having violating the user's belief that the system lacked the moral agency, distinguishing the offence as a moral violation as opposed to a simple system malfunction. While this is not a promising result for the use of moral admission in AI apology intended to trust repair, there may be some benefits to this mechanism. Appropriately calibrated trust is better for the long-term relationship between the system and the user than unduly inflated trust~\citep{deVisser2020}. Furthermore, this trust inhibiting effect would not necessarily be out of place in human-human situations: if an offender admits to immoral behaviour, the result must be that the recipient either believes the offender to be reformed (encouraged by an expression of \textit{reform}), or they do not, and thus retain some expectation of the offender's future immoral behaviour\footnote{Another alternative could be that the recipient rejects the offender's admission in the first place, by refusing to conceive of the offender's capacity to have behaved immorally, such as if they believed the offender lacked the moral agency to do so.}. The investigated scenario also combines the effect of the user realising the moral violation with the apology condition, but does not demonstrate the isolated effect of a moral admission on recovery for a moral offence of which the user is already aware.

\paragraph{Regret} \label{reg} 
The component of \textit{regret} in AI apology inherits the challenge of an imprecise description from the human case described in Section \ref{apol:intcomp}. Some self-identified examples of regret used in the AI apology literature include an intensified cue \citep[e.g.``I am really sorry'', ``I'm very sorry'', and ``I sincerely apologise'';][]{Kox2021, Shen2022, Zhang2023a}, an emotional expression \citep[e.g. ``Please forgive my performance in the previous trial. I had no intention of making a mistake'';][]{Rogers2023, XuJ2022, YuS2022}, direct reference \citep[``I came to regret…'';][]{XuX2022}, and physical gestures \citep[e.g. covering the face;][]{Cahya2021, Pompe2022}. The subsequent results for these studies are mixed. Some studies found expressing regret could be beneficial under the right conditions, balancing \textit{affect} with \textit{regulatory} and \textit{informative} support. \citet{Kox2021} found the regret condition enabled the trust repair to be significant as a standalone approach, but more substantially when combined with an explanation which was not influential alone. \citet{Shen2022} also found a significant positive effect on customer satisfaction, and \citet{Pompe2022} noted a positive but non-significant trend in perceived intelligence and self-reported trust, as well as a significant effect in the behavioural trust measure. \citeta{XuJ2022} indicated a possible positive trend, but the data was inconclusive. This result was attributed to a small sample size ($n=128$, $N\approx32$), although other related studies have reported significant findings with similar group sizes \citep[e.g.][$n=66$, $N\approx17$]{Kox2021} or sample sizes \citep[e.g.][$n=140$ repeated measures]{Cahya2021}. Other studies reported a negative trend, attributed to predetermined perception of insincerity; \citeta{YuS2022} found that the regret condition was evaluated more negatively than the non-emotional alternative, and \citeta{XuX2022} found the regret condition was received less positively than the humorous alternative except for severe offences. Finally, \citet{Cahya2021} used a physical gesture wherein the robot looked down and covered its face with its hands to indicate regret. However, it was not favoured over verbal approaches as a stand-alone recovery strategy. 

These results are varied and inconclusive, as is the expression of regret employed in each. As no other suitable reason has been proposed for these mixed results, we suggest that this inconsistency could be a possible explanation. In order to make meaningful inferences about the inclusion of regret in apologies by AI systems, further clarity on what defines this component is required.

\paragraph{Reform} \label{ref} 
An expression of \textit{reform} is a commitment by the \textit{system} (\textit{offender}) to abstain from the behaviour that caused the \textit{offence} in the future. It is often represented in the AI apology literature in the form of a promise to do better, either as part of an apology \citeeg{Feng2022,Karli2023} or as an alternative approach \citep[e.g.][see Section~\ref{other}]{Esterwood2022, Nesset2023}. 

A number of studies have directly investigated the influence of reform in the context of AI apology. \citet{Feng2022} found that expression of a commitment to reform improves the perceived competence of the system, leading to better outcomes regarding trust repair and intention to use. Similarly, \citet{Karli2023} reported that the approach was more effective at repairing trust than the explanation condition. 

The opposite approach, wherein the system expresses how it will not change following an offence, also often appears in this research. A preliminary study by \citet{Perkins2021} suggests that the use of reform as part of a response to an offence that includes either a commitment to improve or an admission of a limitation has an influence over the future trust behaviour of the user. \citet{Albayram2020} compared an optimistic reform condition (``I promise to do better next time'') with a realistic non-reform (``I cannot do better than this'') indication of future performance. The study found that users in the optimistic condition showed a significantly higher initial trust increase. However, after the system failed to follow-through on its promise to improve, participant trust was substantially reduced. Conversely, the realistic condition did not demonstrate such a significant trust benefit, but it did correspond to much lower frustration than the optimistic case, which was only marginally better than none at all. It was also found to be the most efficacious and believable approach, whereas the optimistic condition was less so than even none at all. Similarly, \citet{Jensen2022} considered a trust dampening non-reform condition (``In the next round, poor image quality may further reduce my accuracy''), reporting a marginally significant trend for better trust appropriateness and significant benefits in the agent's perceived integrity. These results indicate that the positive effects of reform are conditional on whether the system can follow through with a \textit{demonstration}, described by its capability to \textit{adapt}. However, none of these studies have explored the influence of a reform condition after which the system genuinely improves its behaviour. Furthermore, the expressions used in these studies to describe reform are vague, without providing any practical indication of how the system might adjust its behaviour to improve. Without either a meaningful description of the system's intentions or the underlying functionality to enact these intentions, an expression of reform is essentially an empty promise. 

Many other outstanding questions emerge from the findings of the existing research on reform in AI apology, for which answers are needed to clarify when and how to incorporate it into a recovery approach. For example, what is the boundary condition for determining that the agent has failed to fulfil the committed reform? Furthermore, how does a reform condition perform in explicitly one-off interactions, where the user would neither have the opportunity to benefit from the agent's professed improvement nor observe any failure to do so? These questions represent gaps in knowledge that highlight the need for user-centred, relationship-based and longitudinal studies in future AI apology research.

\paragraph{Repair} \label{Rep} 
\textit{Repair} in an apology describes the \textit{system's} (\textit{offender's}) commitment to undertake actions to restore what has been damaged in the \textit{offence}. It is an alternative form of a promise to \textit{reform}, focused on counteracting or mitigating the present offence rather than avoiding future offences. \citet{Benner2021} list repair as a core principle for effective design to support user experience in conversational agents, describing it as an opportunity to correct miscommunications, request help or gain clarification. \citet{Tewari2022} propose a similar need for developing intelligent agents capable of seeking proactive repair. 

Repair has been used by a number of studies in the AI apology literature, in a few different forms. The most commonly used expression of repair is through the agent's offering of next steps that enable the user to progress out of the present situation \citep[e.g. ``Please try again'';][]{Fan2020, Lv2021, Lv2022, Song2023, Wang2021}. While expressions of this form can be helpful for recovering an interaction, it still requires the user to self-administer the suggested repair. Other examples of repair are more proactive, including replacing and refunding dropped food~\citep{Okada2023}, replacing incorrect food~\citep{XuX2022}, offering a voucher for poor service~\citep{Hu2021, Zhu2023}, picking up and passing the correct item after the wrong item was selected~\citep{Karli2023}, requesting help~\citep{Cahya2021}, or updating an incorrect estimation~\citep{Lu2022}. In circumstances where it has been varied, studies have reported positive effects on customer satisfaction~\citep{Fan2020}, likeability~\citep{Cahya2021} and perceived intelligence~\citep{Cahya2021}. 

In order to meet the needs of future users of AI systems that demonstrate autonomous apology, future AI apology research could consider more advanced repair conditions that can address the specific context of the given offence. While some of the aforementioned studies used proactive agent behaviour for repair, many of the existing demonstrations of repair require the user to perform the action. If future systems were better equipped to recognise and resolve the cause of various issues without user intervention, it could substantially reduce user burden, leading to better experience.

\paragraph{Petition} \label{Pet} 
A \textit{petition} for forgiveness in an apology is a request from the \textit{system} (\textit{offender}) to the \textit{user} (\textit{recipient}) to relinquish any resentment regarding the \textit{offence}~\citep{Schlenker1981}. The AI apology literature demonstrates a sparse use of petition, especially as a point of focus or variation between compared approaches. Among an assortment of other components altered between the three strategies considered, \citeta{XuJ2022} exchanged the cue (``I'm sorry about…'') for a petition (``Please forgive…'') in an \textit{emotional} apology condition that was received favourably by the respondents. However, as both the use of the petition was not isolated as a variable and the study results were inconclusive in general, no meaningful conclusions can be drawn. 
 
There may be a number of factors that contribute towards explaining the low representation of petition in AI apology. One may be how, in some cases, the petition is conflated with a cue~\citep{Aydin2013}. This is especially plausible when also considering the cultural and linguistic differences that alter how an apology is phrased from the Anglo-centric representations that populate the English-written apology literature. Specific examples are discussed with the \textit{identity} moderator in Section~\ref{Id}. However, it may be possible that a petition is highly motivating for the user, or could be used in place of the cue in scenarios that elicit negative responses. Further clarity on whether the under-representation of this component is justified through some preliminary research on the use of a petition might reveal substantial value, or would at least enable completeness.

\paragraph{Dialogue} \label{dia} 
The function of \textit{dialogue} in an apology is to enable the \textit{system} (\textit{offender}) to understand and address the conflict in collaboration with the \textit{user} (\textit{recipient}). We have identified the presence of dialogue in the AI apology literature as being where the recipient is given the opportunity to respond during the apology process, such that this response could then be factored into the system's own subsequent response. There are only a few studies that considered dialogue in the AI apology literature. \citetp{Fratczak2021} experiment included an ongoing dialogue in the apology condition by having the robot both ask the user a question (``are you okay?'') and follow-up with an explanation of its behaviour (``it seems like I can work faster if I move like that'') at a later stage. While the study only investigated the one apology condition, the time-series data suggests that the latter explanation echoed the effect of the initial apology towards recovering indicators of the user's comfort. 

\citet{Morimoto2020} developed a customer service complaint management model that extends the baseline of ``confirm, apologise, explain'' to include a ``waiting, listening, and asking questions'' mode. The model was proposed to meet the user's emotional needs for expression and the desire to be listened to, and recognise the trigger point for progressing the conversation through the customer's behavioural cues. In a successful demonstration of the intended goal, users reported that the cyclical dialogue model was more satisfying, demonstrated better diligence and listening behaviours than the baseline, and was preferred by two-thirds of the respondents. Criticisms of the approach noted frustrations with the agent repeating itself, and not using any of the information they provide in subsequent responses. Other extended interaction studies find that users appreciate the proactive communication, but dislike the repetitive nature of the responses from a system with limited expressive flexibility~\citep{Albayram2020}. \citet{Tewari2022} found similar results in a scenario investigating communication breakdowns in interactions with a healthcare-supporting conversational agent. This issue also appears in the algorithmic demonstration of an apologetic system, which can reach an unresolvable state under particularly complicated user requirements that leave it with no option other than to repeat the apology describing its inability to perform~\citep{Harland2023}. 

These results suggest that if the system were able to fully demonstrate dialogue capabilities, with technological advancements that adequately mitigate the issues with repetition and enable the system to meaningfully incorporate new information provided by the user into its responses, this component could represent substantial benefit to the user. However, unless this component at least approximates a responsive dialogue experience, benefits are unlikely to be observed. While users appreciate the opportunity to feel heard, repetitive and inflexible dialogue options do not adequately meet this need. 

Reflecting on the persistent sub-theme of users' rights, \citet{Rakova2023} provide some related conceptual advancement for addressing harm caused by advanced AI systems. Further on from just disclosure, they require that such systems include \textit{disclosure-centred mediation} via apology, and that users of a system must still have the opportunity to refuse further participation. Collectively, we suggest that these ideas describe a user's `right to response' to an apology offered in fulfilment of their `right to reparation' and `right to explanation'. Within the context of the apology, this requires dialogue.

\paragraph{Engagement} \label{Eng} 
In an apology, \textit{engagement} is an expression of the \textit{system's} (\textit{offender's}) vested interest regarding the impact of the \textit{offence} on the \textit{user} (\textit{recipient})~\citep{Roschk2013, Slocum2011}. Similarly to the human case, AI systems have been designed to present emotionally expressive content through existing verbal channels, and some research has expanded into non-verbal communication channels~\citep{Stock-Homburg2021}. While this component often arises in the AI apology literature, perceptions can be mixed. 

Verbal engagement is often described as \textit{empathetic} or \textit{emotional}~\citep{Liu2020, XuJ2022} apology. Engagement in the AI apology literature is largely represented through the use of emotive language, with some examples including recognising the user's expressed emotions and reflecting them back~\citep{Liu2020}, expressing an understanding of the emotional cost of the offence~\citep{Zhang2023} and describing negative emotions experienced by the agent~\citep{XuX2022}. Asking if the user is okay~\citep{Fratczak2021}, or telling them to take care~\citep{Feng2022} are other ways in which an agent may show engagement. Despite commonly held lay beliefs that technological systems do not have emotions~\citep{Airenti2015}, engagement can have a positive influence on the user's reception of an apology and their perception of the system in general~\citep{Chandra2022, Stock-Homburg2021}. \citet{Shen2022} suggests that the importance of engagement in apology derives from the underlying mechanism of enacting an empathetic response and connection with the user, which then leads to a sense of satisfaction. 

The emotional strategies employed by \citeta{Liu2020}, including affirmations of \textit{affect reflection} and \textit{content reflection} as well as regret as in the \textit{affect expression}, collectively represent emotional engagement between the system and the user. In combination with the interactive dialogue strategy, this approach was found to be helpful for comforting the emotional needs of the user and thus resulting in improvements to the user's overall service satisfaction. The appropriateness of emotional engagement in this manner was also explored with regard to the channel modality, between a social media exchange or a website-based chatbot. The study reported that the emotional strategies were more highly favoured in the case of a social media exchange, whereas the website context corresponded to an increased preference towards a formal professional performance. 

\citeta{Lv2022} found that a high empathy response resulted in higher trust and closeness with the agent and improved continuous usage intentions, echoing similar findings from other studies~\citep{Aliasghari2021, Feng2022, Rogers2023}. \citeta{XuJ2022} presented a simple emotional apology as a contrastive case to an informative apology containing an expression of self-responsibility and an explanation. Although the final results in this study were not statistically significant, the data indicated that the emotional strategy likely had the greatest positive influence on the participant's behavioural trust following an error. Conversely, \citeta{YuS2022} found that an extensive and emotive expression was less favourable than its brief counterpart. These conflicting results likely indicate the influence of context-specific factors. 

The use of humour in trust recovery has also been suggested as a mechanism for demonstrating engagement and relatability with the user. \citeta{XuX2022} found that humorous responses encouraged customer tolerance, the relationship between them was mediated by the perceived warmth and competence of the agent. \citeta{Yang2023} reported similar results, identifying the mediating role of the perceived sincerity and intelligence of the agent. 

Non-verbal expressions of engagement include demonstrating emotion through body language~\citep{Cahya2021, Okada2023, Pei2023, Pompe2022} as well as humorous emojis~\citep{Liu2023}. The results of studies that explored different expressions of engagement via physical gestures suggest the approach may be useful in a supporting role. \citet{Pei2023} and \citet{Pompe2022} both found that the use of facial expressions and gestures enhanced the effectiveness of an apology. However, \citet{Cahya2021} found that a demonstration of regret though a physical gesture was not an effective standalone recovery strategy. Humorous emojis also demonstrated a supporting effect in the attempted service recovery, helping to preserve customer's willingness to use the system by supporting their perception of the agent's intelligence~\citep{Liu2023}. 
 
Beyond this exploration of users' receptiveness to emotive apology in AI systems, \citet{Porra2020} raise concerns that these behaviours could actually be damaging to users, based on the authors' established theories of \textit{humanness} and human-computer social bonding. \textit{Humanness} describes the unique qualities and characteristics that define the state of being human: our experiences of consciousness, self-awareness, emotions, and social bonding. The author argues that using emotive expressions such as apologies carries an inherent \textit{humanness}, and that their use by AI systems is misleading. Moreover, they assert that an AI system that uses an expression referencing their physiological state is lying and a trivialisation of this important facet of human communication. The risks of endorsing this misrepresentation of a system's ability to feel emotions or fill social roles as humans do is the potential for users to develop harmful and unfulfilling emotional attachments as a substitute for genuine human interaction, or otherwise to see the system as dishonest. \citet{Lajante2023} raise similar concerns, noting that the process of \textit{decoding} human emotional states as to facilitate this form of engagement is the same premise that is understood to form the socially manipulative behaviours demonstrated by humans with antisocial personality disorders.

\paragraph{Demonstration} \label{Dem} 
\textit{Demonstration} regards the actualisation of the standard of behaviour, reform, or repair that was committed to by the \textit{system} (\textit{offender}) in the content of the apology. Demonstration is necessary for an apology that creates an expectation for improvement to avoid `cheap talk'~\citep{deVisser2020, Jelinek2023}. Thus, it is important for a genuine apology to be accompanied by an appropriate change in behaviour. 

Demonstration is not well represented in the AI apology literature, with this review having identified no human research studies that directly investigated its effects. However, it is possible to derive some insights on the topic from the existing literature through analysis of the interactions between the component of \textit{reform} and the agent's performance and their subsequent effects on apology outcomes. In \citetp{Fratczak2021} study, the robot was designed to behave differently based on the user's response when the robot asked if they were okay. If the user gave an affirmative response, the robot's operating mode continued in the same manner as before the accident occurred. Conversely, a negative response would trigger the robot to change its behaviour to work more slowly. However, as all the participants responded positively to the agent's query, this behavioural change was never enacted. 

A selection of studies used repeated subtasks to position the offence(s) between instances of acceptable performance. By using accuracy of around $70\%$~\citep{Esterwood2021} or $75\%$~\citep{Kim2023}, the system reflects an acceptable but imperfect performance similar to what would be expected in a real-world system. We can provide some insight as to the possible influence of demonstration through these studies, as the sporadic `mistakes' made by these systems are interspersed by successes such that they could be briefly interpreted as a demonstration of corrected behaviour. The approach seemed to work best for studies using many subtasks, as studies with fewer repetitions were susceptible to ordering effects based on the failure/success configuration. For example, participants in the experiments by \citet{Kox2021, Kox2022} experienced successful completion of the first task before encountering the failure condition, which resulted in a state of overtrust at the time of the first trust measurement through until when the failure occurred. The subsequent trust measurements, taken following the violation and repair, could not meet that initial trust level. However, as the initial trust level was over-inflated due to perfect performance, it is not a good baseline for comparison. \citet{Esterwood2022, Esterwood2023, Esterwood2023c} considered a repeated errors experiment, and evaluated the participants' trust in comparison to the pre-violation level. In the study design, the agent completes the task successfully twice prior to making an error. It then temporarily corrects its behaviour for the following two tasks before making another error, for a total of three errors over ten attempts. In a manipulation check with no intervention applied, it was shown that each error served to decrease the participant's trust~\citep{Esterwood2022, Esterwood2023c}. Interventions in the form of an apology, explanation, or promise can mitigate some of this impact, but this effect appears to be influenced by the number of errors that has occurred~\citep{Esterwood2022}. Moreover, for a system that has repeatedly shown no improvement, neither apology nor any of the other investigated approaches were able to restore user trust to its initial level. 

Revisiting the analysis above in relation to \textit{reform} (Section \ref{ref}), it is clear that users respond poorly to systems that express a promise to improve but did not follow through. \citet{Albayram2020} study compared two different levels of system reliability ($90\%$ vs. $60\%$ accuracy) over a series of interactions. The participants' trust initial initially increased in the case of reform, but when the agent's performance did not change and the commitment was not fulfilled, they became frustrated and provided a poor evaluation of the apology messages. Similar negative user responses were given in other studies investigating unethical behaviour~\citep{Perkins2022, Schelble2022, Textor2022}, or miscommunication during collaboration~\citep{Nesset2023}. Comparatively, \citetp{Jensen2022} study considered the same scenario as \citet{Albayram2020}, but further included a `trust dampening' expression that indicated worsening performance overtime. Although the agent's performance still remained the same throughout the study, this expression corresponded to a significantly greater perception of integrity in the agent. However, aspects of the results indicated that the trust dampening manipulation was not fully successful. The most notable being that a majority of participants in the trust dampening reported the expectation that the agents' future performance would improve, which was also more than any other condition. 

The apologetic system developed by \citet{Harland2023} emphasised the importance of demonstration in apology to support the long-term social relationship between an AI system and its user. The study investigated an approach to implementing apology in a learning system in a manner that directly linked the problematic result to its preceding behaviours, and amplified the effect of the associated penalty to discourage but not preclude those behaviours from future interactions. This algorithm was evaluated using simulated users. However, if this algorithm or a similar implementation of an adaptive apologetic system could be used as part of user study in AI apology, subsequent research could expand the literature on demonstration substantially.

\paragraph{Other approaches} \label{other} 
Other, non-apology approaches arise from research that explored AI apology as one possible option from a set of proposed strategies. Contrasted approaches explored in the present literature include denials, explanations, promises, and gratitude. According to the definitions and boundary conditions for apology described in the literature~\citep{Gill2000, Lazare2004, Lewicki2016}, these approaches often describe another way of structuring an apology, and have been discussed as such, above, where they overlap with components of apology (e.g. explanations and promises). However, these individual components do not necessarily describe an apology when used alone~\citep{Gill2000, Smith2008}. Furthermore, other contrasted approaches may be unapologetic \citep[e.g. denials;][]{Kim2004} or completely unrelated \citep[e.g. gratitude;][]{You2020}. We briefly discuss the merits of these alternative approaches here, for completeness. 

The aforementioned research articles by \citet{Esterwood2021, Esterwood2022, Esterwood2022b, Esterwood2023, Esterwood2023c} represent a series of works that investigated apologies, denials, explanations, and promises as trust repair strategies for human-robot interaction. Between the various studies undertaken, it appears that no significantly advantageous approach was found, although the results do vary slightly. Explanations appeared to be more effective than either apology or promises in an initial study~\citeyearpar{Esterwood2021}, but these findings were not replicated in later works~\citeyearpar{Esterwood2022, Esterwood2023} and were later excluded~\citeyearpar{Esterwood2023c}. However, as was previously noted, denials appeared to be consistently the least effective approach. 

For service recovery settings with human service providers, an expression of gratitude (``Thank you for using this service'') is a common substitute for apology~\citep{You2020}. This approach can be effective despite drastically differing forms, as it functions in a similar affective manner, to satisfy the customers' emotional needs and make them feel valued~\citep{Ahmadi2021}. This approach has also been considered in AI systems as an alternative to apology \citep[e.g. ``Thank you for using me...'' vs. ``Sorry for the unsatisfactory service...'';][]{LvL2022, Song2023} or as an addition~\citep{Babel2022, Shen2022}, and has been positively received. \citeta{LvL2022} reported that in the specific context of rejection failures, described as when the agent is unable to complete the user's request due to functional limitations, an expression of gratitude for the user's patronage was more successful in obtaining the user's forgiveness than one of apology. However, the contributing effects of other aspects of the expression must also be taken into consideration. In this study, the agent expressed a commitment to further improvement in both the apology and gratitude case. We suggest that the presence of this reform condition, despite the lack of a cue, serves a regulatory function in an analogical manner to an apology. The difference in affect is that gratitude uplifts the recipient, but is not humbling for the offender in the same manner as apology, thus allowing the offender to restore the recipient's self-image without cost to their own~\citep{Song2023}.

\subsubsection{Moderators of the interaction} \label{res:intmod} 

\paragraph{Sincerity} \label{sin} 
The \textit{sincerity} of an apology is a subjective description encapsulating an array of interacting effects between its content, delivery, and context. As a moderator within AI apology, sincerity can describe both the approach used to define how the apology is expressed~\citep{Mahmood2022} and the resulting measure of how successfully the apology was employed~\citep{Zhang2023}. The way that it is represented can vary; \citet{Pereira2020} describe sincerity as having the intent to collaborate. Conversely, \citet{Mahmood2022} interchanged sincerity as a descriptor of the manipulated variable \textit{seriousness}, as a contrast to a `casual' apology. Regardless of how it has been represented, the label of \textit{sincerity} is consistently associated with the most efficacious strategy evaluated. Sincerity enhances the likeability and perceived intelligence of the agent, and improves user satisfaction~\citep{Mahmood2022}. The singular noted exception is \citetp{Albayram2020} report that their measurement of sincerity had no main or interaction effects. The unlikeliness of this result might be more indicative of an ineffective measurement. 

The repeating theme of humour is also relevant to the demonstration and perception of sincerity for trust recovery in AI. Both \citeta{Yang2023} and \citeta{XuX2022} found that users perceived an apology to be more sincere when it included self-depreciating humour (e.g. ``I probably got rocks in my head''). However, humour is not a suitable substitute for an apology; humorous statements without apology are perceived as insincere~\citep{Yang2022}. The relationship between humour and sincerity can be explained through increased \textit{relatability}. Increasing how relatable the system feels to the user, through the use of casual and emotive language such as humour, can enhance their perception of the system's sincerity through human-like cues~\citep{Lv2022, Song2023}. This effect is mediated by perceptions of anthropomorphism~\citep{Yang2022}. 

Other studies have also proposed explanations for precursors to sincerity. \citeta{Zhang2023} identified a significant sequential mediation pathway associated with the perceived sincerity of the apology, i.e., it is dependent on how natural the user found the interaction. This relationship suggests that users' perceptions of sincerity decrease when a recovery approach contradicts their preconceptions of the system. The study examined the use of apologies for symbolic recovery by chatbots, and is one of only a few notable examples in the literature that present evidence of a negative effect. The respondents found the chatbot's apology unsatisfying because they perceived the gesture as unnatural. However, the apologetic expression itself included some uncommon language and phrasing that may have aggravated this perception. \citeta{XuX2022} also found a mediating pathway to sincerity via perceptions of warmth. \citet{Esterwood2023c} noted that prior human-centered research identified a link between perceptions of sincerity and mind perception, to emphasise the importance of research on the latter. However, neither this study nor any others encountered in this review have explored this relationship in an AI context.

\paragraph{Intensity} \label{intr} 
The \textit{intensity} of an apology can be influenced by a broad range of aspects including phrasing (e.g. amplifiers, downtoners), delivery (e.g. use of metaphor, formality), and length of an expression~\citep{Blum-kulka1984, Bowers1964}. Although the inclusion was not explicitly reported, we observed that \citeta{Zhang2023a} included an assortment of both amplifiers and downtoners in their repair strategies: in the internally-attributed apology, ``I \textit{probably} made a mistake, and I am \textit{very} sorry about that; it was \textit{totally} my fault. I promise I won’t make the same mistake again next time''; in the externally-attributed apology, ``I \textit{sincerely} apologize for this mistake; but it was not fully my responsibility—there was some disturbance in the environment. It won’t happen again in the future'', and in the denial condition, ``\textit{Something may have} happened, but I do not take responsibility. I don’t know what happened, but you shouldn't worry about my future performance''. The focus of the study was the varied expressions of \textit{responsibility}, but as the comparative intensity of these three conditions is unclear, this may have contributed to the mixed results observed. 

The effect of an apology of varying degrees of intensity is, as with other moderators, intrinsically tied to the context of the interaction. Other strategies that serve to reduce the level of intensity in the present literature include using reductionist statements, or casual rather than serious language. In the case of preemptive apologies, \citet{Babel2022} found that including a diminution of the imposition (e.g. ``it will not take long!'') lead to greater reported acceptability and trustworthiness. \citet{Mahmood2022} used casual expressions (e.g. ``Sorry for the mishap'') in contrast to a more serious apology (e.g. ``I am sorry for the inconvenience''), in a manner acknowledged to be very similar to a prior study that investigated humorous responses~\citep{Ge2019}. This study found that the serious expression was preferred over the casual one, resulting in higher reported service recovery satisfaction, perceived intelligence, and likeability. In contrast, a number of other studies found the casual approach to be preferred~\citep{Liu2023, Lv2022, Yang2022}. For example, \citeta{Lv2022} used casual expression (e.g. ``I'm sorry I did not catch'') as part of a high-empathy condition in contrast to a low-empathy condition represented by a simple but formal apology (e.g. ``I'm sorry I didn't hear clearly''), whereas \citeta{Liu2023} incorporated emojis as a humour condition. Although the results from these studies cannot be directly compared due to the many other confounding factors, they serve to indicate the mixed results observed. 

The manner in which these concepts are discussed in the literature indicates that the distinction between sincerity and intensity is currently not well understood. Along with similar related concepts such as seriousness as in~\citep{Mahmood2022} or low empathy as in~\citep{Lv2022}, these ideas are often conflated and interchanged or distinguished from one another in an inconsistent manner, and there does not appear to be any present theory that provides a clear direction for conformity. At the root of this issue is the secondary nature of these moderators: they are non-orthogonal concepts and recursively interact with each other and with their antecedents in an imprecise manner. For example, a more casual apology may be less intense than a serious expression in terms of linguistic force through the use of more formal language, or may be more intense for the use of emotive expression or metaphor lacking in a sterile apology. Conversely, the same may be perceived as more sincere due to a mediating pathway via anthropomorphism, or less sincere due to a lack of thoroughness or consideration. 

A more tangible measurement of intensity could be the quantity of components included; in the human case, the general effect of an apology comprising more components has been found to be more substantial than one with fewer~\citep{Lewicki2016}. The AI apology literature also supports a positive correlation between the number of components and the extent of trust repair~\citep{Kox2021, Shen2022}. \citet{Shen2022} investigated an apology vs. control condition with a service robot, where the former involved a far more extensive expression (``I'm very sorry. Thank you for your tolerance and patience. I hope my mistake does not affect your mood. Wish you have a pleasant journey'') than the latter (``Wish you have a pleasant journey''). The study investigated the mediating effect of these expressions on the sense of empathy elicited from the respondents, which contributed towards explaining the reported preference for the apology condition. While the study did not directly consider intensity, both the additional components (gratitude, affirmation, and emotive engagement) and the increased length of the apology condition likely contributed to this effect. Conversely, a similarly extensive and emotive expression was less favourable than its brief counterpart, further indicating the influence of context-specific factors~\citep{YuS2022}.

\paragraph{Specificity} \label{Spec} 
In AI apology, the \textit{specificity} of components such as \textit{affirmation} (Section \ref{Aff}) and \textit{explanation} (Section \ref{Expl}) describes the quality and depth of the information provided to the \textit{user} (\textit{recipient}), such as the \textit{system's} (\textit{offender's}) intentions and understanding of the \textit{offence}. Within each of these components, a substantial breadth of information can be communicated (or not communicated), thus ensuring this information is of a high quality is important for demonstrating the competence of the system~\citep{Liew2021}. Furthermore, information of a greater depth may improve the resilience and duration of the trust that was repaired~\citep{Pak2023}. 

Specificity of an apology has not been directly studied in the present literature, although there are some relevant observations that can be drawn from indirect applications. For example, \citet{Kraus2023} considered three different presentations of an explanation in one study, and found that the technical explanation that demonstrated the highest specificity (``…my sensor information was unclear'') was the only form of the three (empty explanation``…something went wrong'', and anthropomorphic explanation ``…I was confused for a moment'') that gained a significant advantage over the baseline apology. Similar studies also lend some support for approaches in which the agent provides specific information about its limitations~\citep{Jelinek2023, Jensen2022, Kox2022, Kreiter2023}, with the added benefit of more appropriately calibrated trust~\citep{Jensen2022}. In one sub-study within the article by \citeta{Lv2022}, high and low empathy conditions were described by a sequence of responses provided while the system processed the query. The high empathy condition included the expressions ``don’t worry, it is being customized, please wait'', followed by ``waiting makes you impatient, but in order to make the route meet your requirements as much as possible, please wait'', and concluding with ``thank you for your patience. The route has been accurately planned according to your needs. I wish you a pleasant journey''. Conversely, the low empathy condition included the expression ``the line is being customized, please wait'' twice over, concluding with ``successfully planned your route, I wish you a pleasant journey''. In this case, the more empathetic expression also provided the user with both a more specific affirmation, conveying a detailed awareness of the negative consequences of the delay, and a more specific explanation of the reason for the wait. The increased specificity may have contributed to the reported preference for this condition.

\paragraph{Compensation} \label{Comp} 
\textit{Compensation} in AI apology describes that which is given by the \textit{system} (\textit{offender}) to the \textit{user} (\textit{recipient}) to rebalance the relationship after what was lost due to the \textit{offence}. In the AI apology literature, compensation has been largely addressed as a binary variable demonstrated through the provision of financial benefits or relief. \citet{Rebensky2021} discuss compensation in this form, and suggests that it can be effective but is only a valid option if there is a tangible cost. While financial compensation is not an uncommon inclusion in experimental studies, it is not often varied or directly addressed. In \citetp{Galdon2020, Galdon2020b} research, survey participants reported approximately equal preference between compensation-based reparations in comparison to apology, in response to unintended errors resulting in severe consequences. In service recovery, financial compensation has shown a strong positive effect on customer's repurchase intentions~\citep{Fota2022}. Regarding the underlying mechanisms of these results, \citet{Fota2022} suggest that compensation helps users' perceptions that they have been treated fairly. The study found a complementary mediating effect on the influence of the compensation through the respondent's evaluation of the resolution, level of engagement, and perception of anthropomorphism in the agent, suggesting that human-like features support positive perceptions of the whole interaction. Another possible contributing effect explaining the positive influence of compensation is as a form of validation, as a demonstration of an understanding of the full cost of the offence~\citep{Rebensky2021}. We also put forth a possible additional mechanism that may explain the support towards repurchase intentions as a reduction of the perceived risk in the interaction. By providing financial compensation for a failure related to a purchased service, the recipient both regains the financial means for a second attempt and has evidence to suggest that they would be compensated again if the failure reoccurred. Thus, lessening the financial risk of future use. 

One study in the AI apology literature also investigated non-monetary compensation. \citet{Okada2023} represented a more costly apology for a service agent error by including a secondary robot in the interaction, effectively doubling the socio-emotional compensation provided. The study found that the additional robot significantly enhanced subsequent measures of competence and forgiveness. These results encourage future research on this moderator, especially further exploration of possible other forms of non-monetary compensation, as well as investigation of varied levels of compensation. If other moderators can influence how a given compensatory action is perceived, service providers may be able to identify inexpensive adjustments to the system that could enable similar successes at a lower financial cost.

\paragraph{Timing} \label{Tim} 
The \textit{timing} of an apology relates to the length of the delay between the \textit{offence} and the expression of the apology. Timing has been a key focus for early AI apology research. \citet{Robinette2015} investigated how the timing of an apology influenced the likelihood of users trusting a robot that had previously made a mistake in an emergency scenario. This research found that respondents were more likely to follow directions given by a robot that delayed its apology than one that apologised immediately after exhibiting poor performance, almost to the same extent as the robot that consistently performed well. Other early research found similar results~\citep{Nayyar2018}, tentatively attributing the result to the effect of memory decay. \citet{Pak2023} propose that this effect can be explained by reframing the timing as the length of time that passes between the apology and the subsequent trust measurement, rather than the original offence. They theorise that as the effect of the apology dissipates over time, an initially equivalent effect would appear to be less substantial at the time of the subsequent trust measurement if associated with a longer delay. 

This moderator has been less central in the recent AI apology literature. \citet{Zhu2023} found that a delayed apology could result in a lower level of customer satisfaction and revisit intention, subject to interaction effects. This effect was observed for the chatbot agent using an apology for symbolic recovery, but not for either the human agent or in the case of economic recovery (i.e. a coupon). Conversely, the introduction of a time delay between speaking and subsequent responses was one aspect of the improvements to \citetp{Morimoto2020} cyclical model of apology that enabled customer dialogue, with beneficial results for customer satisfaction. 

Aside from considerations of the timing of an apology following an offence, the literature also presents the case for apologies outside this timeline. \citet{Wischnewski2023} define three time points at which an intervention may take place as prior, during, and after the interaction. A preemptive apology takes place prior to the offence, and represents a distinct case wherein the offender is aware of the likelihood of the offence before it occurs and seeks to excuse the impending offence. The offence may be certain to occur but essential to the agent's goals and thus excusable. Alternatively, the offence might be a possible outcome of an exploratory action or learning process. In both cases, the offence may be excusable in light of the potential benefits that might be incurred as a result. The research shows that pre-warning the user of possible offences strengthens positive effects of service recovery~\citep{Weiler2022, XuX2022}, and can aid in gaining user cooperation for successful navigation of goal conflicts~\citep{Babel2021, Babel2022}. A traditional apology is an intervention that takes place after the interaction, whereas there is no real equivalent for during an offence except to abort the action. 

The mixed results and otherwise unclear recommendations arising from this research warrant further investigation. The proposed explanation by \citet{Pak2023} presents a promising school of theory for describing the observed results, but has not yet been validated through any direct studies and thus could be investigated in future research. For example, a possible future study design could look at varying the delay between the apology and the offence, but ensuring the subsequent delay before taking the later measurements are held constant.

\paragraph{Mode} \label{Mode} 
The \textit{mode} of an apology is the format through which it is perceived by the \textit{user} (\textit{recipient}). In AI apology, relevant modes of expression generally include text-based and voice-based. The distinction between vocal expression via a system interface compared to a physically present robot is described by the related moderator \textit{embodiment}. 

Only a few studies in the AI apology literature have sought to investigate the influence of mode on apology reception. \citet{Pei2023} found that a multi-modal apology involving both text and voice expression was correlated with lower physiological stress and lower distrust than the equivalent single-modality approach, although they did not provide a reason for or discuss the implications of this result. Conversely, \citet{Hu2021} found that their text condition did not differ from the voice expression. However, this study used an exclusively text-based format including a text based representation of voice, so it is unlikely to have triggered the same underlying cognitive processes. \citet{SharifHeravi2020} also had both a voice and a text-only mode in their study, using a verbally expressive avatar to represent anthropomorphism in an air-traffic control support simulation. This study also did not lend support for the vocal mode, finding no significant effect related to the apology on the advice acceptance rate, and two-thirds of the participants reported a preference for the text-only system. However, as the main intended manipulation of anthropomorphism that was demonstrated through the talkative agent failed to satisfy a manipulation check, it is possible that these results are better explained by confounding variables. Finally, \citeta{XuX2022} considered both a text only and combined text and voice approach, but did not investigate these as a variable. Rather, these different modes were used to provide a slight contextual variation between repeated studies to demonstrate the robustness of the findings, which did not differ between the investigated modes. 

While there is opportunity to undertake further research on the influence of the mode of apology on participant reception, the present literature demonstrates the importance of careful study design. In order to effectively investigate the influence of the mode as distinct from other environment variables, future research should ensure that the study scenario and format can natively support each of the modes considered.


\subsection{The Task and Context} \label{res:tac} 
A prerequisite for apology is the existence of an \textit{offence}: a misalignment between the \textit{system} (\textit{offender}) and the \textit{user} (\textit{recipient}) represented as a mistake, misunderstanding, malicious action, disagreement, incapacity, inconvenience, or some other misaligned behaviour that gives rise to an \textit{offender}-\textit{victim} relationship. The \textit{task and context} in AI apology describe the circumstances of the \textit{offence}. In this section, we describe the tasks and contexts considered in current AI apology research, as well as address the two moderators that describe this element: the \textit{offence type} and \textit{severity}.

The present literature has represented this scope under two dominant case examples: service agents, such as interacting with a customer in a business setting, and collaborative agents, where interactions occur as part of mutual pursuit of a common or correlated goal. Service contexts adopted by studies in which AI apology has been explored include customer service robots~\citep{Morimoto2020} and support chatbots for businesses~\citep{Fota2022, Liu2020, Liu2023, Lv2022, Weiler2022, Yang2023, YuS2022, Zhang2023, Zhu2023}, as an online shopping assistant~\citep{Mahmood2022, Mahmood2023, Song2023}, guidance robot~\citep{Cameron2021}, food service~\citep{Choi2021, Lv2022, Okada2023, Shen2022, Song2023, Wright2022, XuX2022, Yang2022}, hotel service~\citep{Choi2021, Hu2021, Lv2021, Lv2022, LvL2022, Shen2022, Wang2021, XuX2022, Yang2022} and other traveller support~\citep{Fan2020, Lv2021, Lv2022, LvL2022, XuX2022}. In collaborative settings, studies have specifically considered industrial~\citep{Esterwood2021, Esterwood2022, Esterwood2023, Esterwood2023b, Esterwood2023c, Fratczak2021}, student-teacher~\citep{Aliasghari2021, Karli2023}, military~\citep{Kox2021, Kox2022, Kreiter2023, Schelble2022, Textor2022} dynamics (simulated through video games), air traffic control~\citep{SharifHeravi2020}, and other workplace~\citep{Kraus2023} or education~\citep{Harris2023} settings. Collaborative game-based settings were also used, including with physical robots~\citep{Jelinek2023, Natarajan2020, Nesset2023, Pompe2022} and virtual teammates~\citep{LiM2023, Natarajan2020, Pei2023, Perkins2021, Perkins2022}. Hybrid contexts in which the system is both in a service role and required to undertake collaborative-style interactions are also common, especially in home-based or personal assistant settings. These include home-based cleaning~\citep{Babel2021, Babel2022, Na2023, Harland2023} and health management~\citep{Tewari2022, Zhang2023a, ZhangX2023b} robots, voice~\citep{vanOver2020, XuJ2022, Xu2023} or text-based~\citep{Feng2022, Lu2022, Rogers2023} assistants embedded within an autonomous vehicle, specially engineered interactive settings providing support or advice to a user~\citep{Ahn2023, Albayram2020, Jensen2022, Cahya2021, Kim2021, Kim2023, Pei2023}, or other forms of personal virtual assistants~\citep{Galdon2020, Galdon2020b}. 
Some edge case examples in the supporting literature regard the use of AI systems to develop, personalise, analyse, or perform grammatical revisions of apologies on behalf of a human offender~\citep{Brown-Devlin2022, Glikson2023, Yu2023}.

\subsubsection{Moderators of the task and context} \label{res:taskmod} 

\paragraph{Offence Type} \label{type} 
An \textit{offence} can be classified based on characteristics such as its underlying cause or area of effect, described as the \textit{offence type}. In AI apology research, classifications of different types of offences are used to distinguish the context of an offence and enable the investigation of patterns that exist in users' responses to that offence or subsequent repair attempts used by the agent. 

The seed work for many classification schemes for \textit{offence type} in the AI apology literature is a study on trust in human organisations by \citet{Mayer1995}. The three categories of violations designated in this work align with the three dimensions of trustworthiness: ability, integrity and benevolence (Section \ref{def:apol}). In relation to offences of AI systems, ability (or competence) violations occur due to limitations of the system, such as mistakes. However, a mistake can only be a trust violation if it is unaligned with the user's expectations upheld by that trust. Integrity and benevolence violations also occur due to misalignment between the user and the system. For an integrity violation, the resulting behaviour contradicts the user indirectly as to appear indifferent, whereas a benevolence violation is in direct contradiction as to appear malicious. Reflecting on the works discussed~\citep{Tomlinson2020} and our earlier analysis in Section \ref{apol:out} aligning these concepts of trustworthiness with our three core areas, these violation types emphasise informative, regulatory, and affective effects, respectively. An alternative taxonomy describes four types of human errors~\citep{Reason1990}. Errors arising in the execution of otherwise correctly intended actions are either \textit{slips}, the incorrect performance of the intended action, or \textit{lapses}, the incorrect part- or non-performance of the intended action. When the action was executed as intended yet itself was incorrect, the error is either a \textit{mistake}, if it was unintentional, or a \textit{violation}, if it was intentional. \citet{Marinaccio2015} proposed a trust repair framework that integrated these two concepts in relation to human-automation trust; describing \textit{slips}, \textit{lapses} due to attention failure, and \textit{violations} as integrity-based; and \textit{lapses} due to memory failure and \textit{mistakes} as competency-based. Further, it integrates other work in human trust repair by \citet{Kim2013} that recommended apology as a suitable strategy for recovering from competency violations, but denial for integrity violations. 

\citeta{Zhang2023a} explored AI apology in the context of \textit{offence type}, deriving three technical error types from the examples of \textit{slips} and \textit{lapses} described in \citetp{Marinaccio2015} framework: \textit{logic} (i.e. relevant but incorrect action), \textit{semantic} (i.e. irrelevant or meaningless action), and \textit{syntax} (i.e. failure to execute action) errors. This study suggested that these errors are all competency violations, and that users generally do not perceive integrity violations in robotic systems. The study found that internally-attributed apology was the most effective approach for recovering from logic and semantic errors, validating their assertion that these slips are interpreted by the user as competence-based errors. The results corresponding to the syntax error type were non-significant, possibly due to a lack of statistical power. 

\citet{Tolmeijer2020} proposed categorisations of system failures that lead to trust violations in human-robot interactions, and the associated recovery strategies. This research distinguished four categories based on cause, further considering the implications of the user's perception of the system: \textit{design}, \textit{system}, \textit{expectation}, and \textit{user} failures. A system failure occurs when the system fails to do what it intended to do. If the system is behaving as it was designed, a design failure occurs when that behaviour is actually not appropriate due to a misunderstanding on the designer's behalf, and an expectation failure occurs when the behaviour is perceived as inappropriate due to a misunderstanding on the user's behalf. The final error type, user failure, occurs when the user causes inappropriate system behaviour through misuse. The paper proposes that apologies and explanations are suitable and appropriate mitigation strategies for all but user failure types. Furthermore, it explores prospects of autonomous trust repair, presenting a description of system capabilities required for failure detection and the subsequent generation of an explanation that includes facets of planning, counterfactual reasoning, and user awareness. 

In service contexts, a two-item classification system is common, consisting of \textit{outcome} and \textit{process} failures. An outcome failure has occurred if a service has been performed, but was done incorrectly as to produce the wrong outcome. A process failure describes an issue that occurs during the service process, such as where the system is unable to perform the requested task. Sometimes a third classification is also included; an \textit{interaction} failure arises from issues in the service interaction, such as difficulties with communication and understanding~\citep{Hoffman1995}. These error types are highly similar to \citetap{Zhang2023a} \textit{logic}, \textit{syntax}, and \textit{semantic} classification. However, the distinction of the last from the two is less clear between the description and the corresponding examples, indicating possible differences arising from the social vs. technical perspectives of these frameworks. Two studies considered all three of these offence types in their investigation for completeness~\citep{Lv2021, XuX2022}, with a further one study considering only outcome and process failures~\citep{Lv2022}, but they did not report any significant observations. Two further studies investigated outcome and process failures as independent variables~\citep{Choi2021, Yang2023}. These studies found that the study foci of robot anthropomorphism and self-depreciating humour, respectively, had a significant effect on perceived service recovery for process failures, but was not significant for outcome failures. These results indicate process failures may have a greater \textit{affective} impact than outcome failures. 

\citeta{LvL2022} provides a link between process and interaction errors and areas of effect, describing two types of errors based on the psychological effect on the user of being rejected or being ignored. Both errors describe limitations of the agent's capabilities, the former relating to an inability to fulfil the user's request (low flexibility), and the latter relating to an inability to receive or process the user's request (low comprehension). This study reported that an apology, as opposed to gratitude, was the recommended strategy to fulfil the user's efficacy needs following a comprehension error (being ignored)~\citep{LvL2022}. \citet{Benner2021} further break down understanding-based errors into misunderstandings (incorrect interpretation) and non-understandings (no interpretation). This work also further clarifies the recommendation of apology for recovering from a misunderstanding, but emphasises the need to solve and inform for recovery from a non-understanding. 

Other categorisations of agent errors include a three-item classification used by \citet{Cahya2021}, consisting of self-descriptive groupings of social norms violations (SNV), execution errors (EE), and planning errors (PE). This study found that these error types are associated with significantly different results in the Godspeed subscales~\citep{Bartneck2009} of anthropomorphism (SNV, EE rated lower than PE; an \textit{affective} effect), likeability (SNV rated lower than PE and EE; also an \textit{affective} effect), and perceived intelligence (SNV rated lower than PE and EE; an \textit{informative} effect). Similarly, \citet{Tewari2022} describe four offence types relevant to communication breakdown in conversational agents due to a lack of factual knowledge, of procedural knowledge, of social norms, and having conflicting goals. To resolve these forms of errors, the study describes an apology process that incorporates information-seeking, inquiry, elaboration, and persuasion dialogue, respectively. These schemes both explicitly designate the case of social norm violations, which is both an important consideration for possible offences of AI systems and a major challenge for this research. Such violations may be highly subjective and largely invisible, thus difficult to effectively identify and address. 

Considering the types of the offences used in the previously discussed studies reveals a trend that may explain some of the contradicting results. Unethical behaviour, as demonstrated by the linked studies by \citet{Schelble2022} and \citet{Textor2022}, is an example of an integrity violation. The former study provides a quantitative analysis and reports that both apologies and denials were unsuccessful in repairing trust following unethical behaviour. The latter study provides a qualitative analysis; the participants described the \textit{ethical apology} condition as ingenuine and the \textit{ethical denial} condition to be irresponsible, but the \textit{unethical denial} condition was understandable. Lying is another example of an integrity violation, for which \citet{Rogers2023} found that non-acknowledgement was the most effective resolution for preserving trust. \citetp{Rebensky2021} recommendations support these findings, suggesting that apology is best for competency errors rather than integrity violations. This is a reflection of the human case, which similarly recommends denial as an effective repair strategy for integrity violations~\citep{Kim2013}. This approach, which \citet{Rogers2023} described as ``lying about lying'', raises some concerns regarding both its questionable ethicality and its consequences for long-term trust. However, these studies, as a reflection of human research studies in the AI apology literature more broadly, do not address the ethical implications of their proposed approaches. \citet{Perkins2022} also investigated an integrity violation, represented by the demonstration of selfish behaviour, in contrast to a competence violation. However, each of the trust repair conditions considered in this study included an excuse to circumvent moral responsibility, so it is unlikely the results are representative of the case. 

In total, the AI apology literature has collated a substantial set of classification schemes for defining the types of offences that may arise and made good headway towards understanding the most appropriate way to handle their recovery. However, while these various classification schemes are not mutually exclusive, they are disjoint. It is not currently clear from the literature where these concepts align and diverge, and thus it is unclear to what extent they each contribute. An exception to this would be the theory development work of \citeta{Zhang2023a}, which represents a substantial contribution towards aligning these concepts within a cohesive theory. Future research could consider continuing this work through a combination of experimentation and theoretical analysis in a similar manner.

\paragraph{Severity} \label{sev} 
The \textit{severity} of an \textit{offence} describes the extent to which it negatively impacts the \textit{user} (\textit{recipient}). The effects of the different strategies explored in AI apology research depends on the severity of the offence~\citep{Rebensky2021}. A more severe violation with a greater impact on the recipient will require a more substantial recovery effort to achieve the same final result~\citep{Lu2022}. 

\citeta{Lv2021} investigated the moderating effects of offence severity using a scenario where the consequences were missing a train resulting in being late for work, or merely missing a morning workout, in a study that explored the influence of `cuteness' on apology reception. \citeta{Yang2022} employed the same severity manipulation in a highly similar scenario, in a study that investigated the use of a humorous apology compared to a rational expression. In both of these studies, the atypical apology (high `cuteness' and humorous conditions, respectively) received a more positive evaluation when applied to trivial errors, but was poorly received in comparison to the classical apology baseline when the severity was high. Furthermore, \citet{Esterwood2022b} report that the more severe the violation, the less effective explanations were at restoring trust.

Aside from direct manipulation, the perceived severity of an offence can have an association with how the offender is perceived by the recipient. \citeta{Zhang2023} and \citeta{Zhang2023a} both examined how the perceived severity of a failure relates to other mediating factors. Both studies revealed a significant negative correlation between the perceived competence of the system and the perceived severity of the offence. Furthermore, the former also identified a significant negative correlation with perceived warmth, and the latter found similar results with respect to post-interaction competence-based trust and integrity-based trust measures. 

An alternative approach for manipulating the perceived severity of an offence that is used in the AI apology literature is through the introduction of time pressure. The perceived time pressure of an offence scenario can be varied by referencing a pending deadline \citep[e.g. ``There are only five minutes left before the meeting'';][]{Lv2021, Song2023, XuX2022, Yang2022} or through perceived social pressure by describing other waiting customers~\citep{Lv2021}. While this manipulation does not demonstrate a significant main effect on the reported severity of the offence according to the failure severity measure described by \citet{Hess2008}, the approach demonstrates an otherwise indistinguishable intensifying effect on the perceived offence~\citep{Lv2021}. In comparison to other severity conditions, time-pressure variation poses a substantial benefit for streamlining and simplifying scenario manipulation in that all other features of the offence can be held constant, and thus does not have any impact on the type of offence. There does appear to be some correlation between these measures, and it is possible that the measurement instrument used cannot adequately capture the aspect of the offence that influences these results. Future research could investigate these considerations in greater depth towards possible broad improvements in the measurement and manipulation of perceived offences. 


\subsection{The user} \label{res:user} 
The \textit{user} (\textit{recipient}) is the penultimate element of AI apology, as the victim or a representative for the victim of the \textit{offence}, and as the one who receives the apology. In this section, we address the two moderators that describe this element: \textit{disposition} and \textit{identity}. This discussion also encompasses consideration of methodological biases introduced through the user groups recruited as study recipients and linguistic variations.

\subsubsection{Moderators of the user} \label{res:usermod} 

\paragraph{Disposition} \label{dis} 
The \textit{disposition} of a \textit{user} (\textit{recipient}) describes the characteristics that dictate the individual idiosyncrasies in how they interact with the world. This includes their attitudes, values, and beliefs, and how these influence their perception of others and their interactions with them. In their review, \citet{Benner2021} recognise and draw attention to the influence of individual differences in human-agent interaction, and the criticality of understanding these effects towards robust and effective social AI systems. Thus, while disposition describes an incredibly broad and imprecisely defined set of considerations, it is also one of the most well-researched moderators in AI Apology research. 

One key area of investigation within user disposition in AI apology research is of attitudes and beliefs regarding AI systems. Four relevant measures to this effect include the attitude towards working with robots \citep[AWOR;][]{Robert2021}, tendency to anthropomorphise technology \citep[ANTEN;][]{Kraus2023}, affinity for technological interaction \citep[ATI;][]{Franke2019}, and negative attitudes towards robots \citep[NARS;][]{Nomura2008, Syrdal2009}. \citet{Esterwood2022} found that a participant's AWOR had a significant effect depending on the type of repair strategy used and the number of errors. The main effect of the AWOR measure was not significant. However, an examination of the interaction revealed a significant effect, with the data indicating a positive correlation between attitude and trust repair for apology, explanation, and promise conditions. Conversely, denials appeared to follow an inverse trend, to the extent of being the most effective approach for participants with strong negative attitudes. \citet{Kraus2023} found a positive association between ANTEN and the users' perceptions of casual and emotive apology, and between ATI and the more technical expression. Conversely, a high \textit{Need for Human Interaction} can lessen the perceived sincerity of a robot apology~\citep{Hu2021}. Expectedly, ATI is associated with higher initial trust and NARS is associated with lower trust~\citep{Kraus2023}, but no interaction effects with the latter on apology have been reported. 

Aside from using questionnaires to assess participants' bias towards AI systems, some studies also observed distinct respondent groups identifiable through behavioural data. \citet{Fratczak2021} grouped respondents according to the magnitude of their reaction to sudden unexpected movements by an industrial robot in a VR scenario. Data analysis controlling for this introduced factor revealed that apology significantly enhanced rates of task continuance for reactive participants, with success similar to if the robot was idle. Conversely, participants who were unreactive to the offence were similarly unreactive to the apology. A similar effect was seen in \citetp{Perkins2022} research, in a group of respondents described as `stalwart trustors', who never demonstrated any trust loss behaviour. The authors of the former study suggest that these participants were simply not effected by the trust violation and thus the subsequent recovery was redundant, whereas the latter study did not propose any explanation or analysis. These results highlight the importance of understanding the user's needs when designing an apologetic system. For offences that do not constitute genuine technical errors or objectively problematic behaviours, the user's response to an agent's unexpected behaviours may be a suitable predictor of both the presence of an offence and the user's receptiveness to a subsequent apology. 

The user's perception of an agent's mental capabilities also influences the effectiveness of an apology. One study used a priming technique to temporarily influence respondents' beliefs, to demonstrate that users who believe that AI systems have low emotional competence find service recovery via apology to be less satisfying~\citep{Zhang2023}. Conversely, users who report a greater perception of the agent's conscious experience, ascribing it capacity for emotions, experienced an enhanced effectiveness of apology for trust repair~\citep{Esterwood2023c}. Perceptions of intentional agency, describing the agent's capacity to pursue a goal, were not significant~\citep{Esterwood2023c}. Two studies investigated the influence of respondents' implicit theories on apology reception, also temporarily influencing the user's characteristics using priming techniques~\citep{Kim2023, Liu2023}. The two conditions were designed to elicit a bias towards either entity or incremental theory. Entity theorists believe that intelligence and capability are fixed characteristics, whereas incremental theorists believe that these traits can be improved upon over time. Participants who subscribe to incremental theory reported significantly higher on measures including perceived intelligence, likeability and intention to use than the entity theory group~\citep{Kim2023, Liu2023}. 

Another factor is the user's personality, which can also influence their responses. One investigation reported that participants that scored lower in traits of extroversion, agreeableness, and conscientiousness tended to perceive lower likeability and intelligence in the agent across all employed recovery strategies~\citep{Cahya2021}, and lower scores in open-mindedness also corresponded to lower likeability scores. Another study found a positive correlation between openness and conscientiousness with the participants’ initial trust level, indicating a risk for initial overtrust~\citep{Lu2022}. Subsequently, the study found a negative correlation with trust change as well as an interaction effect with the severity of the offence~\citep{Lu2022}, representing the correction of that initial overtrust and a high reactivity associated with those characteristics. Other studies report no significant effects~\citep{Babel2021, Aliasghari2021} related to personality measures. \citet{Pak2023} propose that cognitive capability increases the likelihood of longstanding trust repair due to the ability to gain a greater and deeper understanding. They suggest that a high \textit{Need for Cognition}, an individual's positive bias towards cognitively challenging tasks, will be correlated with a willingness to be cognitively engaged in a manner that supports a more robust trust perception. However, this association was not observed in \citetp{Kraus2023} research. 

The context of the interaction and the relationship between the recipient and the offender can also reveal varying effects~\citep{Rebensky2021}. An individual's social identity combined with their technological self-efficacy has been found to influence their perception of anthropomorphic AI systems, and subsequently, perception of their use of apology~\citep{Fan2020}. An individual with a highly interdependent self-construal, meaning a highly social and externalised sense of identity, is more susceptible to the influence of a more anthropomorphic agent, as is an individual with a lower technological self-efficacy\citep{Fan2020}. \citeta{Liu2020} found that users' needs and expectations from an interaction with a customer support chatbot differ depending on whether the conversation occurs over social media or directly on the website. Similarly, \citet{Babel2021} found some differences in how users interact with imposing cleaning robots in public compared to private contexts, although there was not a clear or significant trend. A user's sense of power, describing their self-perceived influence over others, also plays a mediating role in their perception of an offence and the subsequent apology. Specifically, sense of power mediates the relationship between perceived sincerity~\citep{Yang2023} and customer empathy~\citep{Shen2022} on recovery satisfaction. Furthermore, \citet{Karli2023} found that users had a greater perceived trust in a robot supervisor that apologised than the same robot in a student role, to the extent of demonstrating inappropriate compliance. 

One consideration and possible confounding factor not discussed in the current AI literature regards the influence of the experimental context on the measurement of the users' characteristics. The timing of measurements used in these studies influences the balance of two potential biases of opposing origins. Users interacting with unfamiliar technologies may present novelty bias. Over longer interactions, the natural decay of this novelty bias may counteract or exaggerate the underlying relational changes that the study intended to measure. Conversely, measurements taken after the interaction may be biased towards the user's short-term thoughts and feelings brought about by the study, rather than the intended long-term characteristics. The successful manipulation of user beliefs demonstrated in a number of the studies discussed above illustrates the susceptibility of users towards external influences. Some studies have already used approaches that help mitigate this risk, such as \citeta{XuX2022} who used the respondent's mood as a control variable. However, future research could benefit from investigating the possible effects of this ordering bias.

\paragraph{Identity} \label{Id} 
The \textit{identity} of a \textit{user} (\textit{recipient}) describes their intrinsic, fixed characteristics that define them as a person. In the AI apology literature, identity characteristics of study participants are generally obtained as part of demographic information. Studies commonly record information about the age and gender distribution of their sample, with some also collecting basic data about participants' prior experiences related to the context of the scenario~\citeeg{Albayram2020, Aliasghari2021, Feng2022, Fratczak2021, Jensen2022, Kim2021, Kim2023, Kox2021, Liu2023, Lv2022, Schelble2022, Xu2023, Zhang2023, Zhu2023} or type of agent used~\citeeg{Albayram2020, Babel2022, Esterwood2022, Fan2020, Fratczak2021, Karli2023, Liu2020, Natarajan2020, Pompe2022, Rogers2023, Shen2022, Song2023, Yang2022, Zhang2023a, Zhu2023}. For example, one study asked participants' prior experience with financial trading for a study using a finance game scenario~\citep{Kim2023}, and another study that investigated use of emojis in apologetic chatbots asked participants how often they used emojis~\citep{Liu2023}. Salary and occupation~\citeeg{Albayram2020, Babel2021, Kox2022, Liu2023, Song2023, XuX2022, Zhang2023}, education history~\citeeg{Albayram2020, Aliasghari2021, Babel2021, Babel2022, Fan2020, Fratczak2021, Jensen2022, Karli2023, Liu2023, Mahmood2022, Pei2023, Song2023, XuX2022, Zhu2023}, and cultural information~\citeeg{Albayram2020, Aliasghari2021, Galdon2020, Galdon2020b, Jensen2022, Kreiter2023, Rogers2023, Tewari2022, Zhang2023a} were also commonly requested. 

In most cases, this demographic data was not used for any purpose beyond being reported as to demonstrate approximate homogeneity between participant groups. However, a few studies did investigate possible implications for reception of the apology. Gender was found to have significant effects in some studies; \citet{Cahya2021} reported that male respondents tended to give lower likeability and perceived intelligence scores across all employed recovery strategies. \citet{Mahmood2023} report an interaction effect wherein male respondents (vs. female) were more responsive to apologies (vs. compensation without apology) by feminine (vs. masculine) voice assistants, also noting that male respondents were more than twice as likely to interrupt the voice assistant while it was speaking. Other studies reported no significant effects~\citep{Albayram2020, Aliasghari2021, Lu2022}. The literature has also revealed some differences in participants' responses to AI apology related to age, although the results are conflicting. Two studies report a positive correlation between age and reception of AI apology~\citep{Tewari2022, Wang2021}, supported by the suggestion that users of advanced age have a greater affinity for natural and straightforward language~\citep{Tewari2022}. Conversely, this result was flipped for another study~\citep{Lu2022}, with a number of other studies reporting that the effect of age was not significant~\citep{Albayram2020, Aliasghari2021, Cahya2021}. 

One possible cause of mixed results related to poor representation is using a sample acquisition process that is biased towards specific characteristics. The manner in which study participants are recruited can influence how closely the sample represents the population it intends to represent. For example, \citet{Fratczak2021} recruited students from the local university as study participants, all of whom were engineers. In the AI apology literature, experimental participants were recruited largely through online research or work platforms, or through research connections such as university students~\citeeg{Ahn2023, Babel2022, Cameron2021, Fratczak2021, Jelinek2023, Kim2021, Kim2023, Kox2021, Kreiter2023, Liu2023, Lv2022, LvL2022, Mahmood2023, Natarajan2020, Pei2023, Rogers2023, Schelble2022, SharifHeravi2020, Textor2022, Yang2023, Zhang2023a, ZhangX2023b} and military organisations~\citeeg{Kox2022, Textor2022}. Survey platforms include Amazon Mechanical Turk~\citeeg{Albayram2020, Aliasghari2021, Choi2021, Esterwood2021, Esterwood2022, Esterwood2023, Esterwood2023c, Fan2020, Harris2023, Hu2021, Jensen2022, LiM2023, LvL2022, Perkins2021, Perkins2022, Rogers2023, Wang2021, YuS2022, XuJ2022}, Survey Monkey~\citeeg{Lu2022}, Survey Swap~\citeeg{Kox2022}, Prolific~\citeeg{Babel2022, Cahya2021, Nesset2023, Weiler2022, ZhangX2023b}, CrowdWorks~\citeeg{Okada2023}, PollPool~\citeeg{Kox2022}, Sojump~\citeeg{Feng2022, Lv2021}, and Credamo~\citeeg{Song2023, XuX2022, Yang2022, Zhang2023, Zhu2023}. Other approaches included email and social media~\citeeg{Babel2021, Babel2022, Cameron2021, Feng2022, Fota2022, Galdon2020, Galdon2020b, Kox2022, Kreiter2023, Liu2023, Tewari2022}, online forums~\citeeg{Fota2022, Galdon2020, Galdon2020b}, live approach~\citeeg{Lv2021}, flyers~\citeeg{Babel2021, Karli2023, Kreiter2023}, and employees~\citeeg{Liu2020, Yang2023}. While studies commonly used some form of remuneration to compensate participants for their time, some studies provided additional compensation to incentivise good performance~\citeeg{Albayram2020, Jelinek2023, Jensen2022, Kim2021, Kim2023}. It is possible that these factors influence how respondents interacted with the various scenarios. However, these influences are largely unexplored. One study investigated the influence of identity bias by undertaking the study twice, once with a civilian sample and again with a military sample~\citep{Kox2022}. The study reported that the use of apology as well as another variable, \textit{uncertainty communication}, demonstrated significant main effects in the civilian sample that were not replicated in the military group. This result presents a compelling case towards encouraging further future research as to what groups of people might be less receptive to an apologetic approach, and whether there are specific features or alternative strategies that can be adjusted to better suit their needs. 

Another possible source of individual differences in responses is through the influence of cultural and linguistic variations. Differences in meaning arising from linguistic variations can impact the interpretation of a given expression between members of distinct sociolinguistic groups. For example, the phrases ``I am sorry'' and ``forgive me'' are equivalent expressions in Russian, whereas they perform distinct separate actions in English~\citep{Ogiermann2008}. For this reason, it can be superficial to directly equate apologies performed in similar contexts but delivered in two different languages. We have not found any studies that directly address this phenomenon in the AI apology literature. However, a number of the present studies have reported on the use of expressions performed in a language other than English, having translated the expressions and the findings to English for the purpose of publication. The human literature has not presented a comprehensive theory for the variance between cultural groups~\citep{Blum-kulka1984} so we don't expect to achieve that in the AI apology case either. However, some exploration to identify key areas of deviation could provide a great resource for understanding and predicting the variations that might arise when translating an approach between cultural contexts. Similarly, for linguistic differences, if future research could provide visibility and transparency as to apology scenarios that have been translated from their original language, it may be possible to advance our understanding of how these factors interact. 


\subsection{The system} \label{res:sys} 
The \textit{system} represents the final element of AI apology, as the \textit{offender} (or a proxy for them) and as the one who performs the act of apologising. In this section, we address the two moderators that describe this element, \textit{embodiment} and \textit{anthropomorphism}, and thus conclude our list of moderators described by the framework of AI apology. Finally, we address the system's capabilities represented through technical works in AI apology and relevant theoretical contributions, as well as refer to some supporting works that may help bridge the present capability gaps.

\subsubsection{Moderators of the system} \label{res:sysmod} 

\paragraph{Embodiment} \label{emb} 
The moderator \textit{embodiment} describes the physical form of the AI system. The influence of the embodiment of a system on how it perceived by the user is an area of active research in HCI~\citep{Almeida2022}. \citet{Liew2021} describe four types of artificial agents as embodied virtual agents, voice user interfaces, chatbots and social robots. Additionally, AI systems may simply be algorithms or have some minimal interface, or they may be non-social robots, including industrial robots and drones. We simplify this categorisation into three broad groups: physical robots, conversational agents, and other virtual systems (Table~\ref{tab:emb}). 

\begin{table}[h]
\setlength{\tabcolsep}{2.5pt}%
\caption[System Embodiment]{A brief overview of the systems represented in the AI apology literature, in terms of the described embodiment of the system and the format of the study}\label{tab:emb}
\begin{tabular*}{\textwidth}{@{\extracolsep} p{15pt} p{350pt}}
\toprule%
\multicolumn{2}{l}{Embodiment Study Format} \\ \midrule
\multicolumn{2}{l}{{Physical Robot}} \\
 & {Live:}~\citet{Jelinek2023, Karli2023, Morimoto2020, Natarajan2020, Pei2023, Pompe2022} \\ \cmidrule{2-2}
 & {Virtual Interactive:}~\citet{Aliasghari2021, Babel2022, Esterwood2021, Esterwood2022, Esterwood2023, Esterwood2023c, Fratczak2021, Kox2021, Perkins2021, Perkins2022, Schelble2022, Textor2022} \\ \cmidrule{2-2}
 & {Video:}~\citet{Babel2021, Babel2022, Cahya2021, Cameron2021, Choi2021, Kox2022, Kraus2023, Kreiter2023, Na2023, Nesset2023, Okada2023, Tewari2022, Zhang2023a, ZhangX2023b} \\ \cmidrule{2-2}
 & {Text:}~\citet{Fan2020, Harris2023, Hu2021, Lv2022, Shen2022, Wang2021, Yang2022}\\\midrule
\multicolumn{2}{l}{{Conversational Agent}}\\
 & {Virtual Interactive:}~\citet{Ahn2023, Liu2020, Weiler2022} \\ \cmidrule{2-2}
 & {Text:}~\citet{Fota2022, Liu2023, LvL2022, Song2023, Yang2023, YuS2022, Zhang2023, Zhu2023}\\\midrule
\multicolumn{2}{l}{{Other Virtual Systems}} \\
 & {Live:}~\citet{Kim2021, Kim2023, Mahmood2023, Xu2023} \\ \cmidrule{2-2}
 & {Virtual Interactive:}~\citet{Albayram2020, Jensen2022, LiM2023, Mahmood2022, Rogers2023, SharifHeravi2020, vanOver2020, XuJ2022} \\ \cmidrule{2-2}
 & {Video:}~\citet{Feng2022} \\ \cmidrule{2-2}
 & {Text:}~\citet{Galdon2020, Galdon2020b, Lu2022, Lv2021, XuX2022}\\
\botrule
\end{tabular*}
\end{table} 

Physical robots represent one of the most substantial use cases for AI apology in the literature. A physically embodied AI system is capable of directly interacting with and impacting its environment, which has direct implications for the type of offences that it may cause. The behaviour of a physical robot, such as approaching the user during an interaction, can cause the user to react defensively~\citep{Na2023}. Furthermore, tangibility, such as via a physical presence, can encourage feelings of trust in the user~\citep{Rebensky2021}. Examples of physical robots used in the current research include humanoid robots such as Softbank's \textit{Pepper}~\citeeg{Aliasghari2021, Babel2021, Esterwood2021, Esterwood2022, Esterwood2023c, Hu2021, Natarajan2020, Nesset2023, Okada2023, Yang2022}, mechanoid robots such as a robotic manipulator arm~\citeeg{Cahya2021, Esterwood2021, Esterwood2023, Fratczak2021, Karli2023, Kraus2023}, and small navigational systems such as drones and vacuum cleaners~\citeeg{Babel2021, Kox2022, Kreiter2023, Perkins2021, Perkins2022, Schelble2022, Textor2022}. However, most studies that investigated physical robots used virtual representations, video or text-based formats (see Table~\ref{tab:emb}), which may mitigate the influence of embodiment effects. Some in-person studies have been undertaken, using scenario formats such as games~\citeeg{Jelinek2023, Karli2023, Natarajan2020, Pei2023, Pompe2022} or a simulated customer service experience~\citeeg{Morimoto2020}. However, these studies commonly present with limitations, including low sample sizes and loss of data due to technical issues. Furthermore, each of these studies follows a predetermined static program or uses a Wizard of Oz (WoZ) manipulation, where the system is controlled by a human overseer with no actual system capabilities demonstrated. While these formats are helpful for ensuring consistent user experience, they are often not a realistic representation of the system's capabilities, demonstrating situational awareness, responsiveness, and flexibility beyond what is currently possible in real-world systems (see Section \ref{dis:capa}). Furthermore, they increase the risk that the types of offences presented in the experimental scenarios will poorly represent the future and present-day experiences of real-life users of these systems. 

Conversational agents, colloquially described as chatbots, are another common presentation of interactive AI systems. The primary purpose of these systems is to converse with a user, and thus they are generally equipped with a substantial capacity for human language. In the AI apology literature, studies have largely investigated these systems through text-based scenarios using transcripts or screenshots of simulated conversations. One exception is a customer service chatbot shown in \citetap{Liu2020} research, which was performed using an existing commercial chatbot, \textit{Moli}, but still was required to use a WoZ approach in order to demonstrate the intended behaviour. While Moli may not have been equipped with existing apologetic capabilities, chatbots are becoming increasingly advanced in more recent years and use apologies as a regular part of human language~\citep{Azaria2023, Yu2023}. 

The remaining representations of AI systems are what we have described as other virtual systems, which include virtual assistants~\citeeg{Galdon2020, Galdon2020b, Lv2021, XuX2022}, autonomous vehicles~\citeeg{Feng2022, Lu2022, Rogers2023, XuJ2022, Xu2023} and other interfaces~\citeeg{Albayram2020, Jensen2022, Kim2021, Kim2023, Mahmood2022, Mahmood2023, SharifHeravi2020}. Some studies using virtual systems have been conducted in an in-person format~\citeeg{Kim2021, Kim2023, Mahmood2023, Xu2023}, although the distinction from an online virtual format~\citeeg{Albayram2020, Jensen2022, Mahmood2022, Rogers2023, SharifHeravi2020, XuJ2022} with respect to agent embodiment effects seems insubstantial. 

Embodiment was only considered as a directly manipulated variable in two of the present studies~\citep{Natarajan2020, Pei2023}. \citet{Natarajan2020} considered four robot forms with varying humanoid (i.e. \textit{Pepper} and \textit{Nao}) through to mechanoid features (i.e. \textit{Sawyer} and \textit{Kuri}) between two presence conditions: an in-person experience and a virtual representation. The study was unable to establish statistical significance for the embodiment manipulation on either perception of the robot's anthropomorphism or user trust, which the authors accredited to the lack of physical closeness and interaction with the user and may also be impacted by the lost data or the within-subjects format. \citet{Pei2023} explored the interaction between system embodiment and AI apology using a physical robot, virtual robot, or map interface representation. The study found that the virtual and physical embodiment levels were associated with higher trust, lower distrust and lower physiological stress than the map interface, although it did not report any direct differences between the two. These insignificant results are at odds with the trends observed in broader human-machine interaction research~\citep{Almeida2022}, and warrant further investigation. 

One contributing factor to the limited number of live physical robot studies presented is the COVID-19 global pandemic, which limited opportunities for in-person experiments. Coupled with the existing challenges associated with live experiments, this has resulted in an increasingly sparse representation of embodied robotics studies in AI apology. Although the limitations against in-person experiments are largely repealed at the time of writing, the limited representation of this work is at risk of self-reinforcement. Difficulties arising from live experimentation can prevent researchers from obtaining sufficient high-quality data to support their hypotheses, which can then impede access to the additional resources required to resolve the issues. Insignificant results greatly increase the possibility that the research is not published, or, if it is, that it will have a low impact, leading to limited growth in that area.

\paragraph{Anthropomorphism} \label{ant} 
\textit{Anthropomorphism} is the presence and extent of human-like characteristics in non-humans, which can impact the expectations of the \textit{user} \citep[\textit{recipient};][]{Liew2021}. It can be represented in a broad number of ways; Figure \ref{fig:ant} describes a few examples from the AI apology literature. Higher anthropomorphism in technology has been shown to increase positive user perceptions and support adoption, and can help preserve user trust in the wake of an \textit{offence}~\citep{Fan2020, Rebensky2021}. 

\begin{figure}[!h]
\centering
\includegraphics[width=73.36mm]{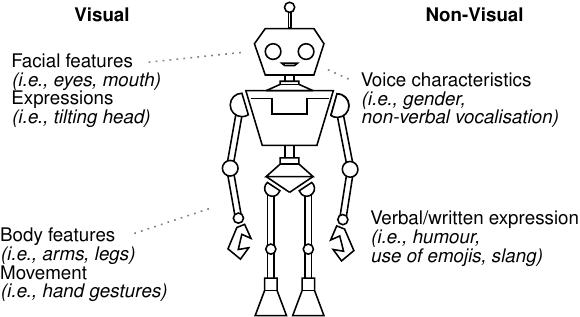} 
\caption[Anthropomorphism in AI apology]{Examples of how anthropomorphism has been represented in AI apology}
\label{fig:ant}
\end{figure}

The underlying mechanism for the beneficial effects of anthropomorphism is the tendency of humans to prefer interactions with other humans. Research has shown there is an innate disadvantage in efficacy of apologies and similar social interactions by AI systems compared to their human counterparts~\citep{Babel2021, Choi2021, Hu2021, Xu2023, Zhang2023, Zhu2023}. Even for an apology from a human offender, simply notifying the user that an AI algorithm was involved in the process can significantly impact its effect~\citep{Brown-Devlin2022, Glikson2023}. For example, \citet{Glikson2023} revealed how human apologies that have been refined using AI language tools, even expressly designed for the purpose of supporting effective communication, are perceived as less authentic by the receiver. This effect was seen in an experiment using a text-based scenario that provides no evidence of any effect, or any other differences in the prompt excepting a brief statement in the scenario context stating that the message was written with the support of an AI program. This is not to say that AI apology is ineffective, especially in comparison to no apology (see Section \ref{res:out}). Furthermore, there is one study~\citep{YuS2022} that reported a scenario in which AI's apology was preferred over that of the human; in the case of a non-emotional service rejection. This aligns with the central theory that users will prefer the configuration that most closely resembles their pre-existing expectations. In this case, it appears that the presumed lower flexibility of the AI service agent partially excuses the offence of the rejection. 

In a series of four studies, \citet{Choi2021} investigated the effects of anthropomorphism on robot service recovery through a comparison of two different representations of a humanoid (i.e. \textit{Nao}) and a non-humanoid robot (a visually simplified and largely featureless equivalent). This research found that an apology delivered by a humanoid robot facilitated greater satisfaction and perceived warmth than either the non-humanoid or the control strategy (e.g. ``Hope you enjoy the evening''). \citet{Fota2022} reported a positive effect of anthropomorphism on customers' repurchase intention, and \citet{Natarajan2020} found a positive correlation between the perceived anthropomorphism and the user's trust. Conversely, \citet{Jensen2022} found the high-anthropomorphic agent was associated with lower trust. However, as the context of the study was trust dampening, this is not a contradicting result. Rather, these findings suggest that the high-anthropomorphic agent may have been able to better communicate the intended trust dampening effects to the user, as participants under that condition were more closely aligned with the expected result. \citeta{Yang2022} also confirmed previous results for the supporting effect of anthropomorphism on service recovery, with respect to a boundary condition. In the case of the poorly received humour expression arising from the high-severity condition, the amplifying effect of anthropomorphism resulted in a greater negative result for the high-anthropomorphism condition. A final note regarding the anthropomorphism manipulation used in this study is that the low-anthropomorphic robot was depicted as a Pepper robot, which has generally been used to represent the high-anthropomorphic condition in other studies. Instead, the high-anthropomorphic robot (discussed here using the exact same label) was depicted as a young Chinese businesswoman. The most notable possible implication of this change is the increased proximity to the uncanny boundary condition for the latter depiction. 

The literature also includes some exploration of less typical expressions of anthropomorphism; \citet{Mahmood2023} found that perceptions of warmth were higher for an apologetic virtual assistant with a feminine (vs. ambiguous) voice. A series of studies by \citeta{Lv2021} found that `cuteness', in particular, had a substantial effect on consumer's responses to failing service agents. Investigated over an array of offences of varying types, severity and time pressure conditions, as well as an array of presentation forms including voice, human and non-human-like visual features, and language, this study found that a `cute' system presentation almost always was associated with higher tolerance of failure and higher tenderness, yet lower performance expectancy from the recipient. The only circumstances for which this is no longer the case are in high-severity or high-time pressure scenarios. In these cases, the effect is flipped such that the low-`cuteness' agent was preferred. 

While the effect of anthropomorphism on the users' perception of the system is generally positive, two studies from the literature were not supportive. The first of these studies is by \citet{SharifHeravi2020}, who investigated a verbal, avatar-based, and modal anthropomorphism manipulation, but found two thirds of the participants preferred the low-anthropomorphism case and thought that it was more reliable and competent. However, there are a number of possible explanations for why this result was observed. Firstly, the manipulation check for the anthropomorphism of the system was not successful, indicating that the study would not be able to draw significant conclusions from the data based on this claim. Secondly, the verbal aspect of the anthropomorphism manipulation involved additional expressions not balanced in the other case, which included bringing attention to the agent's poor performance regarding less serious mistakes, as well as pointing out when the user made a mistake which could be seen as disrespectful. The second study is by \citet{Lu2022}, who explored an emoji-based anthropomorphism manipulation but did not find any significant effect. Moreover, some experimental participants provided responses that suggested they found it annoying, especially in high-criticality situations. Other studies in the AI apology literature have also explored the use of emojis to positive effect~\citep{Liu2023}, although this study did not address anthropomorphism even as a possible confounding variable, thus cannot provide any more specific insights to explain this result. 

Some studies have identified interaction effects between anthropomorphism and other attributes of apology, although possible mechanisms for these effects are not fully clear. \citet{Esterwood2021} reported an interaction between anthropomorphism and the trust repair strategy used, specifically regarding the trustworthiness dimension of benevolence. This study found that explanations were the most effective at restoring perceptions of benevolence for an anthropomorphic robot, but that promises were more effective if the robot was mechanoid. However, this result was not expected, and the study did not present any possible explanations beyond the prospect of a complex and non-linear relationship. \citet{Kim2021} investigated the relationship between anthropomorphism and responsibility attribution, reporting user preferences for internally attributed fault from a human-like agent but external attribution for a machine-like agent. Regarding dimensions of trust, the study found that cognitive trust patterns differed between the scenarios for machine-like agents but did not for human-like agents, whereas the inverse was true for behavioural trust. The study did not elaborate on the possible implications of this relationship, but emphasised the need to carefully consider the use of anthropomorphic cues in different contexts and to further investigate the complex influence of anthropogenic agents on user perception. \citet{Porra2020} raises a related concern, arguing that anthropomorphic presentation allows the user to attribute greater intelligence and agency to the system than is justified, leading to an over inflated sense of their capabilities. This theory is supported in \citetp{Ahn2023} study, which established a positive link between anthropomorphic characteristics and the perceived freewill of the agent, and subsequently translates to a preference towards an apologetic request refusal (vs. an excuse). 

In general, these results suggest that the anthropomorphism of an AI system can influence both the manner and extent of the effect of an apology. However, it is not clear if there is any pattern to these effects. More research will be needed before the influence of anthropomorphism on a specific apology can be made clear. However, if this relationship can be understood, it may be able to inform the appropriate structure for an apology for a system of a given form. This could enable apologetic capabilities retrofitted to existing AI systems to gain access to the wealth of value that AI apology represents. 

As a final note, we highlight one particular technical contribution of interest to this topic, regarding the development and distribution of a simulation environment specifically designed for research in AI apology. \citet{Esterwood2023b} published the platform used for their associated human research studies~\citeyearpar{Esterwood2021, Esterwood2022, Esterwood2023}, entitled `The Warehouse Robot Interaction Sim'. The platform facilitates an interactive, collaborative, sorting task scenario with a robot. The participant's task is to classify boxes that the robot will then move accordingly, observing and interacting with the environment from the human worker's perspective through on-screen controls. The simulation includes three alterable features; the robot form (humanoid or machine-like), performance (with errors or without), and communication (customisable), and is pre-programmed with the approaches used in the author's prior research (apology, denial, promise, explanation and none)~\citeyearpar{Esterwood2021, Esterwood2022}. As noted by the authors, the distribution of this platform is an asset for future research in AI apology, providing researchers access to an existing, publication-quality virtual experimental environment and facilitating reproducible and extendable experimental designs.

\subsubsection{Capabilities of the system} \label{res:syscap} 
In section \ref{def:apol}, we proposed that a self-initiated apology requires four key capabilities: detection, attribution, explanation, and adaptation (Figure \ref{fig:capa}). In this section, we address research that considers these capabilities in the pursuit of practical implementation of autonomous AI apology, including some technical works and additional theoretical contributions from the literature. 

Our review of the field reveals very few applied technical studies that explicitly work towards real-world implementation of AI apology. \citet{Harland2023} propose the earliest identified attempt at autonomous AI apology, through the novel application of a multi-objective reinforcement learning (MORL) algorithm inspired by AI safety techniques for human-alignment~\citep{Vamplew2018}. This work describes a step-wise process of generating an apology, `Act, Assess, Apologise', that enables the agent to identify misaligned behaviour and self-correct. In addition to its primary task, the agent pays attention to other changes in the environment caused by its actions, and assesses which of these changes may be upsetting to the user by their responses. It uses this information to determine if an offence has occurred (detection) and how to apologise for the offence, with regard to appropriately identifying (attribution), understanding (explanation), and correcting (adaptation) the misalignment. The authors propose that their algorithm may help to bridge the gap between the solely social perspective of apology that is representative of the field and the technical functionality of AI systems. While the study has some limitations that could be addressed in future research, such as having used simulated (rather than real-life) users and a deterministic, tabular (rather than stochastic, continuous) environment, it serves as a starting point for integrating apologetic capabilities in AI systems. 

Conversational agents are another technical area prominent in the AI apology literature, as an example of apologetic systems commonly used in the real-world. Large Language Models (LLMs) and Natural Language Processors (NLPs), such as those that drive autonomous interactive chatbots including \textit{ChatGPT}~\citep{Azaria2023} and \textit{Copilot} (previously known as the \textit{Bing chatbot}), are AI algorithms that describe the intersection between human communication and AI systems, of which apology is a core part. During interactions with human users, the agent may apologise to express a limitation of its function \citep[e.g. ``I'm sorry, but as an AI language model, I cannot say...'';][]{Wester2023}, or to provide a correction for a mistake \citep[e.g. ``I apologize for the mistake. Here is the correct solution...'';][]{Azaria2023}. It is unclear to what extent these agents represent the capabilities for detection, attribution, explanation, and adaptation; these ideas remain largely unaddressed in the literature. However, a few analyses regarding such algorithms have emerged~\cite[e.g.][]{Azaria2023, Koc2023, Wester2023, Yu2023}, providing some initial insights and enabling some linkages which shall be discussed below.

\paragraph{Detect} \label{detect}
Following the apology prerequisite of an \textit{offence}, in order for the system to gain awareness of the need to apologise, it must be able to recognise that an offence has occurred. This capability is distinct from error detection in general, as offences do not necessarily translate to reputable errors, but rather describe a broad set of possible misalignment: such as social norm violations \citeeg{Cahya2021, Tewari2022}, intentional unethical actions \citeeg{Rogers2023, Schelble2022, Textor2022}, or conflicting goals \citeeg{Babel2022, Tewari2022}. 

\citet{Tolmeijer2020} discuss capability requirements for autonomous trust repair as part of their review on trust-relevant failures and mitigation strategies, including detection and explanation. The study describes some theoretical approaches for devising that a failure has occurred, such as built-in model checks and simulations that assess the deviation of the system from a desired state. For less clear failure conditions, they refer to research on anomaly detection and knowledge derived from generating explanations. 

An alternative approach to detection proposed by a number of studies is via recognition of loss of trust by the user. \citet{Mahmood2022} provided some thoughts on how an effective apologetic AI system could be designed following their study on apology in voice assistants. They suggest that the capability of the AI agent to recognise when an apology is required could be facilitated through user feedback, both via the user's explicit indication of the error or via behavioural cues, in addition to through system-recognised errors. 

\citet{Stiber2023} presented a framework for identifying offences in complex HRI situations using social signals from users. They describe four main aspects of error management that correspond closely to the four capabilities proposed in this review: error detection, classification, mitigation, and recovery. By collecting a dataset of implicit social signals from experimental participants, during collaborative interactions with a robotic arm across task contexts such as cooking and object assembly, the authors developed an error recognition system to identify offences as they occur. The social signals used in this dataset were encoded as facial action units (AU) by an existing program, \textit{OpenFace}, as well as explicit verbal error notifications expressed by the participants upon recognition. The model architecture was an assembly comprising a classifier for each time step (i.e., error or no-error occurred) and with sequence validation based on labelled cases (i.e., subsequent verbal notification of error). Specifically related to the present topic, the authors describe how this model could be used to automatically generate apologies for such errors, and could distinguish between different types of errors within the scenarios with an accuracy of $92\%$. 

\citetp{Fratczak2021} post-hoc clustering analysis describes an advantage to this approach, in that it identifies reactive and unreactive user groups and identifies a correlation in the unreactive users with subsequent disinterest in the agent's apology. The study uses these findings to emphasise the importance of understanding the user's needs when designing an apologetic system. We also submit that this result describes how, for offences that do not constitute genuine technical errors or objectively problematic behaviours, the user's response to an agent's unexpected behaviours may be a suitable predictor of both the presence of an offence and the user's receptiveness to a subsequent apology.

\paragraph{Attribute} \label{attribute}
Extending from the system's requirement to have recognised an \textit{offence}, in order for an apology to be applicable, the system must \textit{attribute} itself with responsibility for the offence. In order for the system to recognise its causal influence in an offence, it must be able to understand the impact of its behaviour in general. This is the problem that is sought to be addressed through the study of causal learning: a field of research that seeks to uncover the underlying data generating process inherent to an AI algorithm. 

Further on the review by \citet{Tolmeijer2020}, the agent may derive understanding of the impacts of its various behaviours through the process of generating explanations, as explored in eXplainable AI (XAI) research~\citep{Dazeley2021a}. There are two major aspects to this process: the development of information relevant to the explanation, and the expression of this information in an explanation, or in this case, an apology (see Section \ref{explain}). \textit{Attribution} relies on the former aspect; a suitably accurate explanation of the cause of an event should identify any actions that had causal effect, and if these actions were taken by the agent, then it can determine that it is at fault. \citet{Tolmeijer2020} summarise the process of formally describing the actions, features and preconditions relevant to the agent's environment, and the relationship between them. Subsequently, they describe how this formalism can be used to define the plans and expectations of the system which can be compared to the outcome state to identify discrepancies. 

\citet{Cheng2021a} surveyed a subset of approaches in causal learning, motivated by the pursuit of \textit{socially responsible AI}, and proposed a taxonomy consisting of seven tools inherently related to this task. The seven tools include the encoding of causal assumptions, counterfactual analysis, mediation analysis for identifying direct and indirect effects, adaptability in changing environments, imputation of missing data, and causal discovery. The application of these approaches may assist in developing the underlying reasoning capabilities required for autonomous apology. Functionally, this requirement resolves to a need for the agent to demonstrate dynamic situational awareness, having access to enough knowledge and sufficient capability to enable logical reasoning in cause and effect. The survey notes some current applications of causal modelling in mitigating gender bias in natural language processing, recognising structures in image processing, and inflating agent knowledge by extrapolating from known experiences. However, these approaches are not widespread and are still subjected to many early challenges, especially relevant to flexibility, including accounting for hidden social biases, and catering for non-static and heterogeneous environments.

\paragraph{Explain} \label{explain}
Given that the system has detected that an \textit{offence} has occurred and recognised its causal influence on it, it must be able to translate this knowledge into an apology. In the articulation of the apology, the system must construct each of the components that will be included. These components may reveal information related to the agent's understanding of what offence has occurred and why, and provide an explanation of its decision process leading to that behaviour. Conversely, it may take a more minimal form. 

The process of producing interpretable information from the knowledge structures of AI systems is largely addressed in the study of eXplainable AI (XAI). \citet{Dazeley2021a} provide a recent review of the field, and, moreover, present a model for explainability in AI based on increasing detail through an interactive conversational framework, thus adapting the depth of explanation (or alternatively, apology) to the needs of the user. The article presents two entwined contributions relevant to the progression of AI apology; the conversational model of explanation, and the framework that delineates the levels of explanations and establishes the vocabulary required to discuss the concepts in-depth. The conversational approach may help address the previously noted limitations of \textit{dialogue} as it presently appears in apologetic systems, by providing responses based on a dynamic model of the users' current knowledge and informational needs. The framework is built on an idea of \textit{intentionality} in behaviour adapted from Animal Cognitive Ethology research, and introduces five levels of explanations relating to increasingly broadened perspectives of factors in play. The levels include reactive (0\textsuperscript{th} order), dispositional (1\textsuperscript{st} order), social (2\textsuperscript{nd} order), cultural (3\textsuperscript{rd} order), and reflective (meta) explanations. Each level describes the contextual awareness or interpretation of influences relevant to that level of abstraction; from the agent's isolated internal processes of interpretation and motivation, through the systemic processes underpinning immediate and broader interactions with external agents, to the processes governing how the agent is able to interpret and represent its own cognition in order to form these explanations. 

Other contributions include work by \citet{Tsakalakis2022} that presented a taxonomy for explanations driven by the principle of `Explainability-by-Design'. The taxonomy describes nine dimensions representing different perspectives that may be consulted to develop a comprehensive explanation: source (e.g. origin of requirement), perspective (e.g. timing in relation to event), autonomy, trigger, content, scope, explainability goal, intended recipient and priority. 

A prominent exemplar of recent technologies demonstrating the use of apologies is LLMs, with some recent explorations providing some insight as to the linguistic comprehension of these systems in relation to apology. \citet{Yu2023} explored the use of LLMs to automate the process of linguistic analysis, using apology as a case study. The study investigates the task of local grammar annotation, usually a gruelling manual process, that regards the study of a specific speech act and identification of its functional elements in a body of text. The pre-identified functional tags for local grammar analysis of apology in English included \textit{apologiser} (one who apologises), \textit{reason} (for apologising, was \textit{specification}), \textit{apologising} (expression that realises the act), \textit{apologisee} (one who receives the apology), and \textit{intensifier} (calibration of the degree) labels. Two LLM-based chatbots, \textit{ChatGPT-3.5} and the then-named \textit{Bing chatbot} (now \textit{Copilot}) in \textit{More Precise} mode, were provided with a description of the task and these functional tags before their annotation skills were tested against a corpus of 1000 apologetic utterances. While the core motivation for the study was the investigation and demonstration of LLM capabilities for linguistic analysis in general, it does provide some insight regarding the capability of these models to understand the context of an apology, necessary for \textit{attribution}. Overall, the performance of the leading model was impressive ($92.7\%$ accuracy, compared to $95.4\%$ for the human annotator), although it demonstrated some predictable limitations. Specifically, the study noted the weaknesses of the chatbot annotators as a high rate of false-positive identifications of keywords in irrelevant contexts compared to a human annotator, which were amplified in non-conventional expressions. Similarly, \citet{Koc2023} evaluated the capability of ChatGPT-4 to generate responses to customer complaints, reporting that the system was able to generate highly satisfactory apology letters to example reviews (rated 4.78/5 by a panel of 40 industry experts). However, this language capability is not a representation of the system's own capability to apologise. As \citet{Wester2023} point out, many instances of LLMs' use of apologies are hard-coded by designers to suppress responses of moral significance; an approach they criticise as suppressing the systems' transparency and projecting the designers' moral bias onto the user. In combination, these result suggests that, while LLMs may be able to reproduce suitable apologetic expressions for a given prompt, these models are still limited with respect to demonstrating the level of abstract thought required for an apology.

\paragraph{Adapt} \label{adapt}
The final capability requirement for an apology to fulfil the definition set out in the literature is to \textit{adapt} its behaviour to align with the expectations set out in the apology. This is the functionality necessary to realise the \textit{demonstration} of the system's \textit{reform} and exaction of \textit{repair}. Adaptivity is not limited to functional behaviour, but also includes non-functional behaviours, such as communication style. \citet{Tewari2022} describe how socially intelligent agents need the capability to develop and maintain an in-depth understanding of the user and the broader interaction context, and act in alignment with the users' expectations, including autonomously resolving conflicts that arise. 

Behavioural adaptation aligned with an apology was a major feature of the apologetic algorithm by \citet{Harland2023} previously introduced (Section \ref{dis:capa}). The algorithm dynamically adjusts the agent's priorities to avoid undesirable actions for a pre-learned task by leveraging the representation of the agent's behavioural policy given a multi-objective expression of utility. The user becomes upset when the agent interferes with specific features of the environment, each of which are mapped to a set of auxiliary objectives. When an offence occurs, having demonstrated each of the prior steps, the agent updates its policy by adjusting the threshold values used to evaluate its expected utility, amplifying the penalty for impacts against the associated objective. Subsequently, the agent follows this updated policy, balancing the priority of the main task with the user’s needs, such that its performance is slightly less efficient overall but successfully avoids the undesirable result.

\citet{Ahmad2017} reviewed adaptivity in social robots, identifying a broad range of adaptation techniques used in Human-Robot interactions to improve user alignment through behaviour. The theoretical capabilities of adaptive robots as described include understanding and showing emotions, dialogue communication with a user, learning and adapting based on user responses, and engaging in social behaviour suited to the present context, relationship, and role. While these capabilities have substantial overlap with those described and proposed as relevant for apologetic agents, this review does not directly address adaptivity in the context of apology. However, the ethical and social implications identified are worthy of translation. Specifically, this review raises concerns around the appropriate management of user privacy given the possible creation of sensitive data that may arise from modelling user preferences and patterns of interaction, and subsequent user discomfort regarding adaptive behaviours that violate the expectation that the system's behaviour is fixed. Recent work by \citet{Ahmad2022} explored principles for designing personality-adaptive conversational agents in the context of mental health services through a qualitative study including consultation with a panel of experts. In brief, the principles suggest that such systems should also be adaptive to the users' needs in terms of the social role (DP4), degree of anthropomorphism (DP5), and level of personality adaptation (DP6) employed. Moreover, they should be proactive in their support (DP1), competent (DP2), and transparent (DP3). 

The conversational AI systems also previously discussed (Section \ref{dis:capa}) demonstrate a degree of adaptation throughout the process of a conversation. Many live chatbots will use apologies to politely retract a position stated or provide a rebuff, after the user has indicated a discrepancy or mistake~\citep{Azaria2023}. The agreeable design of the algorithm and lack of functions to verify the accuracy of a specific statement lends itself to readiness to concede its prior statements to conform to a user's correction. This could be interpreted as a demonstration of \textit{adaptation} in AI apology. However, this use case presents an intriguing social dynamic, especially with regard to how the conversation might continue given the updated context. In order to continue the conversation, the agent might hallucinate statements to agree with the correction or demonstrate similarly undesirable behaviour that utterly undermines its use. Beyond the case of language models, analogous behaviour in a physical system may go beyond inconvenience to represent serious danger for the user. However, over the course of this review, we did not encounter any research directly addressing the implications of using apologies in this way, and advocate for further exploration of the topic. 


\section{Discussion} \label{disc}
Based on the findings of our review described in Section \ref{res} and the framework for AI apology proposed in section \ref{def:apol}, we now present a discussion summarising the state of AI apology research. We have identified the main gaps and limitations of the field as belonging to four key areas: the composition of apologies (Section~\ref{dis:comp}), challenges for human-alignment (Section~\ref{dis:hum}), research designs and measures (Section~\ref{dis:res}), and system capabilities (Section~\ref{dis:capa}) (Figure \ref{fig:dis}). For each of these areas, we discuss these gaps and propose opportunities for future research. 

\begin{figure}[!h]
\centering
\includegraphics[width=91.28mm]{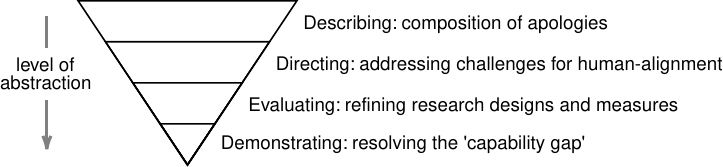} 
\caption[Areas to be addressed]{Four key areas to be addressed in future AI apology research}
\label{fig:dis}
\end{figure}


\subsection{Composition of apologies in AI}\label{dis:comp}
The framework of AI apology describes the composition of an apology according to twelve possible components: \textit{cue}, \textit{responsibility}, \textit{affirmation}, \textit{explanation}, \textit{moral admission}, \textit{regret}, \textit{reform}, \textit{repair}, \textit{petition}, \textit{dialogue}, \textit{engagement}, and \textit{demonstration} (Table \ref{tab:comp}). The representation of each of these components in the AI apology literature is varied. While a substantial amount of work has been undertaken towards identifying which of these are important components of an apology, and the effects and interactions between the subsequent moderators, there is still a lot left unexplored. Furthermore, which components and moderators to include in which forms to suit a given context is also an underexposed area. 

Reflecting on Table \ref{IntStudies}, we note that the most commonly used components are \textit{cue}, \textit{affirmation}, \textit{explanation}, and \textit{responsibility}. Similarly, apologies that are composed of either three or four of these components are the most common configuration\footnote{The most commonly appearing compositions for apologies used in the literature are the combination of \textit{cue} and \textit{affirmation} with: an \textit{explanation} (seen in 6 studies), \textit{responsibility} (5 studies), both (4 studies), and both also including \textit{reform} (4 studies).}. However, it is \textit{responsibility} and \textit{engagement} that have garnered the greatest research focus. Conversely, the components of \textit{petition} and \textit{moral admission} have been explored very little, followed closely by \textit{dialogue} and \textit{repair}. Where they have been studied, almost all of these components show mixed effects that are not well explained by theory. The exceptions to this are the components of \textit{affirmation} and \textit{repair}, which appear to be generally well received, whereas the inverse is true for \textit{moral admission}. \textit{Reform} and \textit{dialogue} might also have mostly supporting evidence, if not for interference from limited system capabilities. We suggest that the central limiting theme of this field of research is the lack of congruence between studies as to the use and expression of various components of apology, and recommend that resolving this be a goal of future research. 

We propose three areas for exploration as future research avenues. The first proposal regards the distinction between those components of apology imprecisely defined and commonly interchanged in the literature. The components of \textit{cue}, \textit{responsibility}, \textit{regret}, and, to a lesser extent, \textit{petition} are each described by phrases of ``I'm sorry'', or ``Forgive me'', yet also appear as distinct concepts. Establishing whether a clear distinction can be drawn, and determining how it can be defined if so, is a question that could be explored in future research. 

The second proposal further regards the component of \textit{responsibility}. This component was one of the most thoroughly examined in the literature, yet the findings remain conflicted. A possible cause for this conflict is the inconsistency in the way that responsibility (or non-responsibility) is expressed between studies using the same description. In order to establish clear relationships between other contexts and features of apology and the manner of responsibility expression, there must be continuity in how these expressions are labelled. We propose that, for the sake of consistency, the term `[taking] responsibility' is only used to describe a forthright and internally attributed expression of responsibility. All other forms of the expression constitute a non-responsibility condition, which could be described using an appropriate label such as `by proxy' (i.e., as a representative of the responsible party, e.g. from a service agent), `externally attributed' (i.e., redirected to some identified external party, e.g. blaming the user), `deflected' (i.e., redirected away, so the attribution is unclear, e.g. ``I am not at fault''), or `sympathetic' (i.e., attribution is unclear as there is no responsible party, e.g. ``it is not your fault'') responsibility. Furthermore, future research may benefit from further exploration of these types of expressions to distinguish the influence of a particular phrasing from that of anthropomorphism~\citep{Kim2021, Kim2023} or other contextual features~\citep{Esterwood2022b} on the perceived responsibility of the system. We also recommend that future research consider validating their intended manipulations using approaches such as sentiment analysis and manipulation checks, especially when varying features of the expression between two conditions. 

The third proposal regards the sparsely considered component of \textit{moral admission}. There were only a few instances in the reviewed literature that explored the topic of immoral agent behaviour, and only one study involved the agent's admission of immoral behaviour~\citep{Rogers2023}. In contrast to the mixed results observed in the case of \textit{responsibility}, this study indicates a distinctly poor reception of the moral apology. Future research might consider validating these findings in other scenarios. Furthermore, other research has investigated the influence of users' implicit beliefs on willingness to accept an apology from an AI system. As a moral admission inherently describes the undertaking of knowingly problematic behaviour and its subsequent renunciation, implicit beliefs could have a moderating role on its perception. 


\subsection{Challenges for human-alignment}\label{dis:hum}
In order for the system to be aligned with the needs of its human users, it is necessary to understand the nature of the evolving relationship between them. A longitudinal perspective on any repeated interaction is crucial for understanding how trust is developed, maintained and calibrated in human-AI relationships~\citep{deVisser2020}. Humans engaging in repeated interactions with social robots over extended periods allows for the attenuation of novelty effects and the development of increased familiarity~\citep{Leite2013}. Some research has begun to investigate how robots influence their human users over long-term interactions in non-apologetic contexts~\citep{Sagheb2023}. However, research that considered longitudinal effects of apology in the human-robot relationship is limited, where only a few \citep{Babel2022, Esterwood2021, Esterwood2022, Esterwood2023} reported repeating the applied strategies multiple times, and none over an extended period. Furthermore, studies that have investigated the effects of repeated errors only appear to have considered repeating the same repair strategy for each error. The implications of this are twofold. Firstly, if the expression that is used by the system does not vary at all between instances, the respondent may perceive it as repetitive, which may undermine its effects. Secondly, the impact of a repeated error following a given recovery approach is likely to differ based on the expression itself. Different verbal expressions can each have distinct and explicit meanings, which may have differing effects on the user's expectations of the agent's future behaviour. For example, if the agent expresses an intent to reform, the user may expect the agent not to make the error again. Conversely, the user may expect the error to be repeated if the agent explains that the cause of the error is something beyond their control. In the former case, the repeated error is a violation of both the user's expectations of the agent's ability to perform and the agent's expressed intent, whereas it is aligned with the user's expectations in the latter. Future research could investigate the influence of these factors in human-agent interactions, and consider longitudinal studies to gain greater understanding about the long-term effects of AI apology. Such studies can also reveal expectations that humans have from AI apology that differ from human apology, considerations that are underexplored in the current literature, supporting the extension of the framework developed in this review as well. 

A derivative prospect regards the boundary conditions for instances of user resistance, as observed in some studies~\citep{Albayram2020, Lu2022, Rogers2023, SharifHeravi2020, Textor2022}. When given the opportunity to provide open-ended comments, user feedback often contains a proportion of complaints about the system's use of apology in general~\citep{Albayram2020}. However, the proportion of respondents that express this position is not substantial enough to undermine the general positive findings associated with it~\citep{Cahya2021}. One possible model for this discrepancy can be derived from theories of human cognitive processes. The elaboration likelihood model suggests that when users interact with a system, their perception of that interaction is determined by an array of central and peripheral processes~\citep{Petty1981}. Habitual communication is mostly processed in the peripheral~\citep{Petty1981}. However, when an aspect of communication is unexpected, it may instead require conscious intervention to process and resolve~\citep{Harmon-Jones2019}. In a central cognitive process, ideas may be evaluated more critically as the listener actively applies their conscious beliefs~\citep{Petty1981}. \citet{Pak2023} discussed similar ideas, and emphasised how the central route to persuasion is enacted for trust repair processes that are information heavy and seek to convince the recipient through cognition. Conversely, the peripheral route to persuasion is information light but appeals to the emotions of the recipient and relies on subconscious processing. According to the examples used to describe this model, apology was listed as a strategy relevant to the peripheral route. We counter that this is true of a \textit{shallow} apology, one lacking in substantive content or meaning, but not of apology as holistic approach. However, these ideas have not yet been empirically evaluated. Furthermore, these constituent theories represent a simplified account for the complex and dynamic process of human cognition~\citep{Kitchen2014}. There remains a substantial opportunity for these ideas to be expanded, validated, or refuted. 

Finally, a few studies in this review structured their arguments around the rights of the user. Recalling the aforementioned `right to reparation'~\citep{Galdon2020, Galdon2020b} and `right to explanation'~\citep{Kim2022} in contrast with the preservation of \textit{humanness}~\citep{Porra2020}, it can appear as if these rights are contradictory. However, we propose that these can be acknowledged simultaneously under the appropriate framework, and, that such a framework would present a solution to many of the outstanding questions on the topic. The crucial overlap between these perspectives lies in the emphasis and prioritisation of human needs in the implementation of technological systems with which they may interact. Features of human communication, especially but not limited to apologies, have an important function in sustaining healthy interactions. Communicating ideas and expressing reasoning and intent behind behaviour, especially when it is not self-evident, is crucial for effective collaboration and coexistence between distinct individuals. Furthermore, using subjective, self-referencing language to describe these states and ideas is both the natural form of human communication and often the only way to effectively caveat the inherent bias of any one perspective. Expressive speech acts such as apology serve as meta-commentary, giving the user a glimpse into the internal processes that brought about past behaviour and from which future behaviour will be derived. While human behaviour is largely governed by the dynamic interplay of norms, morals, emotions and needs, and these concepts are not directly mirrored in the AI case: AI systems themselves have analogical internal frameworks and processes governing them. Thus, the processes of communication that have been developed to facilitate the functional coexistence of humans with each other are, in the least, the most promising starting place for their equivalent processes in AI systems. However, these processes are also complex and nuanced in themselves, and thus any efforts towards their automation must be properly examined and weighed against these same aforementioned risks. 


\subsection{Research designs and measures}\label{dis:res}
Beyond the themes and concepts explored in the AI apology literature, there are also some features related to how this research has been undertaken that are worth discussing. Low statistical power resulting in insignificant results is one of the most prominent limitations to present AI apology research. Despite a good possibility of an underlying effect, a number of studies reported non-significant results~\citeeg{Feng2022, Jensen2022, Kim2023, Kreiter2023, Lu2022, Natarajan2020, Perkins2022, SharifHeravi2020, XuJ2022, Zhang2023a}. One likely contribution to this effect is the small sample sizes used for some of these works. Some studies had a sufficiently small and varied sample that they found clear interference due to random differences between the groups~\citep{Kox2021}. The most effective recommendation that can be made to address the issue of small sample sizes is to gather more results. It may not be possible to gain access to additional resources to resolve this. However, it is still possible to improve the power of an experiment without extensive additional resources, by minimising the influence of other limitations. These will be discussed below. 

Aside from sample size, another possible limitation stems from the non-specific description of apologies and related manipulations present in AI apology research. Given the broad array of variables under investigation in addition to the individual differences in the interpretation of these variables between these studies, it can be difficult to identify consistent trends and to distinguish the effects of different features. For example, in Section \ref{resp} we described a study wherein the agent expressed confusion in both the apology condition and the contrasted denial~\citep{Perkins2022}. Similarly, in Section \ref{ref}, a study comparing an apology to an expression of gratitude included used an expression of reform in both cases~\citep{LvL2022}. Furthermore, some studies did not report the apologies that were used at all~\citeeg{Fota2022, Jelinek2023}. A lack of visible and well-labelled apologies may also limit the reproducibility of the study, or its comparison against other similar works. Reproducibility is a key principle of scientific research, yet is recognised as a weakness in HRI~\citep{Gunes2022}. To protect the reproducibility of research in AI apology, we recommend that future works fully report the approaches used and explicitly state any scripted communication provided to participants. We also suggest the inclusion of a baseline condition without an apology to clarify the value of different approaches. 

Effect sizes can be increased without increasing sample size through improvements to the validity of the experimental design~\citep{Meyvis2018}. In order to address limitations related to experimental validity, some studies have sought to verify that their proposed approach successfully represents the concepts that they intend to explore through the use of pretests~\citep{LvL2022} and manipulation checks~\citep{Albayram2020}. For example, using sentiment analysis to select the most effective phrasing for describing a regretful expression~\citep{XuJ2022}. These approaches can be effective for guiding study design~\citep{LvL2022}, and their use is encouraged. However, researchers should keep in mind that they are not an assured solution. The use of pretests does not guarantee that the study will find significant results~\citep{XuJ2022}, even for measurement instruments and experimental conditions that have been found to provide significant results in similar studies. Conversely, a manipulation check can not influence an insignificant result as it is performed in parallel with the main study~\citep{SharifHeravi2020}, although it does provide insight as to the validity of the design~\citep{Yang2022}. 

A broad array of concepts are used to evaluate the results of these studies (see Table \ref{tab:meas}), with a similarly broad array of both established and novel measurement instruments. While we acknowledge the need for measurement instruments to be varied to some degree, to be fitted to suit their present context and purpose, the present literature appears to have little consistency in measurement instruments, even beyond this requirement. With such an extensive array of concepts and related instruments, with minimal oversight as to their similarities and differences, it is difficult to be confident that the measurement appropriately describes the intended concepts and the same concepts as described by a similarly labelled instrument. Potential risks include difficulties comparing or generalising results between studies due to inconsistent use of labels and measurements, or inappropriately aggregating results for differing concepts that share a label. A related consideration regards how these instruments are applied in current research. Many studies elect to only use a subset of items from an existing measure, with the intent of reducing participant fatigue and facilitating the inclusion of additional variables. However, it is possible that this approach is contributing towards some of the difficulties observed. Questionnaires that use only a few questions from the original instruments may point to weaker results~\citep{Lu2022}. 

An individual assessment and commentary on each of these instruments is beyond the scope of this review. However, some recent research has been undertaken towards collating and cross-referencing these measures. \citet{Krausman2022} summarised a number of instruments used in the present studies~\citeeg{Jian2000, Mayer1999, Schaefer2016, Singh1993} as part of a proposed \textit{conceptual toolkit} for the measurement of trust in human-agent teams, in addition to listing various behavioural and physiological indicators utilised in the present research. They suggest that there is no single suitable approach, and thus this broad variety of measurements has been developed to each address specific cases. However, this study does not encompass the full extent of the observed approaches, which still threatens to exceed this justification. Nevertheless, collating works such as this are crucial for identifying suitable use cases for these instruments and aligning measurement techniques between similar studies. In the meantime, researchers should be careful to ensure the viability of their proposed evaluation as an integral part of their study preparation, by verifying similarities with effective, previously demonstrated methodologies, or by independently assessing the validity of the measures with applicable checks and controls. 

Beyond existing measures, guidelines to aid in development of good quality and robust measurements for new variables are also needed. Continual exploration of new ideas is essential and inherent to the pursuit of knowledge, and so it is reasonable to expect that future research will measure entirely new variables. Researchers who wish to consider variables beyond existing measurements will need to seek out novel approaches. Thus, the need to have some consistency and validity assurance for variable measurements goes beyond explicitly identifying various possible instruments, to facilitating dynamic checks for these qualities in instruments as novel contexts arise. In practical terms, this means both clearly denoting and justifying the state-of-the-art, and ensuring that researchers are equipped with the knowledge and capabilities to perform the necessary adjustments without compromise. 


\subsection{Resolving the `capability gap'}\label{dis:capa}
The overarching motivation that justifies the pursuit of apology in AI is to understand and reduce negative impacts, and improve the usability and human-alignment of AI systems. However, while a substantial volume of the present AI apology research has investigated considerations for when and how an AI system might find apologising to be useful, there is a much lesser focus on how AI apology might be useful to the user. 
Enabling AI systems to apologise is both a promising and precarious prospect for the self-governance of these systems. 
As artificially intelligent machines continue to reach further prevalence, the need to establish contemporary methods for articulating and addressing their limitations in the social sphere gains urgency~\citep{Kim2022}. Apology is one way in which this can be achieved~\citep{deVisser2020, Wischnewski2023}. However, misuse of this approach, to manipulate, deceive or undermine the autonomy of the user, risks potentially irreversible damage to the reputation of human-oriented AI, and to the user themselves~\citep{Porra2020}. Apology is also one way in which a system may misrepresent itself to gain advantage, by falsely representing capabilities that it does not have. For example, if it were to make a commitment to reform without the ability to alter its behaviour, or initiate dialogue with the user without the capability to listen to and take on board their response. Relatedly, the literature reports that participants find these interactions frustrating~\citep{Albayram2020, Morimoto2020, Tewari2022}. 

In order for the use of these types of expressions to not be misleading, the system would need to be able to demonstrate the capability that it has claimed. We propose that this discrepancy represents a `capability gap' in human-AI interaction. Moreover, we suggest that if an artificially intelligent system is intended to be truly apologetic, it must be automated at every step of the generation of the apology: in action, in its determination of the need for apology, and in the delivery. It requires that the system meet the capabilities for apology described: \textit{detect}, \textit{attribute}, \textit{explain}, and \textit{adapt}. Yet, as this review has revealed, there is little progress within the AI apology literature towards these capabilities. As such, there stands a substantial gap between the social possibilities of what AI might be able to do, and the practical implementation of a socially observant, self-correcting agent. This review is not the first to note this increasing gap, with other papers calling for research into related areas of dynamic answerability~\citep{Tigard2021} and explainability~\citep{Dazeley2021a}.

We propose that this `capability gap' represents a critical research priority for AI apology. While the field of AI apology specifically is still sparse in regard to technical implementation~\citep{Harland2023}, there is an opportunity for rapid progress in this space. We have identified a number of relevant links to research outside this direct field that show that these capabilities are likely achievable, with connections that can be made to research in highly correlated areas, such as behaviour-adaptive AI agents~\citep{Ahmad2017, Hellou2021}, eXplainable AI~\citep{Dazeley2021a}, and human-robot collaboration~\citep{Arents2021, Semeraro2022}.  There is also substantial overlap with areas calling for greater work to be done, such as dynamic answerability~\citep{Tigard2021}, safety~\citep{Amodei2016}, and social responsibility~\citep{Cheng2021b}. Our suggestion and intent is to leverage these technical and theoretical contributions in combination with the extant literature to establish a multidisciplinary paradigm for future AI apology research. 




\section{Conclusion}\label{conc} 
As a field of research, AI apology is still in its infancy. Collating this research, as well as establishing the necessary language, frameworks, and principles to support comparative and analytical discussions of contributions, is essential for the further development of the field. This review has identified and catalogued the contemporary research on the use of apology by AI systems, as to provide a clear image of the present state of the field. This article represents three key contributions to this research: a theoretical framework, vigorous synthesis and critique, and recommendations for future directions in 4 main areas. 

We have devised a framework for AI apology to support the investigation and articulation of key findings relevant to the study of this field, comprising five elements described by three areas of effect, twelve components, and twelve moderators (Figure \ref{framework}). This framework represents the integration of prescriptive and descriptive perspectives from the human apology literature, including both philosophical and empirical research, with insights from HCI theory. In addition to enabling the first structured review of AI apology as a distinct area of research, it has also identified four key capabilities that such systems must have in order to achieve autonomous apology. 

We have provided a synthesis and critique of the contemporary literature on AI apology, integrating the findings of the various theoretical contributions, existing reviews, human research studies, and technical works. This survey addresses an important gap in the developing field of AI apology, by linking each of these contributions together via their attributes to a central reference point described by the framework of AI apology (Table \ref{inc}). Through a topical analysis of the reviewed literature with respect to each of the components and moderators identified in the framework, we have emphasised how the different aspects of the content and expression of an apology by an AI system contribute to its effect on the recipient. We also describe the impactful features of the offence, the system itself, and the user receiving the apology, as well as the interactions between them. While this review has revealed a substantial amount of work on the topic, there is still a lot left unexplored. Thus, not only has the present discussion organised the existing knowledge as to be readily applied and built upon in future works, but it has also provided an extensive list of possible areas for future research, to resolve outstanding inconsistencies and progress towards a cohesive field of research. 

Moreover, this article has sought to bridge the growing rift in functional and affective approaches to service recovery by technologies, by presenting a view on AI apology encompassing how it can be leveraged for both practical and social benefits in Human-AI interaction for human alignment. Specifically, we have embraced the that AI apology is inherently interlinked with themes of human-alignment. In order for the system to be aligned with the needs of its human users, it is necessary to understand the nature of the evolving relationship between them. Thus, we derived four capabilities required for autonomous apology: detect, attribute, explain, and adapt (Figure \ref{fig:capa}). We identified a `capability gap' between the claimed capabilities of AI systems and that demonstrated in the literature, and provided recommendations on how this could be addressed (Section \ref{dis:capa}). In so doing, we have provided a firm basis to facilitate the design and implementation of effective human-aligned AI apology. 

Further research is needed in order to ensure the next generation of interactive AI systems are able to meet the needs and expectations of their users. Future work in the areas highlighted within will be essential to ensuring the appropriate and effective application of AI apology, towards the realisation of autonomous, human-aligned interactive systems. 

\newpage

\section*{Backmatter}

\subsection*{Author Contributions}
All authors contributed to the study conception and design. The literature search, data analysis, and preparation of the manuscript was performed by Hadassah Harland. All authors commented on previous versions of the manuscript. All authors read and approved the final manuscript.

\subsection*{Compliance with Ethical Standards}
This research did not involve any human participants. 

\subsection*{Competing Interests}
The authors have no competing interests to declare that are relevant to the content of this article.

\subsection*{Funding}
This research was supported by an Australian Government Research Training Program (RTP) Scholarship and a Commonwealth Scientific and Industrial Research Organisation (CSIRO) Top-Up Scholarship.


\newcounter{bodyfigures}
\setcounter{bodyfigures}{\thefigure}

\begin{appendices}

\renewcommand{\thefigure}{\arabic{figure}} 
\setcounter{figure}{\thebodyfigures}

\section{Retrieval of Records} \label{records}


\subsection{Source Overlap} \label{sources}

\begin{figure}[!h]
\centering
\includegraphics[width=0.38\textwidth]{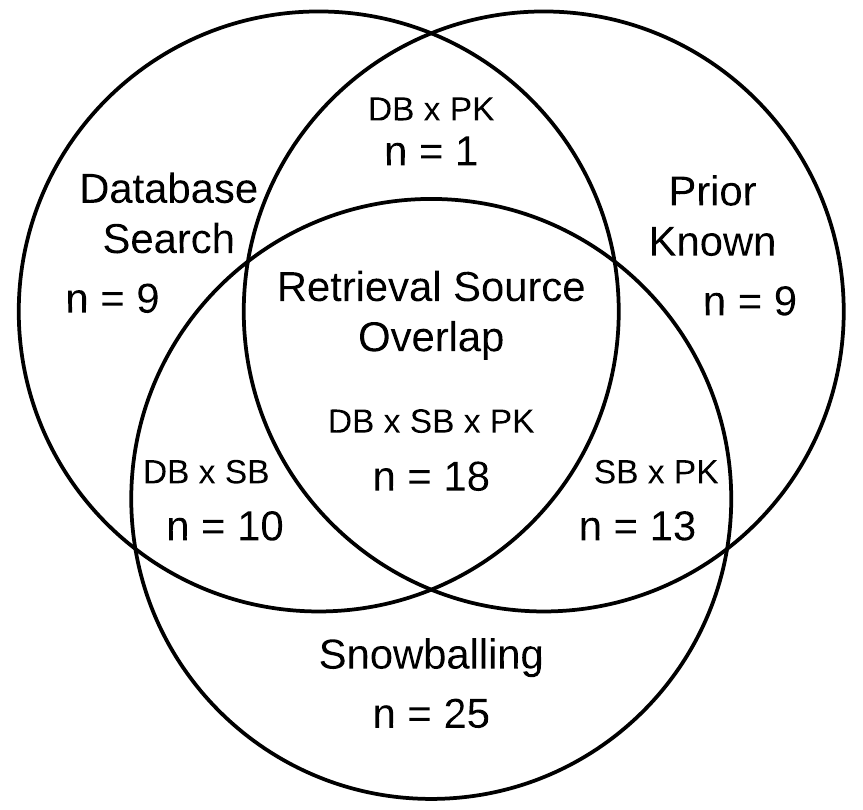}
\caption[Record Retrieval Overlap]{This diagram shows the overlap in the record retrieval methods used, being database search (DB), prior known (PK) and snowballing (SB), between the 85 included records }
\label{fig:ret}
\end{figure}

\clearpage

\begin{table}[!h]
\setlength{\tabcolsep}{2.5pt}%
\caption[Record Retrieval]{This table lists the studies relevant to each of the groups described by the prior retrieval diagram (Figure \ref{fig:ret})}\label{tab:ret}
\begin{tabular*}{\textwidth}{@{\extracolsep} p{65pt} p{300pt}}
\toprule%
Retrieval & Records \\ \midrule
Database Search & {\citet{Cahya2021, Kureha2023, Liu2020, Morimoto2020, Pei2023, Pereira2020, Perkins2022, Porra2020, SharifHeravi2020}} \\ \cmidrule{2-2}
Prior Known & {\citet{Ahn2023, Fota2022, Harris2023, Jelinek2023, Kim2022, LiM2023, Rakova2023, Tewari2022, YuS2022}} \\\cmidrule{2-2}
Snowballing & {\citet{Albayram2020, Aliasghari2021, Babel2021, Benner2021, deVisser2020, Esterwood2021, Esterwood2023b, Esterwood2023c, Fan2020, Feng2022, Galdon2020b, Karli2023, Khavas2021, Na2023, Natarajan2020, Nesset2023, Perkins2021, Rebensky2021, vanOver2020, Weiler2022, Wischnewski2023, XuX2022, Yang2022, ZhangX2023b, Zhu2023}} \\
\midrule
DB x PK & {\citet{Lu2022}} \\\cmidrule{2-2}
DB x SB & {\citet{Esterwood2022b, Jensen2022, Kox2022, Kraus2023, LvL2022, Mahmood2023, Okada2023, Rogers2023, Wright2022, Xu2023}} \\\cmidrule{2-2}
SB x PK & {\citet{Babel2022, Cameron2021, Galdon2020, Kreiter2023, Lajante2023, Liew2021, Liu2023, Lv2021, Lv2022, Pak2023, Shen2022, Tolmeijer2020, Yang2023}} \\
\midrule
DB x SB x PK & {\citet{Choi2021, Esterwood2022, Esterwood2023, Fratczak2021, Harland2023, Hu2021, Kim2021, Kim2023, Kox2021, Mahmood2022, Pompe2022, Schelble2022, Song2023, Textor2022, Wang2021, XuJ2022, Zhang2023, Zhang2023a}} \\
\botrule
\end{tabular*}
\end{table} 


\subsection{Literature Database} \label{allrecords}
A dynamic database repository for the literature used in this article is available through the project webpage at \url{https://hadassahharland.au/ai-apology/literature}. 

\clearpage


\section{Coding Scheme} \label{code}
\begin{table}[!h]
\caption{Relevance coding scheme}\label{tab:code}
\begin{tabular*}{\textwidth}{@{\extracolsep\fill} p{32pt} p{320pt}}
\toprule%
\multicolumn{2}{l}{Code and Description} \\
\midrule
\multicolumn{2}{l}{\textbf{(9)\ \ \ Direct}} \\
 & explicitly and directly addressing apology in AI systems (eg. How, when, why a robot or AI system may apologise) \\
\multicolumn{2}{l}{\textbf{(8)\ \ \ Indirect}} \\
 & indirectly addressing the use of apology in AI systems (eg. Apology as a mechanism not in response to offence) \\
\multicolumn{2}{l}{\textbf{(7)\ \ \ Analogous}} \\
 & addressing the use of apology components (Cue, Responsibility, Affirmation, Explanation, Moral Admission, Regret, Reform, Repair, Petition for forgiveness) in AI systems in a similar manner to apology, without direct reference \\
\multicolumn{2}{l}{\textbf{(6)\ \ \ Adjacent}} \\
 & Regarding non-apologetic behaviours for mitigating offence after it has occurred (eg. Corrective behaviour, recognition of issues, other trust repair or service recovery) \\
\multicolumn{2}{l}{\textbf{(5)\ \ \ Avoidant}} \\
 & Non-apologetic approaches for avoiding offence before it has occurred (eg. Adaptive behaviour, proactive communication, self awareness, trust calibration, behavioural empathy, commonsense reasoning) \\
\multicolumn{2}{l}{\textbf{(4)\ \ \ Characteristic}} \\
 & Static characteristics regarding how users interact with offence (eg. reducing perception or extent of an offence, via anthropomorphism, empathetic language, etc.) \\
\multicolumn{2}{l}{\textbf{(3)\ \ \ Suitability}} \\
 & Regarding how AI systems are positioned in relation to apology and associated themes (Broad suitability of assigning responsibility and blame, user trust patterns, Ethics, fairness, accountability, transparency, communicative capability) \\
\multicolumn{2}{l}{\textbf{(2)\ \ \ Perception}} \\
 & Regarding how humans percieve and interact with AI systems without offence (eg. Social and Service robotics, Usage intentions, service acceptance, presence, user attitudes) \\
\multicolumn{2}{l}{\textbf{(1)\ \ \ Negligible}} \\
 & Low or highly abstracted relevance to apology (eg. Knowledge capabilities, management of AI systems, relevant to capabilities or positioning of AI systems) \\
\multicolumn{2}{l}{\textbf{(0)\ \ \ None}} \\
 & No relevance (eg. No reference to AI behaviour or perception) \\
\multicolumn{2}{l}{\textbf{(-)\ \ \ Unclassified}} \\
 & Stand-in for when Adherence is unclear and should be revisited, or has been already excluded \\
\botrule
\end{tabular*}
\end{table}

\clearpage

\end{appendices}

\newpage
\bibliography{references}

\begin{thebibliography}{256}
\providecommand{\natexlab}[1]{#1}
\providecommand{\url}[1]{{#1}}
\providecommand{\urlprefix}{URL }
\providecommand{\doi}[1]{\url{https://doi.org/#1}}
\providecommand{\eprint}[2][]{\url{#2}}
 \bibcommenthead

\bibitem[{Ahmad et~al(2017)Ahmad, Mubin, and Orlando}]{Ahmad2017}
Ahmad M, Mubin O, Orlando J (2017) {A Systematic Review of Adaptivity in Human-Robot Interaction}. Multimodal Technologies and Interaction 1(3):14. \doi{10.3390/mti1030014}

\bibitem[{Ahmad et~al(2022)Ahmad, Siemon, Gnewuch, and Robra-Bissantz}]{Ahmad2022}
Ahmad R, Siemon D, Gnewuch U, et~al (2022) {Designing Personality-Adaptive Conversational Agents for Mental Health Care}. Information Systems Frontiers 24(3):923--943. \doi{10.1007/s10796-022-10254-9}

\bibitem[{Ahmadi and Fakhimi(2021)}]{Ahmadi2021}
Ahmadi A, Fakhimi S (2021) {Expressing gratitude versus empathetic apology: which one is better to use as an initial recovery strategy after a service failure?} Journal of Contemporary Marketing Science 4(3):341--361. \doi{10.1108/JCMARS-01-2021-0001}

\bibitem[{Ahn et~al(2023)Ahn, Kim, and Sung}]{Ahn2023}
Ahn J, Kim J, Sung Y (2023) {The role of perceived freewill in crises of human-AI interaction: the mediating role of ethical responsibility of AI}. International Journal of Advertising pp 1--27. \doi{10.1080/02650487.2023.2299563}, publisher Copyright: {\textcopyright} 2023 Advertising Association.

\bibitem[{Airenti(2015)}]{Airenti2015}
Airenti G (2015) {The Cognitive Bases of Anthropomorphism: From Relatedness to Empathy}. International Journal of Social Robotics 7(1):117--127. \doi{10.1007/s12369-014-0263-x}

\bibitem[{Akgun et~al(2010)Akgun, Cagiltay, and Zeyrek}]{Akgun2010}
Akgun M, Cagiltay K, Zeyrek D (2010) {The effect of apologetic error messages and mood states on computer users’ self-appraisal of performance}. Journal of Pragmatics 42(9):2430--2448. \doi{10.1016/j.pragma.2009.12.011}

\bibitem[{Alarcon et~al(2020)Alarcon, Gibson, and Jessup}]{Alarcon2020}
Alarcon GM, Gibson AM, Jessup SA (2020) {Trust Repair in Performance, Process, and Purpose Factors of Human-Robot Ttust}. In: 2020 IEEE International Conference on Human-Machine Systems (ICHMS). IEEE, pp 1--6, \doi{10.1109/ICHMS49158.2020.9209453}

\bibitem[{Albayram et~al(2020)Albayram, Jensen, Khan, Fahim, Buck, and Coman}]{Albayram2020}
Albayram Y, Jensen T, Khan MMH, et~al (2020) {Investigating the Effects of (Empty) Promises on Human-Automation Interaction and Trust Repair}. In: Proceedings of the 8th International Conference on Human-Agent Interaction. ACM, New York, NY, USA, pp 6--14, \doi{10.1145/3406499.3415064}

\bibitem[{Aliasghari et~al(2021)Aliasghari, Ghafurian, Nehaniv, and Dautenhahn}]{Aliasghari2021}
Aliasghari P, Ghafurian M, Nehaniv CL, et~al (2021) {How Do Different Modes of Verbal Expressiveness of a Student Robot Making Errors Impact Human Teachers’ Intention to Use the Robot?} In: Proceedings of the 9th International Conference on Human-Agent Interaction. ACM, New York, NY, USA, pp 21--30, \doi{10.1145/3472307.3484184}

\bibitem[{Allan et~al(2006)Allan, Allan, Kaminer, and Stein}]{Allan2006}
Allan A, Allan MM, Kaminer D, et~al (2006) {Exploration of the association between apology and forgiveness amongst victims of human rights violations}. Behavioral Sciences {\&} the Law 24(1):87--102. \doi{10.1002/bsl.689}

\bibitem[{Almeida et~al(2022)Almeida, Menezes, and Dias}]{Almeida2022}
Almeida L, Menezes P, Dias J (2022) {Telepresence Social Robotics towards Co-Presence: A Review}. Applied Sciences 12(11):5557. \doi{10.3390/app12115557}

\bibitem[{Almuttalibi(2016)}]{Almuttalibi2016}
Almuttalibi N (2016) {The Forms and Functions of Apologies With Indicative References References To the Letters of Keats and Byron}. International Journal of English and Literature 6(February):11--22

\bibitem[{Amodei et~al(2016)Amodei, Olah, Steinhardt, Christiano, Schulman, and Man{\'{e}}}]{Amodei2016}
Amodei D, Olah C, Steinhardt J, et~al (2016) {Concrete Problems in AI Safety}. ArXiv

\bibitem[{Arents et~al(2021)Arents, Abolins, Judvaitis, Vismanis, Oraby, and Ozols}]{Arents2021}
Arents J, Abolins V, Judvaitis J, et~al (2021) {Human–Robot Collaboration Trends and Safety Aspects: A Systematic Review}. Journal of Sensor and Actuator Networks 10(3):48. \doi{10.3390/jsan10030048}

\bibitem[{Austin(1975)}]{Austin1975}
Austin J (1975) {How To Do Things With Words}. Oxford University Press

\bibitem[{Aydin(2013)}]{Aydin2013}
Aydin M (2013) {Cross Cultural Pragmatics : A Study of Apology Speech Acts by Turkish speakers , American English Speakers and Advance Nonnative Speakers of English in Turkey}. PhD thesis, Minnesota State University - Mankato

\bibitem[{Aymerich-Franch et~al(2020)Aymerich-Franch, Kishore, and Slater}]{Aymerich-Franch2020}
Aymerich-Franch L, Kishore S, Slater M (2020) {When Your Robot Avatar Misbehaves You Are Likely to Apologize: An Exploration of Guilt During Robot Embodiment}. International Journal of Social Robotics 12(1). \doi{10.1007/s12369-019-00556-5}

\bibitem[{Azaria(2023)}]{Azaria2023}
Azaria A (2023) {Chatgpt more human like than computer like but not necessarily in a good way}

\bibitem[{Babel et~al(2021)Babel, Kraus, and Baumann}]{Babel2021}
Babel F, Kraus JM, Baumann M (2021) {Development and Testing of Psychological Conflict Resolution Strategies for Assertive Robots to Resolve Human–Robot Goal Conflict}. Frontiers in Robotics and AI 7. \doi{10.3389/frobt.2020.591448}

\bibitem[{Babel et~al(2022)Babel, Hock, Kraus, and Baumann}]{Babel2022}
Babel F, Hock P, Kraus J, et~al (2022) {It Will Not Take Long! Longitudinal Effects of Robot Conflict Resolution Strategies on Compliance, Acceptance and Trust}. In: Proceedings of the 2022 ACM/IEEE International Conference on Human-Robot Interaction. IEEE Press, HRI '22, pp 225--235

\bibitem[{Bachman and Guerrero(2006)}]{Bachman2006}
Bachman GF, Guerrero LK (2006) {Forgiveness, Apology, and Communicative Responses to Hurtful Events}. Communication Reports 19(1):45--56. \doi{10.1080/08934210600586357}

\bibitem[{Badampudi et~al(2015)Badampudi, Wohlin, and Petersen}]{Badampudi2015}
Badampudi D, Wohlin C, Petersen K (2015) {Experiences from using snowballing and database searches in systematic literature studies}. In: Proceedings of the 19th International Conference on Evaluation and Assessment in Software Engineering. ACM, New York, NY, USA, pp 1--10, \doi{10.1145/2745802.2745818}

\bibitem[{Bainbridge et~al(2008)Bainbridge, Hart, Kim, and Scassellati}]{Bainbridge2008}
Bainbridge WA, Hart J, Kim ES, et~al (2008) {The effect of presence on human-robot interaction}. In: RO-MAN 2008 - The 17th IEEE International Symposium on Robot and Human Interactive Communication. IEEE, pp 701--706, \doi{10.1109/ROMAN.2008.4600749}

\bibitem[{Bainbridge et~al(2011)Bainbridge, Hart, Kim, and Scassellati}]{Bainbridge2011}
Bainbridge WA, Hart JW, Kim ES, et~al (2011) {The Benefits of Interactions with Physically Present Robots over Video-Displayed Agents}. International Journal of Social Robotics 3(1):41--52. \doi{10.1007/s12369-010-0082-7}

\bibitem[{Baker et~al(2018)Baker, Phillips, Ullman, and Keebler}]{Baker2018}
Baker AL, Phillips EK, Ullman D, et~al (2018) {Toward an Understanding of Trust Repair in Human-Robot Interaction}. ACM Transactions on Interactive Intelligent Systems 8(4):1--30. \doi{10.1145/3181671}

\bibitem[{Bartneck et~al(2009)Bartneck, Kuli{\'{c}}, Croft, and Zoghbi}]{Bartneck2009}
Bartneck C, Kuli{\'{c}} D, Croft E, et~al (2009) {Measurement Instruments for the Anthropomorphism, Animacy, Likeability, Perceived Intelligence, and Perceived Safety of Robots}. International Journal of Social Robotics 1(1):71--81. \doi{10.1007/s12369-008-0001-3}

\bibitem[{Benitez et~al(2017)Benitez, Wyman, Carpinella, and Stroessner}]{Benitez2017}
Benitez J, Wyman AB, Carpinella CM, et~al (2017) {The authority of appearance: How robot features influence trait inferences and evaluative responses}. In: 2017 26th IEEE International Symposium on Robot and Human Interactive Communication (RO-MAN). IEEE, pp 397--404, \doi{10.1109/ROMAN.2017.8172333}

\bibitem[{Benner et~al(2021)Benner, Elshan, Sch{\"{o}}bel, and Janson}]{Benner2021}
Benner D, Elshan E, Sch{\"{o}}bel S, et~al (2021) {What do you mean? A Review on Recovery Strategies to Overcome Conversational Breakdowns of Conversational Agents}. In: International Conference on Information Systems (ICIS), pp 1--17

\bibitem[{Bennet(2008)}]{Bennet2008}
Bennet C (2008) {The Apology Ritual: A Philosophical Theory of Punishment}. Cambridge University Press

\bibitem[{Blum-Kulka and Olshtain(1984)}]{Blum-kulka1984}
Blum-Kulka S, Olshtain E (1984) {Requests and Apologies: A Cross-Cultural Study of Speech Act Realization Patterns (CCSARP)}. Applied Linguistics 5(3):196--213. \doi{10.1093/applin/5.3.196}

\bibitem[{Borboni et~al(2023)Borboni, Reddy, Elamvazuthi, AL-Quraishi, Natarajan, and Azhar~Ali}]{Borboni2023}
Borboni A, Reddy KVV, Elamvazuthi I, et~al (2023) {The Expanding Role of Artificial Intelligence in Collaborative Robots for Industrial Applications: A Systematic Review of Recent Works}. \doi{10.3390/machines11010111}

\bibitem[{Borkin and Reinhart(1978)}]{Borkin1978}
Borkin A, Reinhart SM (1978) {Excuse Me and I'm Sorry}. TESOL Quarterly 12(1):57--69. \doi{10.2307/3585791}

\bibitem[{Bowers(1964)}]{Bowers1964}
Bowers JW (1964) {Some correlates of language intensity}. Quarterly Journal of Speech 50(4):415--420. \doi{10.1080/00335636409382688}

\bibitem[{Brown-Devlin et~al(2022)Brown-Devlin, Lim, and Tao}]{Brown-Devlin2022}
Brown-Devlin N, Lim HS, Tao J (2022) {Examining the Influence of Algorithmic Message Personalization on Source Credibility and Reputation}. International Journal of Business Communication p 232948842211264. \doi{10.1177/23294884221126489}

\bibitem[{Cacioppo et~al(1986)Cacioppo, Petty, Chuan, and Rodriguez}]{Cacioppo1986}
Cacioppo JT, Petty RE, Chuan FK, et~al (1986) {Central and Peripheral Routes to Persuasion. An Individual Difference Perspective}. Journal of Personality and Social Psychology 51(5). \doi{10.1037/0022-3514.51.5.1032}

\bibitem[{Cahya and Giuliani(2021)}]{Cahya2021}
Cahya DE, Giuliani M (2021) {Appropriate Robot Reactions to Erroneous Situations in Human-Robot Collaboration}. In: Lecture Notes in Computer Science, pp 166--177, \doi{10.1007/978-3-030-90525-5{\_}15}

\bibitem[{Cameron et~al(2021)Cameron, de~Saille, Collins, Aitken, Cheung, Chua, Loh, and Law}]{Cameron2021}
Cameron D, de~Saille S, Collins EC, et~al (2021) {The effect of social-cognitive recovery strategies on likability, capability and trust in social robots}. Computers in Human Behavior 114:106561. \doi{10.1016/j.chb.2020.106561}

\bibitem[{Carsten~Stahl(2004)}]{CarstenStahl2004}
Carsten~Stahl B (2004) {Information, Ethics, and Computers: The Problem of Autonomous Moral Agents}. Minds and Machines 14(1):67--83. \doi{10.1023/B:MIND.0000005136.61217.93}

\bibitem[{Chandra et~al(2022)Chandra, Shirish, and Srivastava}]{Chandra2022}
Chandra S, Shirish A, Srivastava SC (2022) {To Be or Not to Be {\ldots}Human? Theorizing the Role of Human-Like Competencies in Conversational Artificial Intelligence Agents}. Journal of Management Information Systems 39(4):969--1005. \doi{10.1080/07421222.2022.2127441}

\bibitem[{Cheng et~al(2021{\natexlab{a}})Cheng, Mosallanezhad, Sheth, and Liu}]{Cheng2021a}
Cheng L, Mosallanezhad A, Sheth P, et~al (2021{\natexlab{a}}) {Causal Learning for Socially Responsible AI}. IJCAI International Joint Conference on Artificial Intelligence pp 4374--4381

\bibitem[{Cheng et~al(2021{\natexlab{b}})Cheng, Varshney, and Liu}]{Cheng2021b}
Cheng L, Varshney KR, Liu H (2021{\natexlab{b}}) {Socially Responsible AI Algorithms: Issues, Purposes, and Challenges}. Journal of Artificial Intelligence Research 71:1137--1181. \doi{10.1613/jair.1.12814}

\bibitem[{Choi and Severson(2009)}]{Choi2009}
Choi JJ, Severson M (2009) {“What! What kind of apology is this?”: The nature of apology in victim offender mediation}. Children and Youth Services Review 31(7):813--820. \doi{10.1016/j.childyouth.2009.03.003}

\bibitem[{Choi et~al(2021)Choi, Mattila, and Bolton}]{Choi2021}
Choi S, Mattila AS, Bolton LE (2021) {To Err Is Human(-oid): How Do Consumers React to Robot Service Failure and Recovery?} Journal of Service Research 24(3):354--371. \doi{10.1177/1094670520978798}

\bibitem[{Coeckelbergh(2020)}]{Coeckelbergh2020}
Coeckelbergh M (2020) {Artificial Intelligence, Responsibility Attribution, and a Relational Justification of Explainability}. Science and Engineering Ethics 26(4):2051--2068. \doi{10.1007/s11948-019-00146-8}

\bibitem[{Cohen and Olshtain(1981)}]{Cohen1981}
Cohen AD, Olshtain E (1981) {Developing a Measure of Sociocultural Competence: the Case of Apology}. Language Learning 31(1):113--134. \doi{10.1111/j.1467-1770.1981.tb01375.x}

\bibitem[{Cohen(2016)}]{Cohen2016}
Cohen AI (2016) {Corrective vs. Distributive Justice: the Case of Apologies}. Ethical Theory and Moral Practice 19(3):663--677. \doi{10.1007/s10677-015-9674-5}

\bibitem[{Cohen(2020)}]{Cohen2020}
Cohen AI (2020) {Apologies and Moral Repair; Rights, Duties, and Corrective Justice}. Routledge Studies in Ethics and Moral Theory, New York

\bibitem[{Colquitt et~al(2007)Colquitt, Scott, and LePine}]{Colquitt2007}
Colquitt JA, Scott BA, LePine JA (2007) {Trust, Trustworthiness, and Trust Propensity: A Meta-Analytic Test of Their Unique Relationships With Risk Taking and Job Performance}. Journal of Applied Psychology 92(4):909--927. \doi{10.1037/0021-9010.92.4.909}

\bibitem[{Correia et~al(2018)Correia, Guerra, Mascarenhas, Melo, and Paiva}]{Correia2018}
Correia F, Guerra C, Mascarenhas S, et~al (2018) {Exploring the Impact of Fault Justification in Human-Robot Trust}. In: Proceedings of the 17th international conference on autonomous agents and multiagent systems

\bibitem[{Dautenhahn(2015)}]{Dautenhahn2015}
Dautenhahn K (2015) {Interaction Studies with Social Robots}. In: Proceedings of the 2015 ACM on International Conference on Multimodal Interaction. ACM, New York, NY, USA, pp 3--3, \doi{10.1145/2818346.2818347}

\bibitem[{Dazeley et~al(2021)Dazeley, Vamplew, Foale, Young, Aryal, and Cruz}]{Dazeley2021a}
Dazeley R, Vamplew P, Foale C, et~al (2021) {Levels of explainable artificial intelligence for human-aligned conversational explanations}. Artificial Intelligence 299:103525. \doi{10.1016/j.artint.2021.103525}

\bibitem[{Duffy(2002)}]{Duffy2002}
Duffy B (2002) {Anthropomorphism and Robotics}. The Society for the Study of Artificial Intelligence and the Simulation of Behaviour p 3–5

\bibitem[{Engelhardt and Hansson(2017)}]{Engelhardt2017}
Engelhardt S, Hansson E (2017) A comparison of three robot recovery strategies to minimize the negative impact of failure in social hri

\bibitem[{Eslami-Rasekh and Mardani(2010)}]{Eslami2010}
Eslami-Rasekh A, Mardani M (2010) {Investigating the effects of teaching apology speech act, with a focus on intensifying strategies, on pragmatic development of EFL learners: The Iranian context}. The International Journal of Language Society and Culture 30(1):96--103

\bibitem[{Esterwood and Robert(2021)}]{Esterwood2021}
Esterwood C, Robert LP (2021) {Do You Still Trust Me? Human-Robot Trust Repair Strategies}. In: 2021 30th IEEE International Conference on Robot {\&} Human Interactive Communication (RO-MAN). IEEE, pp 183--188, \doi{10.1109/RO-MAN50785.2021.9515365}

\bibitem[{Esterwood and Robert(2022{\natexlab{a}})}]{Esterwood2022b}
Esterwood C, Robert LP (2022{\natexlab{a}}) {A Literature Review of Trust Repair in HRI}. In: RO-MAN 2022 - 31st IEEE International Conference on Robot and Human Interactive Communication: Social, Asocial, and Antisocial Robots, \doi{10.1109/RO-MAN53752.2022.9900667}

\bibitem[{Esterwood and Robert(2022{\natexlab{b}})}]{Esterwood2022}
Esterwood C, Robert LP (2022{\natexlab{b}}) {Having the Right Attitude: How Attitude Impacts Trust Repair in Human—Robot Interaction}. In: 2022 17th ACM/IEEE International Conference on Human-Robot Interaction (HRI), vol 2022-March. IEEE, pp 332--341, \doi{10.1109/HRI53351.2022.9889535}

\bibitem[{Esterwood and Robert(2023{\natexlab{a}})}]{Esterwood2023c}
Esterwood C, Robert LP (2023{\natexlab{a}}) {The theory of mind and human–robot trust repair}. Scientific Reports 13(1):9877. \doi{10.1038/s41598-023-37032-0}

\bibitem[{Esterwood and Robert(2023{\natexlab{b}})}]{Esterwood2023b}
Esterwood C, Robert LP (2023{\natexlab{b}}) {The Warehouse Robot Interaction Sim}. In: Companion of the 2023 ACM/IEEE International Conference on Human-Robot Interaction. ACM, New York, NY, USA, pp 268--271, \doi{10.1145/3568294.3580086}

\bibitem[{Esterwood and Robert(2023{\natexlab{c}})}]{Esterwood2023}
Esterwood C, Robert LP (2023{\natexlab{c}}) {Three Strikes and you are out!: The impacts of multiple human–robot trust violations and repairs on robot trustworthiness}. Computers in Human Behavior 142:107658. \doi{10.1016/j.chb.2023.107658}

\bibitem[{Fan et~al(2020)Fan, Wu, Miao, and Mattila}]{Fan2020}
Fan A, Wu LL, Miao L, et~al (2020) {When does technology anthropomorphism help alleviate customer dissatisfaction after a service failure? – The moderating role of consumer technology self-efficacy and interdependent self-construal}. Journal of Hospitality Marketing {\&} Management 29(3):269--290. \doi{10.1080/19368623.2019.1639095}

\bibitem[{Fedoryuk(2019)}]{Fedoryuk2019}
Fedoryuk A (2019) {Pragmatic Aspect of Phraseological Units in the English Language}. SHS Web of Conferences 69:00038. \doi{10.1051/shsconf/20196900038}

\bibitem[{Fehr and Gelfand(2010)}]{Fehr2010}
Fehr R, Gelfand MJ (2010) {When apologies work: How matching apology components to victims’ self-construals facilitates forgiveness}. Organizational Behavior and Human Decision Processes 113(1):37--50. \doi{10.1016/j.obhdp.2010.04.002}

\bibitem[{Feng and Tan(2022)}]{Feng2022}
Feng Y, Tan H (2022) {Comfort or Promise? Investigating the Effect of Trust Repair Strategies of Intelligent Vehicle System on Trust and Intention to Use from a Perspective of Social Cognition}. In: Lecture Notes in Computer Science, vol 13314 LNCS. Springer, Cham, p 154--166, \doi{10.1007/978-3-031-06053-3{\_}11}

\bibitem[{Flavi{\'{a}}n et~al(2006)Flavi{\'{a}}n, Guinal{\'{i}}u, and Gurrea}]{Flavian2006}
Flavi{\'{a}}n C, Guinal{\'{i}}u M, Gurrea R (2006) {The role played by perceived usability, satisfaction and consumer trust on website loyalty}. Information {\&} Management 43(1):1--14. \doi{10.1016/j.im.2005.01.002}

\bibitem[{Fota et~al(2022)Fota, Wagner, Roeding, and Schramm-Klein}]{Fota2022}
Fota A, Wagner K, Roeding T, et~al (2022) {“Help! I Have a Problem” – Differences between a Humanlike and Robot-like Chatbot Avatar in Complaint Management}. In: Proceedings of the 55th Hawaii International Conference on System Sciences, \doi{10.24251/HICSS.2022.522}

\bibitem[{Franke et~al(2019)Franke, Attig, and Wessel}]{Franke2019}
Franke T, Attig C, Wessel D (2019) {A Personal Resource for Technology Interaction: Development and Validation of the Affinity for Technology Interaction (ATI) Scale}. International Journal of Human–Computer Interaction 35(6):456--467. \doi{10.1080/10447318.2018.1456150}

\bibitem[{Frantz and Bennigson(2005)}]{Frantz2005}
Frantz CM, Bennigson C (2005) {Better late than early: The influence of timing on apology effectiveness}. Journal of Experimental Social Psychology 41(2):201--207. \doi{10.1016/j.jesp.2004.07.007}

\bibitem[{Fratczak et~al(2021)Fratczak, Goh, Kinnell, Justham, and Soltoggio}]{Fratczak2021}
Fratczak P, Goh YM, Kinnell P, et~al (2021) {Robot apology as a post-accident trust-recovery control strategy in industrial human-robot interaction}. International Journal of Industrial Ergonomics 82:103078. \doi{10.1016/j.ergon.2020.103078}

\bibitem[{Galdon and Hall(2020)}]{Galdon2020b}
Galdon F, Hall A (2020) {The Right to Reparations: A New Digital Right for Repairing Trust in the Emerging Era of Highly Autonomous Systems}. In: Advances in Intelligent Systems and Computing, pp 538--543, \doi{10.1007/978-3-030-44267-5{\_}81}

\bibitem[{Galdon and Wang(2020)}]{Galdon2020}
Galdon F, Wang SJ (2020) {From apology to compensation: A multi-level taxonomy of trust reparation for highly automated virtual assistants}. In: Advances in Intelligent Systems and Computing, pp 42--46, \doi{10.1007/978-3-030-25629-6{\_}7}

\bibitem[{Gambino et~al(2020)Gambino, Fox, and Ratan}]{Gambino2020}
Gambino A, Fox J, Ratan RA (2020) {Building a Stronger CASA: Extending the Computers Are Social Actors Paradigm}. Human-Machine Communication 1(1):71--85. \doi{10.30658/hmc.1.5}

\bibitem[{Ge et~al(2019)Ge, Li, Guan, Xu, Sun, and Zhou}]{Ge2019}
Ge X, Li D, Guan D, et~al (2019) {Do Smart Speakers Respond to Their Errors Properly? A Study on Human-Computer Dialogue Strategy}. In: Lecture Notes in Computer Science, vol 11584 LNCS. Springer, Cham, p 440--455, \doi{10.1007/978-3-030-23541-3{\_}32}

\bibitem[{Gert and Gert(2020)}]{sep-morality-definition}
Gert B, Gert J (2020) {The Definition of Morality}. In: Zalta EN (ed) The Stanford Encyclopedia of Philosophy, fall 2020 edn. Metaphysics Research Lab, Stanford University

\bibitem[{Gill(2000)}]{Gill2000}
Gill K (2000) {The Moral Functions of an Apology}. The Philosophical Forum 31(1):11--27. \doi{10.1111/0031-806X.00025}

\bibitem[{Glikson and Asscher(2023)}]{Glikson2023}
Glikson E, Asscher O (2023) {AI-mediated apology in a multilingual work context: Implications for perceived authenticity and willingness to forgive}. Computers in Human Behavior 140. \doi{10.1016/j.chb.2022.107592}

\bibitem[{Glikson and Woolley(2020)}]{Glikson2020}
Glikson E, Woolley AW (2020) {Human trust in artificial intelligence: Review of empirical research}. Academy of Management Annals 14(2):627--660. \doi{10.5465/annals.2018.0057}

\bibitem[{Gogoshin(2021)}]{Gogoshin2021}
Gogoshin DL (2021) {Robot Responsibility and Moral Community}. Frontiers in Robotics and AI 8. \doi{10.3389/frobt.2021.768092}

\bibitem[{Gosling et~al(2003)Gosling, Rentfrow, and Swann}]{Gosling2003}
Gosling SD, Rentfrow PJ, Swann WB (2003) {A very brief measure of the Big-Five personality domains}. Journal of Research in Personality 37(6):504--528. \doi{10.1016/S0092-6566(03)00046-1}

\bibitem[{Gunes et~al(2022)Gunes, Broz, Crawford, der P{\"{u}}tten, Strait, and Riek}]{Gunes2022}
Gunes H, Broz F, Crawford CS, et~al (2022) {Reproducibility in Human-Robot Interaction: Furthering the Science of HRI}. Current Robotics Reports 3(4):281--292. \doi{10.1007/s43154-022-00094-5}

\bibitem[{Hamacher(2015)}]{Hamacher2015}
Hamacher A (2015) {Believing in BERT : Making good on bad robot behavior}. PhD thesis, University College London

\bibitem[{Hamacher et~al(2016)Hamacher, Bianchi-Berthouze, Pipe, and Eder}]{Hamacher2016}
Hamacher A, Bianchi-Berthouze N, Pipe AG, et~al (2016) {Believing in BERT: Using expressive communication to enhance trust and counteract operational error in physical Human-robot interaction}. In: 2016 25th IEEE International Symposium on Robot and Human Interactive Communication (RO-MAN). IEEE, pp 493--500, \doi{10.1109/ROMAN.2016.7745163}

\bibitem[{Harland et~al(2023)Harland, Dazeley, Nakisa, Cruz, and Vamplew}]{Harland2023}
Harland H, Dazeley R, Nakisa B, et~al (2023) {AI apology: interactive multi-objective reinforcement learning for human-aligned AI}. Neural Computing and Applications 35(23):16917--16930. \doi{10.1007/s00521-023-08586-x}

\bibitem[{Harmon-Jones and Mills(2019)}]{Harmon-Jones2019}
Harmon-Jones E, Mills J (2019) {An introduction to cognitive dissonance theory and an overview of current perspectives on the theory.} In: Cognitive dissonance: Reexamining a pivotal theory in psychology (2nd ed.). American Psychological Association, Washington, p 3--24, \doi{10.1037/0000135-001}

\bibitem[{Harris and DeChurch(2023)}]{Harris2023}
Harris AM, DeChurch L (2023) {“Alexa, How Can I Trust You Again?” Trust Repair in Human-AI Teams}. PhD thesis, United States -- Illinois

\bibitem[{Hellou et~al(2021)Hellou, Gasteiger, Lim, Jang, and Ahn}]{Hellou2021}
Hellou M, Gasteiger N, Lim JY, et~al (2021) {Personalization and localization in human-robot interaction: A review of technical methods}. Robotics 10(4). \doi{10.3390/robotics10040120}

\bibitem[{Hess(2008)}]{Hess2008}
Hess RL (2008) {The impact of firm reputation and failure severity on customers' responses to service failures}. Journal of Services Marketing 22(5):385--398. \doi{10.1108/08876040810889157}

\bibitem[{Hirshfield et~al(2014)Hirshfield, Bobko, Barelka, Hirshfield, Farrington, Gulbronson, and Paverman}]{Hirshfield2014}
Hirshfield LM, Bobko P, Barelka A, et~al (2014) {Using Noninvasive Brain Measurement to Explore the Psychological Effects of Computer Malfunctions on Users during Human-Computer Interactions}. Advances in Human-Computer Interaction 2014:1--13. \doi{10.1155/2014/101038}

\bibitem[{Ho and MacDorman(2017)}]{Ho2017}
Ho CC, MacDorman KF (2017) {Measuring the Uncanny Valley Effect}. International Journal of Social Robotics 9(1):129--139. \doi{10.1007/s12369-016-0380-9}

\bibitem[{Hoffman et~al(1995)Hoffman, Kelley, and Rotalsky}]{Hoffman1995}
Hoffman KD, Kelley SW, Rotalsky HM (1995) {Tracking service failures and employee recovery efforts}. Journal of Services Marketing 9(2):49--61. \doi{10.1108/08876049510086017}

\bibitem[{Homans(1961)}]{Homans1961}
Homans GC (1961) {Social Behaviour, Its Elementary Forms}. Harcourt, Brace {\&} World, Inc, New York

\bibitem[{Honig and Oron-Gilad(2018)}]{Honig2018}
Honig S, Oron-Gilad T (2018) {Understanding and Resolving Failures in Human-Robot Interaction: Literature Review and Model Development}. Frontiers in Psychology 1:861. \doi{10.3389/fpsyg.2018.00861}

\bibitem[{Hu et~al(2021)Hu, Min, and Su}]{Hu2021}
Hu Y, Min H, Su N (2021) {How Sincere is an Apology? Recovery Satisfaction in A Robot Service Failure Context}. Journal of Hospitality and Tourism Research 45(6):1022--1043. \doi{10.1177/10963480211011533}

\bibitem[{Huang et~al(2023)Huang, Zhang, Mao, and Yao}]{Huang2023}
Huang C, Zhang Z, Mao B, et~al (2023) {An Overview of Artificial Intelligence Ethics}. IEEE Transactions on Artificial Intelligence 4(4). \doi{10.1109/TAI.2022.3194503}

\bibitem[{Hui et~al(2011)Hui, Lau, Tsang, and Pak}]{Hui2011}
Hui CH, Lau FL, Tsang KL, et~al (2011) {The Impact of Post-Apology Behavioral Consistency on Victim's Forgiveness Intention: A Study of Trust Violation Among Coworkers}. Journal of Applied Social Psychology 41(5):1214--1236. \doi{10.1111/j.1559-1816.2011.00754.x}

\bibitem[{Im et~al(2021)Im, Youk, and Park}]{Im2021}
Im WJ, Youk S, Park HS (2021) {Apologies combined with other crisis response strategies: Do the fulfillment of individuals' needs to be heard and the timing of response message affect apology appropriateness?} Public Relations Review 47(1):102002. \doi{10.1016/j.pubrev.2020.102002}

\bibitem[{Jayawickreme et~al(2021)Jayawickreme, Fleeson, Beck, Baumert, and Adler}]{Jayawickreme2021}
Jayawickreme E, Fleeson W, Beck ED, et~al (2021) {Personality dynamics}. Personality Science 2. \doi{10.5964/ps.6179}

\bibitem[{Jel{\'{i}}nek and Fischer(2023)}]{Jelinek2023}
Jel{\'{i}}nek M, Fischer K (2023) {Trust regulation in Social Robotics: From Violation to Repair}. [Unpublished Manuscript]

\bibitem[{Jensen(2021)}]{Jensen2021b}
Jensen T (2021) {Disentangling Trust and Anthropomorphism Toward the Design of Human-Centered AI Systems}. In: Lecture Notes in Computer Science, pp 41--58, \doi{10.1007/978-3-030-77772-2{\_}3}

\bibitem[{Jensen and Khan(2022)}]{Jensen2022}
Jensen T, Khan MMH (2022) {I’m Only Human: The Effects of Trust Dampening by Anthropomorphic Agents}. In: Lecture Notes in Computer Science, vol 13518 LNCS. Springer, Cham, p 285--306, \doi{10.1007/978-3-031-21707-4{\_}21}

\bibitem[{Jian et~al(2000)Jian, Bisantz, and Drury}]{Jian2000}
Jian JY, Bisantz AM, Drury CG (2000) {Foundations for an Empirically Determined Scale of Trust in Automated Systems}. International Journal of Cognitive Ergonomics 4(1):53--71. \doi{10.1207/S15327566IJCE0401{\_}04}

\bibitem[{Joyce(1999)}]{Joyce1999}
Joyce R (1999) {Apologizing}. Public Affairs Quarterly 13(2):159--173

\bibitem[{Kador(2009)}]{Kador2009}
Kador J (2009) {Effective Apology: Mending Fences, Building Bridges, and Restoring Trust}. Berrett-Koehler Publishers

\bibitem[{Karli et~al(2023)Karli, Cao, and Huang}]{Karli2023}
Karli UB, Cao S, Huang CM (2023) {"What If It Is Wrong": Effects of Power Dynamics and Trust Repair Strategy on Trust and Compliance in HRI}. In: Proceedings of the 2023 ACM/IEEE International Conference on Human-Robot Interaction. ACM, New York, NY, USA, pp 271--280, \doi{10.1145/3568162.3576964}

\bibitem[{Khavas(2021)}]{Khavas2021}
Khavas ZR (2021) {A Review on Trust in Human-Robot Interaction}. [Unpublished Manuscript]

\bibitem[{Kim et~al(2004)Kim, Ferrin, Cooper, and Dirks}]{Kim2004}
Kim PH, Ferrin DL, Cooper CD, et~al (2004) {Removing the Shadow of Suspicion: The Effects of Apology Versus Denial for Repairing Competence- versus Integrity-Based Trust Violations}. Journal of Applied Psychology 89(1):104--118. \doi{10.1037/0021-9010.89.1.104}

\bibitem[{Kim et~al(2013)Kim, Cooper, Dirks, and Ferrin}]{Kim2013}
Kim PH, Cooper CD, Dirks KT, et~al (2013) {Repairing trust with individuals vs. groups}. Organizational Behavior and Human Decision Processes 120(1):1--14. \doi{10.1016/j.obhdp.2012.08.004}

\bibitem[{Kim and Hinds(2006)}]{Kim2006}
Kim T, Hinds P (2006) {Who Should I Blame? Effects of Autonomy and Transparency on Attributions in Human-Robot Interaction}. In: ROMAN 2006 - The 15th IEEE International Symposium on Robot and Human Interactive Communication. IEEE, pp 80--85, \doi{10.1109/ROMAN.2006.314398}

\bibitem[{Kim and Song(2021)}]{Kim2021}
Kim T, Song H (2021) {How should intelligent agents apologize to restore trust? Interaction effects between anthropomorphism and apology attribution on trust repair}. Telematics and Informatics 61. \doi{10.1016/j.tele.2021.101595}

\bibitem[{Kim and Song(2023)}]{Kim2023}
Kim T, Song H (2023) {“I Believe AI Can Learn from the Error. Or Can It Not?”: The Effects of Implicit Theories on Trust Repair of the Intelligent Agent}. International Journal of Social Robotics 15(1):115--128. \doi{10.1007/s12369-022-00951-5}

\bibitem[{Kim et~al(2009)Kim, Kim, and Kim}]{Kim2009}
Kim TT, Kim WG, Kim HB (2009) {The effects of perceived justice on recovery satisfaction, trust, word-of-mouth, and revisit intention in upscale hotels}. Tourism Management 30(1):51--62. \doi{10.1016/j.tourman.2008.04.003}

\bibitem[{Kim and Routledge(2022)}]{Kim2022}
Kim TW, Routledge BR (2022) {Why a Right to an Explanation of Algorithmic Decision-Making Should Exist: A Trust-Based Approach}. Business Ethics Quarterly 32(1):75--102. \doi{10.1017/beq.2021.3}

\bibitem[{Kitchen et~al(2014)Kitchen, Kerr, Schultz, McColl, and Pals}]{Kitchen2014}
Kitchen PJ, Kerr G, Schultz DE, et~al (2014) {The elaboration likelihood model: review, critique and research agenda}. European Journal of Marketing 48(11/12):2033--2050. \doi{10.1108/EJM-12-2011-0776}

\bibitem[{Koc et~al(2023)Koc, Hatipoglu, Kivrak, Celik, and Koc}]{Koc2023}
Koc E, Hatipoglu S, Kivrak O, et~al (2023) {Houston, we have a problem!: The use of ChatGPT in responding to customer complaints}. Technology in Society 74. \doi{10.1016/j.techsoc.2023.102333}

\bibitem[{Kok and Soh(2020)}]{Kok2020}
Kok BC, Soh H (2020) {Trust in Robots: Challenges and Opportunities}. Current Robotics Reports 1(4):297--309. \doi{10.1007/s43154-020-00029-y}

\bibitem[{Kort(1975)}]{Kort1975}
Kort LF (1975) {What is an Apology?} Philosophy Research Archives 1:78--87. \doi{https://doi.org/10.5840/pra197515}

\bibitem[{Kox et~al(2021)Kox, Kerstholt, Hueting, and de~Vries}]{Kox2021}
Kox ES, Kerstholt JH, Hueting TF, et~al (2021) {Trust repair in human-agent teams: the effectiveness of explanations and expressing regret}. Autonomous Agents and Multi-Agent Systems 35(2):30. \doi{10.1007/s10458-021-09515-9}

\bibitem[{Kox et~al(2022)Kox, Siegling, and Kerstholt}]{Kox2022}
Kox ES, Siegling LB, Kerstholt JH (2022) {Trust Development in Military and Civilian Human–Agent Teams: The Effect of Social-Cognitive Recovery Strategies}. International Journal of Social Robotics 14(5):1323--1338. \doi{10.1007/s12369-022-00871-4}

\bibitem[{Kraus et~al(2023)Kraus, Merger, Gr{\"{o}}ner, and P{\"{a}}tz}]{Kraus2023}
Kraus JM, Merger J, Gr{\"{o}}ner F, et~al (2023) {'Sorry' Says the Robot}. In: Companion of the 2023 ACM/IEEE International Conference on Human-Robot Interaction. ACM, New York, NY, USA, pp 436--441, \doi{10.1145/3568294.3580122}

\bibitem[{Kraus et~al(2022)Kraus, Wagner, Untereiner, and Minker}]{Kraus2022}
Kraus M, Wagner N, Untereiner N, et~al (2022) {Including Social Expectations for Trustworthy Proactive Human-Robot Dialogue}. In: Proceedings of the 30th ACM Conference on User Modeling, Adaptation and Personalization. ACM, New York, NY, USA, pp 23--33, \doi{10.1145/3503252.3531294}

\bibitem[{Krausman et~al(2022)Krausman, Neubauer, Forster, Lakhmani, Baker, Fitzhugh, Gremillion, Wright, Metcalfe, and Schaefer}]{Krausman2022}
Krausman A, Neubauer C, Forster D, et~al (2022) {Trust Measurement in Human-Autonomy Teams: Development of a Conceptual Toolkit}. ACM Transactions on Human-Robot Interaction 11(3):1--58. \doi{10.1145/3530874}

\bibitem[{Kreiter(2023)}]{Kreiter2023}
Kreiter R (2023) {Can we forgive a robotic teammate? The role of Trust and Trust Repair Strategies in Human-Agent Teams}. PhD thesis, University of Twente

\bibitem[{Kureha(2023)}]{Kureha2023}
Kureha M (2023) {On the moral permissibility of robot apologies}. AI and Society \doi{10.1007/s00146-023-01782-2}

\bibitem[{Kuttal et~al(2021)Kuttal, Sedhain, and AuBuchon}]{Kuttal2021}
Kuttal SK, Sedhain A, AuBuchon J (2021) {Designing a Gender-Inclusive Conversational Agent For Pair Programming: An Empirical Investigation}. In: Lecture Notes in Computer Science, 12797th edn. Springer, Cham, p 59--75, \doi{10.1007/978-3-030-77772-2{\_}4}

\bibitem[{Lajante et~al(2023)Lajante, Remisch, and Dorofeev}]{Lajante2023}
Lajante M, Remisch D, Dorofeev N (2023) {Can robots recover a service using interactional justice as employees do? A literature review-based assessment}. Service Business 17(1):315--357. \doi{10.1007/s11628-023-00525-z}

\bibitem[{Lazare(2004)}]{Lazare2004}
Lazare A (2004) {On Apology}. Oxford University Press

\bibitem[{Lee and Moray(1992)}]{Lee1992}
Lee J, Moray N (1992) {Trust, control strategies and allocation of function in human-machine systems}. Ergonomics 35(10):1243--1270. \doi{10.1080/00140139208967392}

\bibitem[{Lee and See(2004)}]{Lee2004}
Lee JD, See KA (2004) {Trust in Automation: Designing for Appropriate Reliance}. Human Factors: The Journal of the Human Factors and Ergonomics Society 46(1):50--80. \doi{10.1518/hfes.46.1.50{\_}30392}

\bibitem[{Lee et~al(2010)Lee, Kiesler, Forlizzi, Srinivasa, and Rybski}]{Lee2010}
Lee MK, Kiesler S, Forlizzi J, et~al (2010) {Gracefully mitigating breakdowns in robotic services}. In: 2010 5th ACM/IEEE International Conference on Human-Robot Interaction (HRI). IEEE, pp 203--210, \doi{10.1109/HRI.2010.5453195}

\bibitem[{Lee et~al(2011)Lee, Bae, Kwak, and Kim}]{Lee2011}
Lee Y, Bae JE, Kwak SS, et~al (2011) {The effect of politeness strategy on human - robot collaborative interaction on malfunction of robot vacuum cleaner}. In RSS workshop on HRI

\bibitem[{Leech and Thomas(2002)}]{Leech2002}
Leech G, Thomas J (2002) {Language, meaning and context: pragmatics}. In: An Encyclopedia of Language. Routledge, p 105--124, \doi{10.4324/9780203403617-11}

\bibitem[{Leite et~al(2013)Leite, Martinho, and Paiva}]{Leite2013}
Leite I, Martinho C, Paiva A (2013) {Social Robots for Long-Term Interaction: A Survey}. International Journal of Social Robotics 5(2):291--308. \doi{10.1007/s12369-013-0178-y}

\bibitem[{Lewicki and Polin(2012)}]{Lewicki2012}
Lewicki RJ, Polin B (2012) {The Art of the Apology: The Structure and Effectiveness of Apologies in Trust Repair}. In: Restoring Trust in Organizations and LeadersEnduring Challenges and Emerging Answers. Oxford University Press, p 95--128, \doi{10.1093/acprof:oso/9780199756087.003.0006}

\bibitem[{Lewicki et~al(2016)Lewicki, Polin, Lount~Jr., and Lount}]{Lewicki2016}
Lewicki RJ, Polin B, Lount~Jr. RB, et~al (2016) {An Exploration of the Structure of Effective Apologies}. Negotiation and Conflict Management Research 9(2):177--196. \doi{https://doi.org/10.1111/ncmr.12073}

\bibitem[{Li and Zhang(2005)}]{Zhang2005}
Li L, Zhang P (2005) {The Intellectual Development of Human-Computer Interaction Research: A Critical Assessment of the MIS Literature (1990-2002)}. Journal of the Association for Information Systems 6(11):227--292. \doi{10.17705/1jais.00070}

\bibitem[{Li and Lee(2023)}]{LiM2023}
Li M, Lee JD (2023) {Measure and Manage Trust in Human-AI Conversations}. PhD thesis, University of Winconsin, Madison

\bibitem[{Liddle(2017)}]{Liddle2017}
Liddle AC (2017) {Deconstructing the Phenomenon of Apology}. Journal of Applied Hermeneutics 1(3)

\bibitem[{Liew and Tan(2021)}]{Liew2021}
Liew TW, Tan SM (2021) {Social cues and implications for designing expert and competent artificial agents: A systematic review}. Telematics and Informatics 65:101721. \doi{10.1016/j.tele.2021.101721}

\bibitem[{Liu et~al(2023)Liu, Lv, and Huang}]{Liu2023}
Liu D, Lv Y, Huang W (2023) {How do consumers react to chatbots' humorous emojis in service failures}. Technology in Society 73. \doi{10.1016/j.techsoc.2023.102244}

\bibitem[{Liu et~al(2020)Liu, Ding, Zhang, Zhao, Hu, Gong, Xu, Zhang, Zhang, and Wang}]{Liu2020}
Liu M, Ding Q, Zhang Y, et~al (2020) {Cold Comfort Matters - How Channel-Wise Emotional Strategies Help in a Customer Service Chatbot}. In: Extended Abstracts of the 2020 CHI Conference on Human Factors in Computing Systems. ACM, New York, NY, USA, pp 1--7, \doi{10.1145/3334480.3382905}

\bibitem[{Lohan et~al(2014)Lohan, Deshmukh, and Aylett}]{Lohan2014}
Lohan KS, Deshmukh A, Aylett R (2014) {How can a robot signal its incapability to perform a certain task to humans in an acceptable manner?} In: The 23rd IEEE International Symposium on Robot and Human Interactive Communication. IEEE, pp 814--819, \doi{10.1109/ROMAN.2014.6926353}

\bibitem[{Lu et~al(2022)Lu, Konishi, Sano, Hiruta, and Nakagawa}]{Lu2022}
Lu PPY, Konishi M, Sano S, et~al (2022) {What Makes An Apology More Effective? Exploring Anthropomorphism, Individual Differences, And Emotion In Human-Automation Trust Repair}. [Unpublished Manuscript]

\bibitem[{Lucas et~al(2018)Lucas, Boberg, Traum, Artstein, Gratch, Gainer, Johnson, Leuski, and Nakano}]{Lucas2018}
Lucas GM, Boberg J, Traum D, et~al (2018) {Getting to Know Each Other}. In: Proceedings of the 2018 ACM/IEEE International Conference on Human-Robot Interaction. ACM, New York, NY, USA, pp 344--351, \doi{10.1145/3171221.3171258}

\bibitem[{Lv et~al(2022{\natexlab{a}})Lv, Huang, Guan, and Yang}]{LvL2022}
Lv L, Huang M, Guan D, et~al (2022{\natexlab{a}}) {Apology or gratitude? The effect of communication recovery strategies for service failures of AI devices}. Journal of Travel {\&} Tourism Marketing 39(6):570--587. \doi{10.1080/10548408.2022.2162659}

\bibitem[{Lv et~al(2021)Lv, Liu, Luo, Liu, and Li}]{Lv2021}
Lv X, Liu Y, Luo J, et~al (2021) {Does a cute artificial intelligence assistant soften the blow? The impact of cuteness on customer tolerance of assistant service failure}. Annals of Tourism Research 87:103114. \doi{10.1016/j.annals.2020.103114}

\bibitem[{Lv et~al(2022{\natexlab{b}})Lv, Yang, Qin, Cao, and Xu}]{Lv2022}
Lv X, Yang Y, Qin D, et~al (2022{\natexlab{b}}) {Artificial intelligence service recovery: The role of empathic response in hospitality customers’ continuous usage intention}. Computers in Human Behavior 126:106993. \doi{10.1016/j.chb.2021.106993}

\bibitem[{Lyons et~al(2023)Lyons, Hamdan, and Vo}]{Lyons2023}
Lyons JB, Hamdan Ia, Vo TQ (2023) {Explanations and trust: What happens to trust when a robot partner does something unexpected?} Computers in Human Behavior 138. \doi{10.1016/j.chb.2022.107473}

\bibitem[{Madhavan and Wiegmann(2007)}]{Madhavan2007}
Madhavan P, Wiegmann DA (2007) {Similarities and differences between human–human and human–automation trust: an integrative review}. Theoretical Issues in Ergonomics Science 8(4):277--301. \doi{10.1080/14639220500337708}

\bibitem[{Mahmood and Huang(2023)}]{Mahmood2023}
Mahmood A, Huang CM (2023) {Gender Biases in Error Mitigation by Voice Assistants}. [Unpublished Manuscript]

\bibitem[{Mahmood et~al(2022)Mahmood, Fung, Won, and Huang}]{Mahmood2022}
Mahmood A, Fung JW, Won I, et~al (2022) {Owning Mistakes Sincerely: Strategies for Mitigating AI Errors}. In: CHI Conference on Human Factors in Computing Systems. ACM, New York, NY, USA, pp 1--11, \doi{10.1145/3491102.3517565}

\bibitem[{Marinaccio et~al(2015)Marinaccio, Kohn, Parasuraman, and De~Visser}]{Marinaccio2015}
Marinaccio K, Kohn S, Parasuraman R, et~al (2015) {A Framework for Rebuilding Trust in Social Automation Across Health-Care Domains}. Proceedings of the International Symposium on Human Factors and Ergonomics in Health Care 4(1):201--205. \doi{10.1177/2327857915041036}

\bibitem[{Matthias(2004)}]{Matthias2004}
Matthias A (2004) {The responsibility gap: Ascribing responsibility for the actions of learning automata}. Ethics and Information Technology 6(3):175--183. \doi{10.1007/s10676-004-3422-1}

\bibitem[{Mayer and Davis(1999)}]{Mayer1999}
Mayer RC, Davis JH (1999) {The effect of the performance appraisal system on trust for management: A field quasi-experiment}. Journal of Applied Psychology 84(1):123--136. \doi{10.1037/0021-9010.84.1.123}

\bibitem[{Mayer et~al(1995)Mayer, Davis, and Schoorman}]{Mayer1995}
Mayer RC, Davis JH, Schoorman FD (1995) {An Integrative Model Of Organizational Trust}. Academy of Management Review 20(3):709--734. \doi{10.5465/amr.1995.9508080335}

\bibitem[{Meyvis and Van~Osselaer(2018)}]{Meyvis2018}
Meyvis T, Van~Osselaer SM (2018) {Increasing the power of your study by increasing the effect size}. Journal of Consumer Research 44(5). \doi{10.1093/jcr/ucx110}

\bibitem[{Milanovic(2021)}]{Milanovic2021}
Milanovic K (2021) {Artificial Social Constructivism for Long Term Human Computer Interaction}. PhD thesis, Imperial College London, London

\bibitem[{Miller et~al(2017)Miller, Howe, and Sonenberg}]{Miller2017}
Miller T, Howe P, Sonenberg L (2017) {Explainable AI: Beware of Inmates Running the Asylum Or: How I Learnt to Stop Worrying and Love the Social and Behavioural Sciences}. ArXiv

\bibitem[{Morimoto et~al(2020)Morimoto, Even, and Kanda}]{Morimoto2020}
Morimoto D, Even J, Kanda T (2020) {Can a Robot Handle Customers with Unreasonable Complaints?} In: Proceedings of the 2020 ACM/IEEE International Conference on Human-Robot Interaction. ACM, New York, NY, USA, pp 579--587, \doi{10.1145/3319502.3374830}

\bibitem[{Mutlu et~al(2016)Mutlu, Roy, and {\v{S}}abanovi{\'{c}}}]{Mutlu2016}
Mutlu B, Roy N, {\v{S}}abanovi{\'{c}} S (2016) {Cognitive Human–Robot Interaction}. In: Springer Handbook of Robotics, vol Part G. Springer, Cham, Berlin, chap~71, p 1907--1934, \doi{10.1007/978-3-319-32552-1{\_}71}

\bibitem[{Na et~al(2023)Na, Choi, and Kang}]{Na2023}
Na G, Choi J, Kang H (2023) {It’s Not My Fault, But I’m to Blame: The Effect of a Home Robot’s Attribution and Approach Movement on Trust and Emotion of Users}. International Journal of Human-Computer Interaction \doi{10.1080/10447318.2023.2209977}

\bibitem[{Nass et~al(1994)Nass, Steuer, and Tauber}]{Nass1994}
Nass C, Steuer J, Tauber ER (1994) {Computers are social actors}. In: Conference companion on Human factors in computing systems - CHI '94. ACM Press, New York, New York, USA, p 204, \doi{10.1145/259963.260288}

\bibitem[{Natarajan and Gombolay(2020)}]{Natarajan2020}
Natarajan M, Gombolay M (2020) {Effects of Anthropomorphism and Accountability on Trust in Human Robot Interaction}. In: Proceedings of the 2020 ACM/IEEE International Conference on Human-Robot Interaction. ACM, New York, NY, USA, pp 33--42, \doi{10.1145/3319502.3374839}

\bibitem[{Nayyar and Wagner(2018)}]{Nayyar2018}
Nayyar M, Wagner AR (2018) {When should a robot apologize? understanding how timing affects human-robot trust repair}. In: Lecture Notes in Computer Science, \doi{10.1007/978-3-030-05204-1{\_}26}

\bibitem[{Nesset et~al(2023)Nesset, Romeo, Rajendran, and Hastie}]{Nesset2023}
Nesset B, Romeo M, Rajendran G, et~al (2023) {Robot Broken Promise? Repair strategies for mitigating loss of trust for repeated failures}. In: 2023 32nd IEEE International Conference on Robot and Human Interactive Communication (RO-MAN). IEEE, pp 1389--1395, \doi{10.1109/RO-MAN57019.2023.10309558}

\bibitem[{Nomura et~al(2008)Nomura, Kanda, Suzuki, and Kato}]{Nomura2008}
Nomura T, Kanda T, Suzuki T, et~al (2008) {Prediction of human behavior in human - Robot interaction using psychological scales for anxiety and negative attitudes toward robots}. IEEE Transactions on Robotics 24(2). \doi{10.1109/TRO.2007.914004}

\bibitem[{Ogiermann(2008)}]{Ogiermann2008}
Ogiermann E (2008) {On the culture-specificity of linguistic gender differences: The case of English and Russian apologies}. Intercultural Pragmatics 5(3):259--286. \doi{10.1515/IPRG.2008.013}

\bibitem[{Okada et~al(2023)Okada, Kimoto, Iio, Shimohara, and Shiomi}]{Okada2023}
Okada Y, Kimoto M, Iio T, et~al (2023) {Two is better than one: Apologies from two robots are preferred}. PLOS ONE 18(2):e0281604. \doi{10.1371/journal.pone.0281604}

\bibitem[{Okoli(2015)}]{Okoli2015}
Okoli C (2015) {A Guide to Conducting a Standalone Systematic Literature Review}. Communications of the Association for Information Systems 37. \doi{10.17705/1CAIS.03743}

\bibitem[{van Over et~al(2020)van Over, Winter, Molina-Markham, Lie, and Carbaugh}]{vanOver2020}
van Over B, Winter U, Molina-Markham E, et~al (2020) {Apologies in In-Car Speech Technologies}. In: Communication in Vehicles. chap~7, p 103--119, \doi{10.1515/9783110519006-007}

\bibitem[{Pak and Rovira(2023)}]{Pak2023}
Pak R, Rovira E (2023) {A theoretical model to explain mixed effects of trust repair strategies in autonomous systems}. \doi{10.1080/1463922X.2023.2250424}

\bibitem[{Park et~al(2012)Park, MacDonald, and Khoo}]{Park2012}
Park SJ, MacDonald CM, Khoo M (2012) {Do you care if a computer says sorry?} In: Proceedings of the Designing Interactive Systems Conference. ACM, New York, NY, USA, pp 731--740, \doi{10.1145/2317956.2318067}

\bibitem[{Pei et~al(2023)Pei, Huang, Wang, Yang, Xie, and Shen}]{Pei2023}
Pei Y, Huang R, Wang G, et~al (2023) {Multimodal Apology: Using WebXR to Repair Trust with Virtual Companion}. In: Proceedings - 2023 IEEE Conference on Virtual Reality and 3D User Interfaces Abstracts and Workshops, VRW 2023. IEEE, pp 727--728, \doi{10.1109/VRW58643.2023.00206}

\bibitem[{Pereira and Lopes(2020)}]{Pereira2020}
Pereira LM, Lopes AB (2020) {To Grant Decision-Making to Machines? Who Can and Should Apologize?} In: Studies in Applied Philosophy, Epistemology and Rational Ethics, vol~53. Springer Nature Switzerland, p 113--120, \doi{10.1007/978-3-030-39630-5{\_}16}

\bibitem[{Pereira et~al(2022)Pereira, Han, and Lopes}]{Pereira2022}
Pereira LM, Han TA, Lopes AB (2022) {Employing AI to Better Understand Our Morals}. Entropy 24(1). \doi{10.3390/e24010010}

\bibitem[{Perkins et~al(2021)Perkins, Khavas, and Robinette}]{Perkins2021}
Perkins R, Khavas ZR, Robinette P (2021) {Trust Calibration and Trust Respect: A Method for Building Team Cohesion in Human Robot Teams}. [Unpublished Manuscript]

\bibitem[{Perkins et~al(2022)Perkins, Khavas, McCallum, Kotturu, and Robinette}]{Perkins2022}
Perkins R, Khavas ZR, McCallum K, et~al (2022) {The Reason for an Apology Matters for Robot Trust Repair}. In: Lecture Notes in Computer Science, vol 13818 LNAI. Springer Science and Business Media Deutschland GmbH, pp 640--651, \doi{10.1007/978-3-031-24670-8{\_}56}

\bibitem[{Petty et~al(1981)Petty, Cacioppo, and Goldman}]{Petty1981}
Petty RE, Cacioppo JT, Goldman R (1981) {Personal involvement as a determinant of argument-based persuasion.} Journal of Personality and Social Psychology 41(5):847--855. \doi{10.1037/0022-3514.41.5.847}

\bibitem[{Pitardi and Marriott(2021)}]{Pitardi2020}
Pitardi V, Marriott HR (2021) {Alexa, <i>she's</i> not human but{\ldots} Unveiling the drivers of consumers' trust in voice‐based artificial intelligence}. Psychology {\&} Marketing 38(4):626--642. \doi{10.1002/mar.21457}

\bibitem[{Pompe et~al(2022)Pompe, Velner, and Truong}]{Pompe2022}
Pompe BL, Velner E, Truong KP (2022) {The Robot That Showed Remorse: Repairing Trust with a Genuine Apology}. In: 2022 31st IEEE International Conference on Robot and Human Interactive Communication (RO-MAN). IEEE, pp 260--265, \doi{10.1109/RO-MAN53752.2022.9900860}

\bibitem[{Porra et~al(2020)Porra, Lacity, and Parks}]{Porra2020}
Porra J, Lacity M, Parks MS (2020) {“Can Computer Based Human-Likeness Endanger Humanness?” – A Philosophical and Ethical Perspective on Digital Assistants Expressing Feelings They Can’t Have”}. Information Systems Frontiers 22(3):533--547. \doi{10.1007/s10796-019-09969-z}

\bibitem[{Rakova et~al(2023)Rakova, Shelby, and Ma}]{Rakova2023}
Rakova B, Shelby R, Ma M (2023) {Terms-we-serve-with: Five dimensions for anticipating and repairing algorithmic harm}. Big Data {\&} Society 10(2). \doi{10.1177/20539517231211553}

\bibitem[{Reason(1990)}]{Reason1990}
Reason J (1990) {Human Error}. April, Cambridge University Press, \doi{10.1017/CBO9781139062367}

\bibitem[{Rebensky et~al(2021)Rebensky, Carmody, Ficke, Nguyen, Carroll, Wildman, and Thayer}]{Rebensky2021}
Rebensky S, Carmody K, Ficke C, et~al (2021) {Whoops! Something Went Wrong: Errors, Trust, and Trust Repair Strategies in Human Agent Teaming}. In: Lecture Notes in Computer Science, \doi{10.1007/978-3-030-77772-2{\_}7}

\bibitem[{Robert(2021)}]{Robert2021}
Robert L (2021) {A Measurement of Attitude toward Working with Robots (AWRO): A Compare and Contrast Study of AWRO with Negative Attitude toward Robots (NARS)}. SSRN Electronic Journal \doi{10.2139/ssrn.3879544}

\bibitem[{Robinette et~al(2015)Robinette, Howard, and Wagner}]{Robinette2015}
Robinette P, Howard AM, Wagner AR (2015) {Timing is key for robot trust repair}. In: Lecture Notes in Computer Science, pp 574--583, \doi{10.1007/978-3-319-25554-5{\_}57}

\bibitem[{Robinette et~al(2017)Robinette, Howard, and Wagner}]{Robinette2017}
Robinette P, Howard AM, Wagner AR (2017) {Effect of Robot Performance on Human–Robot Trust in Time-Critical Situations}. IEEE Transactions on Human-Machine Systems 47(4):425--436. \doi{10.1109/THMS.2017.2648849}

\bibitem[{Rogers et~al(2023)Rogers, Webber, and Howard}]{Rogers2023}
Rogers K, Webber RJA, Howard A (2023) {Lying About Lying: Examining Trust Repair Strategies After Robot Deception in a High-Stakes HRI Scenario}. In: ACM/IEEE International Conference on Human-Robot Interaction, \doi{10.1145/3568294.3580178}

\bibitem[{Roschk and Kaiser(2013)}]{Roschk2013}
Roschk H, Kaiser S (2013) {The nature of an apology: An experimental study on how to apologize after a service failure}. Marketing Letters 24(3):293--309. \doi{10.1007/s11002-012-9218-x}

\bibitem[{Rzepka and Berger(2018)}]{Rzepka2018}
Rzepka C, Berger B (2018) {User Interaction with AI-enabled Systems: A Systematic Review of IS Research}. ICIS 2018 Proceedings

\bibitem[{Sagheb et~al(2023)Sagheb, Mun, Ahmadian, Christie, Bajcsy, Driggs-Campbell, and Losey}]{Sagheb2023}
Sagheb S, Mun YJ, Ahmadian N, et~al (2023) {Towards Robots that Influence Humans over Long-Term Interaction}. In: 2023 IEEE International Conference on Robotics and Automation (ICRA). IEEE, pp 7490--7496, \doi{10.1109/ICRA48891.2023.10160321}

\bibitem[{Salem et~al(2015)Salem, Lakatos, Amirabdollahian, and Dautenhahn}]{Salem2015}
Salem M, Lakatos G, Amirabdollahian F, et~al (2015) {Would You Trust a (Faulty) Robot?} In: Proceedings of the Tenth Annual ACM/IEEE International Conference on Human-Robot Interaction, vol 2015-March. ACM, New York, NY, USA, pp 141--148, \doi{10.1145/2696454.2696497}

\bibitem[{Schaefer(2016)}]{Schaefer2016}
Schaefer KE (2016) {Measuring Trust in Human Robot Interactions: Development of the “Trust Perception Scale-HRI”}. In: Robust Intelligence and Trust in Autonomous Systems. Springer US, Boston, MA, p 191--218, \doi{10.1007/978-1-4899-7668-0{\_}10}

\bibitem[{Schelble et~al(2022)Schelble, Lopez, Textor, Zhang, McNeese, Pak, and Freeman}]{Schelble2022}
Schelble BG, Lopez J, Textor C, et~al (2022) {Towards Ethical AI: Empirically Investigating Dimensions of AI Ethics, Trust Repair, and Performance in Human-AI Teaming}. Human Factors pp 1--19. \doi{10.1177/00187208221116952}

\bibitem[{Scher and Darley(1997)}]{Scher1997}
Scher SJ, Darley JM (1997) {How Effective Are the Things People Say to Apologize? Effects of the Realization of the Apology Speech Act}. Journal of Psycholinguistic Research 26(1):127--140. \doi{10.1023/A:1025068306386}

\bibitem[{Schlenker and Darby(1981)}]{Schlenker1981}
Schlenker BR, Darby BW (1981) {The Use of Apologies in Social Predicaments}. Social Psychology Quarterly 44(3):271. \doi{10.2307/3033840}

\bibitem[{Schnable et~al(2022)Schnable, DeMattee, Robinson, Brass, and Longhofer}]{Schnable2021}
Schnable A, DeMattee AJ, Robinson RS, et~al (2022) {The Multi-method Comprehensive Review: Synthesis and Analysis when Scholarship is International, Interdisciplinary, and Immense}. Voluntas 33(6):1219--1227. \doi{10.1007/s11266-021-00388-w}

\bibitem[{Schweitzer et~al(2006)Schweitzer, Hershey, and Bradlow}]{Schweitzer2006}
Schweitzer ME, Hershey JC, Bradlow ET (2006) {Promises and lies: Restoring violated trust}. Organizational Behavior and Human Decision Processes 101(1):1--19. \doi{10.1016/j.obhdp.2006.05.005}

\bibitem[{Searle(1969)}]{Searle1969}
Searle J (1969) {Speech Acts: An Essay in the Philosophy of Language}. the Syndics of the Cambridge University Press

\bibitem[{Sebo et~al(2019)Sebo, Krishnamurthi, and Scassellati}]{Sebo2019}
Sebo SS, Krishnamurthi P, Scassellati B (2019) {“I Don't Believe You”: Investigating the Effects of Robot Trust Violation and Repair}. In: 2019 14th ACM/IEEE International Conference on Human-Robot Interaction (HRI). IEEE, pp 57--65, \doi{10.1109/HRI.2019.8673169}

\bibitem[{Semeraro et~al(2022)Semeraro, Griffiths, and Cangelosi}]{Semeraro2022}
Semeraro F, Griffiths A, Cangelosi A (2022) {Human–robot collaboration and machine learning: A systematic review of recent research}. Robotics and Computer-Integrated Manufacturing 79(July 2022):102432. \doi{10.1016/j.rcim.2022.102432}

\bibitem[{SharifHeravi et~al(2020)SharifHeravi, Taylor, Stanton, Lambeth, and Shanahan}]{SharifHeravi2020}
SharifHeravi M, Taylor JR, Stanton CJ, et~al (2020) {It’s a disaster! Factors affecting trust development and repair following agent task failure}. In: Australasian Conference on Robotics and Automation, ACRA

\bibitem[{Shen and Wang(2022)}]{Shen2022}
Shen W, Wang Y (2022) {Facilitation of Customer Empathy: The Effect of Robot Apology on Customer Reaction Following a Service Failure}. Journal of Marketing Development and Competitiveness 16(2). \doi{10.33423/jmdc.v16i2.5254}

\bibitem[{Shiomi et~al(2013)Shiomi, Nakagawa, and Hagita}]{Shiomi2013}
Shiomi M, Nakagawa K, Hagita N (2013) {Design of a gaze behavior at a small mistake moment for a robot}. Interaction Studies Social Behaviour and Communication in Biological and Artificial Systems 14(3):317--328. \doi{10.1075/is.14.3.01shi}

\bibitem[{Shneiderman(2020)}]{Shneiderman2020}
Shneiderman B (2020) {Bridging the gap between ethics and practice: Guidelines for reliable, safe, and trustworthy human-centered AI systems}. ACM Transactions on Interactive Intelligent Systems 10(4). \doi{10.1145/3419764}

\bibitem[{Siau and Wang(2020)}]{Siau2020}
Siau K, Wang W (2020) {Artificial Intelligence (AI) Ethics}. Journal of Database Management 31(2):74--87. \doi{10.4018/JDM.2020040105}

\bibitem[{Singh et~al(1993)Singh, Molloy, and Parasuraman}]{Singh1993}
Singh IL, Molloy R, Parasuraman R (1993) {Automation- Induced "Complacency": Development of the Complacency-Potential Rating Scale}. The International Journal of Aviation Psychology 3(2):111--122. \doi{10.1207/s15327108ijap0302{\_}2}

\bibitem[{Slocum et~al(2011)Slocum, Allan, and Allan}]{Slocum2011}
Slocum D, Allan A, Allan MM (2011) {An emerging theory of apology}. Australian Journal of Psychology 63(2):83--92. \doi{10.1111/j.1742-9536.2011.00013.x}

\bibitem[{Smith(2005)}]{Smith2005}
Smith N (2005) {The Categorical Apology}. Journal of Social Philosophy 36(4):473--496. \doi{10.1111/j.1467-9833.2005.00289.x}

\bibitem[{Smith(2008)}]{Smith2008}
Smith N (2008) {I Was Wrong: The Meanings of Apologies}. Cambridge University Press, Cambridge

\bibitem[{Soares and Fallenstein(2014)}]{Soares2014}
Soares N, Fallenstein B (2014) {Aligning Superintelligence with Human Interests: A Technical Research Agenda}. Tech. rep., Machine Intelligence Research Institute (MIRI) technical report

\bibitem[{Song et~al(2023)Song, Zhang, Xing, and Duan}]{Song2023}
Song M, Zhang H, Xing X, et~al (2023) {Appreciation vs. apology: Research on the influence mechanism of chatbot service recovery based on politeness theory}. Journal of Retailing and Consumer Services 73:103323. \doi{10.1016/j.jretconser.2023.103323}

\bibitem[{Stiber et~al(2023)Stiber, Taylor, and Huang}]{Stiber2023}
Stiber M, Taylor RH, Huang CM (2023) {On using social signals to enable flexible error-aware HRI}. In: ACM/IEEE International Conference on Human-Robot Interaction. IEEE Computer Society, pp 222--230, \doi{10.1145/3568162.3576990}

\bibitem[{Stock-Homburg(2022)}]{Stock-Homburg2021}
Stock-Homburg R (2022) {Survey of Emotions in Human–Robot Interactions: Perspectives from Robotic Psychology on 20 Years of Research}. International Journal of Social Robotics 14(2). \doi{10.1007/s12369-021-00778-6}

\bibitem[{Stowers(2017)}]{Stowers2017}
Stowers K (2017) {The Role of Accounts and Apologies in Mitigating Blame toward Human and Machine Agents}. PhD thesis, University of Central Florida

\bibitem[{Stratton(2016)}]{Stratton2016}
Stratton SJ (2016) {Comprehensive Reviews}. \doi{10.1017/S1049023X16000649}

\bibitem[{Strawson(1974)}]{Strawson1974}
Strawson PF (1974) {Freedom and Resentment}. In: Freedom and Resentment and Other Essays. Methuen

\bibitem[{Strohkorb~Sebo et~al(2018)Strohkorb~Sebo, Traeger, Jung, and Scassellati}]{Sebo2018}
Strohkorb~Sebo S, Traeger M, Jung M, et~al (2018) {The Ripple Effects of Vulnerability}. In: Proceedings of the 2018 ACM/IEEE International Conference on Human-Robot Interaction. ACM, New York, NY, USA, pp 178--186, \doi{10.1145/3171221.3171275}

\bibitem[{Syrdal et~al(2009)Syrdal, Dautenhahn, Koay, and Walters}]{Syrdal2009}
Syrdal DS, Dautenhahn K, Koay KL, et~al (2009) {The Negative Attitudes Towards Robots Scale and reactions to robot behaviour in a live Human-Robot Interaction study}. In: Adaptive and Emergent Behaviour and Complex Systems - Proceedings of the 23rd Convention of the Society for the Study of Artificial Intelligence and Simulation of Behaviour, AISB 2009, pp 109--115

\bibitem[{Tavuchis(1993)}]{Tavuchis1993}
Tavuchis N (1993) {Mea culpa: A sociology of apology and reconciliation}. Stanford University Press

\bibitem[{Tax et~al(1998)Tax, Brown, and Chandrashekaran}]{Tax1998}
Tax SS, Brown SW, Chandrashekaran M (1998) {Customer Evaluations of Service Complaint Experiences: Implications for Relationship Marketing}. Journal of Marketing 62(2):60--76. \doi{10.1177/002224299806200205}

\bibitem[{Taylor et~al(2020)Taylor, Yudkowsky, LaVictoire, and Critch}]{Taylor2020}
Taylor J, Yudkowsky E, LaVictoire P, et~al (2020) {Alignment for Advanced Machine Learning Systems}. In: Ethics of Artificial Intelligence. Oxford University Press, p 342--382, \doi{10.1093/oso/9780190905033.003.0013}

\bibitem[{Tewari and Lindgren(2022)}]{Tewari2022}
Tewari M, Lindgren H (2022) {Expecting, understanding, relating, and interacting-older, middle-aged and younger adults’ perspectives on breakdown situations in human–robot dialogues}. Frontiers in Robotics and AI 9. \doi{10.3389/frobt.2022.956709}

\bibitem[{Textor et~al(2022)Textor, Zhang, Lopez, Schelble, McNeese, Freeman, Pak, Tossell, and de~Visser}]{Textor2022}
Textor C, Zhang R, Lopez J, et~al (2022) {Exploring the Relationship Between Ethics and Trust in Human–Artificial Intelligence Teaming: A Mixed Methods Approach}. Journal of Cognitive Engineering and Decision Making 16(4):252--281. \doi{10.1177/15553434221113964}

\bibitem[{Tigard(2021)}]{Tigard2021}
Tigard DW (2021) {Technological Answerability and the Severance Problem: Staying Connected by Demanding Answers}. Science and Engineering Ethics 27(5):59. \doi{10.1007/s11948-021-00334-5}

\bibitem[{Tolmeijer et~al(2020)Tolmeijer, Weiss, Hanheide, Lindner, Powers, Dixon, and Tielman}]{Tolmeijer2020}
Tolmeijer S, Weiss A, Hanheide M, et~al (2020) {Taxonomy of Trust-Relevant Failures and Mitigation Strategies}. In: Proceedings of the 2020 ACM/IEEE International Conference on Human-Robot Interaction. ACM, New York, NY, USA, pp 3--12, \doi{10.1145/3319502.3374793}

\bibitem[{Tomlinson et~al(2020)Tomlinson, Schnackenberg, Dawley, and Ash}]{Tomlinson2020}
Tomlinson EC, Schnackenberg AK, Dawley D, et~al (2020) {Revisiting the trustworthiness–trust relationship: Exploring the differential predictors of cognition‐ and affect‐based trust}. Journal of Organizational Behavior 41(6):535--550. \doi{10.1002/job.2448}

\bibitem[{Tsakalakis et~al(2022)Tsakalakis, Stalla-Bourdillon, Huynh, and Moreau}]{Tsakalakis2022}
Tsakalakis N, Stalla-Bourdillon S, Huynh TD, et~al (2022) {A taxonomy of explanations to support Explainability-by-Design}

\bibitem[{Tzeng(2004)}]{Tzeng2004}
Tzeng JY (2004) {Toward a more civilized design: Studying the effects of computers that apologize}. International Journal of Human Computer Studies 61(3). \doi{10.1016/j.ijhcs.2004.01.002}

\bibitem[{Ullrich et~al(2021)Ullrich, Butz, and Diefenbach}]{Ullrich2021}
Ullrich D, Butz A, Diefenbach S (2021) {The Development of Overtrust: An Empirical Simulation and Psychological Analysis in the Context of Human–Robot Interaction}. Frontiers in Robotics and AI 8

\bibitem[{Vamplew et~al(2018)Vamplew, Dazeley, Foale, Firmin, and Mummery}]{Vamplew2018}
Vamplew P, Dazeley R, Foale C, et~al (2018) {Human-aligned artificial intelligence is a multiobjective problem}. Ethics and Information Technology 20(1). \doi{10.1007/s10676-017-9440-6}

\bibitem[{de~Visser et~al(2018)de~Visser, Pak, and Shaw}]{deVisser2018}
de~Visser EJ, Pak R, Shaw TH (2018) {From ‘automation’ to ‘autonomy’: the importance of trust repair in human–machine interaction}. Ergonomics 61(10):1409--1427. \doi{10.1080/00140139.2018.1457725}

\bibitem[{de~Visser et~al(2020)de~Visser, Peeters, Jung, Kohn, Shaw, Pak, and Neerincx}]{deVisser2020}
de~Visser EJ, Peeters MM, Jung MF, et~al (2020) {Towards a Theory of Longitudinal Trust Calibration in Human–Robot Teams}. International Journal of Social Robotics 12(2). \doi{10.1007/s12369-019-00596-x}

\bibitem[{Wagatsuma and Rosett(1986)}]{Wagatsuma1986}
Wagatsuma H, Rosett A (1986) {The Implications of Apology: Law and Culture in Japan and the United States}. Law {\&} Society Review 20

\bibitem[{Wagner and Robinette(2021)}]{Wagner2020}
Wagner AR, Robinette P (2021) {An explanation is not an excuse: Trust calibration in an age of transparent robots}. In: Trust in Human-Robot Interaction. Elsevier, p 197--208, \doi{10.1016/B978-0-12-819472-0.00009-5}

\bibitem[{Wang et~al(2018)Wang, Pynadath, Rovira, Barnes, and Hill}]{Wang2018}
Wang N, Pynadath DV, Rovira E, et~al (2018) {Is it my looks? Or something i said? The impact of explanations, embodiment, and expectations on trust and performance in human-robot teams}. In: Lecture Notes in Computer Science, \doi{10.1007/978-3-319-78978-1{\_}5}

\bibitem[{Wang et~al(2023)Wang, Hwang, and Guchait}]{Wang2021}
Wang X, Hwang Y, Guchait P (2023) {When Robot (Vs. Human) Employees Say “Sorry” Following Service Failure}. International Journal of Hospitality {\&} Tourism Administration 24(4):540--562. \doi{10.1080/15256480.2021.2017812}

\bibitem[{Weiler et~al(2022)Weiler, Matt, and Hess}]{Weiler2022}
Weiler S, Matt C, Hess T (2022) {Immunizing with information – Inoculation messages against conversational agents’ response failures}. Electronic Markets 32(1):239--258. \doi{10.1007/s12525-021-00509-9}

\bibitem[{Wester et~al(2023)Wester, Lee, and Van~Berkel}]{Wester2023}
Wester J, Lee M, Van~Berkel N (2023) {Moral Transparency as a Mitigator of Moral Bias in Conversational User Interfaces}. In: Proceedings of the 5th International Conference on Conversational User Interfaces, CUI 2023. ACM, \doi{10.1145/3571884.3603752}

\bibitem[{Weun et~al(2004)Weun, Beatty, and Jones}]{Weun2004}
Weun S, Beatty SE, Jones MA (2004) {The impact of service failure severity on service recovery evaluations andpost‐recovery relationships}. Journal of Services Marketing 18(2):133--146. \doi{10.1108/08876040410528737}

\bibitem[{Wilson et~al(2020)Wilson, Mechenbier, and Melon{\c{c}}on}]{Wilson2020}
Wilson L, Mechenbier MX, Melon{\c{c}}on L (2020) {Affective Investment}. Academic Labor: Research and Artistry 4(6)

\bibitem[{Winkle et~al(2022)Winkle, Jackson, Melsion, Brscic, Leite, and Williams}]{Winkle2022}
Winkle K, Jackson RB, Melsion GI, et~al (2022) {Norm-Breaking Responses to Sexist Abuse: A Cross-Cultural Human Robot Interaction Study}. In: 2022 17th ACM/IEEE International Conference on Human-Robot Interaction (HRI). IEEE, pp 120--129, \doi{10.1109/HRI53351.2022.9889389}

\bibitem[{Wischnewski et~al(2023)Wischnewski, Kr{\"{a}}mer, and M{\"{u}}ller}]{Wischnewski2023}
Wischnewski M, Kr{\"{a}}mer N, M{\"{u}}ller E (2023) {Measuring and Understanding Trust Calibrations for Automated Systems: A Survey of the State-Of-The-Art and Future Directions}. In: Conference on Human Factors in Computing Systems - Proceedings. ACM, \doi{10.1145/3544548.3581197}

\bibitem[{Wright et~al(2022)Wright, Collins, and Cameron}]{Wright2022}
Wright B, Collins E, Cameron D (2022) {When Robots Interact with Groups, Where Does the Trust Reside?} [Unpublished Manuscript]

\bibitem[{Xu and Howard(2022)}]{XuJ2022}
Xu J, Howard A (2022) {Evaluating the Impact of Emotional Apology on Human-Robot Trust}. In: 2022 31st IEEE International Conference on Robot and Human Interactive Communication (RO-MAN). IEEE, pp 1655--1661, \doi{10.1109/RO-MAN53752.2022.9900518}

\bibitem[{Xu and Liu(2022)}]{XuX2022}
Xu X, Liu J (2022) {Artificial intelligence humor in service recovery}. Annals of Tourism Research 95:103439. \doi{10.1016/j.annals.2022.103439}

\bibitem[{Xu et~al(2023)Xu, Wang, Zhai, and Liu}]{Xu2023}
Xu Z, Wang G, Zhai S, et~al (2023) {When Automation Fails: Examining the Effect of a Verbal Recovery Strategy on User Experience in Automated Driving}. International Journal of Human–Computer Interaction pp 1--11. \doi{10.1080/10447318.2023.2176986}

\bibitem[{Yang et~al(2022)Yang, Xu, Zhang, Liang, and Lyu}]{Yang2022}
Yang H, Xu H, Zhang Y, et~al (2022) {Exploring the effect of humor in robot failure}. Annals of Tourism Research 95:103425. \doi{10.1016/j.annals.2022.103425}

\bibitem[{Yang et~al(2023)Yang, Zhou, and Yang}]{Yang2023}
Yang Z, Zhou J, Yang H (2023) {The Impact of AI’s Response Method on Service Recovery Satisfaction in the Context of Service Failure}. Sustainability 15(4):3294. \doi{10.3390/su15043294}

\bibitem[{You et~al(2020)You, Yang, Wang, and Deng}]{You2020}
You Y, Yang X, Wang L, et~al (2020) {When and Why Saying “Thank You” Is Better Than Saying “Sorry” in Redressing Service Failures: The Role of Self-Esteem}. Journal of Marketing 84(2):133--150. \doi{10.1177/0022242919889894}

\bibitem[{Yu et~al(2023)Yu, Li, Su, and Fuoli}]{Yu2023}
Yu D, Li L, Su H, et~al (2023) {Assessing the potential of AI-assisted pragmatic annotation: The case of apologies}. [Unpublished Manuscript]

\bibitem[{Yu et~al(2022)Yu, Xiong, and Shen}]{YuS2022}
Yu S, Xiong JJ, Shen H (2022) {The rise of chatbots: The effect of using chatbot agents on consumers' responses to request rejection}. Journal of Consumer Psychology \doi{10.1002/jcpy.1330}

\bibitem[{Zhang et~al(2023{\natexlab{a}})Zhang, Zhu, Wu, and Yu-Buck}]{Zhang2023}
Zhang J, Zhu Y, Wu J, et~al (2023{\natexlab{a}}) {A natural apology is sincere: Understanding chatbots' performance in symbolic recovery}. International Journal of Hospitality Management 108:103387. \doi{10.1016/j.ijhm.2022.103387}

\bibitem[{Zhang(2023)}]{Zhang2023a}
Zhang X (2023) {“Sorry, It Was My Fault”: Repairing Trust in Human-Robot Interactions}. International Journal of Human - Computer Studies 175(103031):79

\bibitem[{Zhang et~al(2023{\natexlab{b}})Zhang, Lee, Maeng, and Hahn}]{ZhangX2023b}
Zhang X, Lee SK, Maeng H, et~al (2023{\natexlab{b}}) {Effects of Failure Types on Trust Repairs in Human–Robot Interactions}. International Journal of Social Robotics 15(9-10):1619--1635. \doi{10.1007/s12369-023-01059-0}

\bibitem[{Zhou et~al(2020)Zhou, Chen, Berry, Reed, Zhang, and Savage}]{Zhou2020}
Zhou J, Chen F, Berry A, et~al (2020) {A Survey on Ethical Principles of AI and Implementations}. In: 2020 IEEE Symposium Series on Computational Intelligence (SSCI). IEEE, pp 3010--3017, \doi{10.1109/SSCI47803.2020.9308437}

\bibitem[{Zhu et~al(2023)Zhu, Zhang, and Wu}]{Zhu2023}
Zhu Y, Zhang J, Wu J (2023) {Who did what and when? The effect of chatbots’ service recovery on customer satisfaction and revisit intention}. Journal of Hospitality and Tourism Technology 14(3):416--429. \doi{10.1108/JHTT-06-2021-0164}

\end{thebibliography}

\end{document}